\documentclass[lettersize,journal]{IEEEtran}
\usepackage{amsmath,amsfonts}
\usepackage{algorithm}
\usepackage{array}
\usepackage[caption=false,font=normalsize,labelfont=sf,textfont=sf]{subfig}
\usepackage{textcomp}
\usepackage{stfloats}
\usepackage{url}
\usepackage{verbatim}
\usepackage{graphicx}
\usepackage{cite}
\usepackage[table,xcdraw]{xcolor}
\usepackage{algpseudocode}
\usepackage{gensymb}
\usepackage{adjustbox}
\usepackage{tikz}
\usetikzlibrary{arrows.meta, positioning}
\usepackage{amssymb}
\usepackage{booktabs}
\usepackage{threeparttable}
\usepackage{makecell}
\usepackage{enumitem}

\makeatletter
\renewcommand{\Function}[2]{%
  \csname ALG@cmd@\ALG@L @Function\endcsname{#1}{#2}%
  \def\jayden@currentfunction{#1}%
}
\newcommand{\funclabel}[1]{%
  \@bsphack
  \protected@write\@auxout{}{%
    \string\newlabel{#1}{{\jayden@currentfunction}{\thepage}}%
  }%
  \@esphack
}
\makeatother

\begin{document}

\title{Design and Performance Evaluation of Secure RF and WiFi-Based Communication in Drone Swarms via Testbed Implementation}

\author{Bhavya Dixit, Aayushi Rajgor, Subham Kumar, Rushikesh Patil, Ananthapadmanabhan A., Gaurav S. Kasbekar, and Arnab Maity
\thanks{B. Dixit, A. Rajgor, Ananthapadmanabhan A., and
 G. S. Kasbekar are with the Department of Electrical Engineering, Indian Institute of Technology (IIT) Bombay, Mumbai 400076, Maharashtra, India.  S. Kumar, R. Patil, and A. Maity are with the Department of Aerospace Engineering, IIT Bombay. Their email addresses are bdixit9@gmail.com, aayushirajgor98@gmail.com, anantha9102002@gmail.com, gskasbekar@ee.iitb.ac.in,  subhamk642@gmail.com, rishipatil7007@gmail.com, and arnab.maity@iitb.ac.in, respectively. The contributions of the authors have been supported in part by the project with code RGSTC01-001.}
}

\maketitle

\begin{abstract}
Unmanned aerial vehicle (UAV) swarms rely on distributed coordination and cooperative communication to achieve scalability, extend operational range, and support complex applications such as persistent surveillance and real-time message transmission. Although wireless networks such as  Radio Frequency (RF) and WiFi are commonly used to enable UAV-to-UAV and UAV-to-Ground Control Station (GCS) communication, they introduce inherent security challenges. In particular, MAVLink, the predominant communication protocol in UAV systems, provides message integrity and authentication mechanisms, but lacks built-in encryption, leaving the transmitted telemetry messages susceptible to eavesdropping. In our prior work, we proposed the MAVShield cipher, which is  a lightweight encryption framework for MAVLink communications. In this study, five encryption algorithms-- MAVShield, Advanced Encryption Standard in counter mode (AES-CTR), Speck in counter mode (Speck-CTR), ChaCha20, and Rabbit-- are integrated into the communication architecture of four custom built UAVs to establish robust and secure communication links for UAV-to-UAV interactions over both RF and WiFi channels. The performance of the encryption algorithms is evaluated by conducting extensive flight tests using a UAV swarm hardware testbed. The encrypted telemetry data is utilized to achieve autonomous formation control and collision avoidance during flight operations conducted using the testbed. For collision avoidance, we propose a modified artificial potential field (APF) algorithm that computes attractive and repulsive forces directly in geodetic coordinates, eliminating the need for Cartesian transformations, and employs a gradually activated repulsive force to achieve smooth, jitter-free trajectories without local minimum trapping within the operating region of interest. Key performance metrics such as CPU utilization, memory consumption, and packet delivery ratio (PDR) are measured for each encryption scheme.  Our results show that despite incorporating additional security mechanisms, MAVShield maintains performance levels comparable to unencrypted communication, while outperforming the other four encryption schemes in terms of overall efficiency. Furthermore, algebraic cryptanalysis and Wireshark-based traffic analysis confirm its strong resistance to key recovery attacks and its ability to preserve telemetry confidentiality. Overall, the results indicate that MAVShield is an efficient and secure solution for UAV swarm communication.
\end{abstract}


\section{Introduction}
\IEEEPARstart{U}{nmanned} aerial vehicles (UAVs) are evolving as versatile platforms with diverse applications, including precise geographical localization, disaster tracking and monitoring, search and rescue missions, intelligent transportation systems,  relaying for Internet connectivity distribution, emergency response, and remote sensing \cite{shakhatreh2019unmanned,wu2024deep}. UAVs can communicate and fly autonomously without human intervention. Also, the UAVs of a drone swarm communicate via telemetry data links. Telemetry data shared between UAVs includes sensor readings, global positioning system (GPS) coordinates, power status, orientation, velocity, altitude, and other essential flight parameters which are critical for synchronized flight and situational awareness \cite{shakhatreh2019unmanned}. 

UAVs execute pre-planned autonomous missions guided by waypoint navigation, wherein each drone autonomously follows a designated flight path programmed before deployment. For UAV swarms to effectively carry out their missions, maintaining reliable and efficient communication throughout their operation is essential. Achieving this requires not only stable telemetry-based flight formations but also robust collision avoidance mechanisms to guarantee safe and dependable operations \cite{chung2018survey}. 

The primary modes of communication among UAVs include data transmission and reception through radio frequency (RF) devices and WiFi modules \cite{mozaffari2019tutorial}. RF devices provide reliable long-range communication, while WiFi modules enable higher data throughput. Together, these technologies significantly enhance the operational efficiency and flexibility of UAV systems \cite{yanmaz2018drone}.

Micro air vehicle link (MAVLink) is a lightweight, binary, message serialization-based communication protocol widely used in UAV systems \cite{koubaa2019micro,du2024exploiting}. Operating at the application layer, MAVLink serializes messages into compact binary packets, which are then passed to lower communication layers for network transmission. These messages can be transmitted over various media, including WiFi, Ethernet, or sub-GHz serial telemetry links operating at 433 MHz, 868 MHz, and 915 MHz \cite{koubaa2019micro}. In UAV swarm deployments, MAVLink enables inter-UAV and UAV-to-ground control station (GCS) communication for coordination, mission synchronization, and status reporting. Each UAV is assigned a unique system ID (SYSID) and component ID (COMPID), allowing multiple UAVs to operate within the same MAVLink network. The SYSID uniquely identifies a UAV, whereas the COMPID identifies the onboard component responsible for transmitting or receiving MAVLink messages. These identifiers ensure proper source discrimination in a multi-UAV swarm network.

However, securing MAVLink communication is crucial because swarms depend on it for cooperation, formation control \cite{bu2024advancement,lin2026leader}, and collision avoidance \cite{rotta2024secure}. The communication protocol may be either MAVLink v1.0 or MAVLink v2.0, with the latter providing enhancements against cyber threats, including message signing for cryptographic authentication and data integrity. Despite these improvements, MAVLink remains vulnerable to eavesdropping attacks since communication takes place over an unencrypted channel under the protocol  \cite{yaacoub2020cyber,rodday2016exploring, wang2012data,siddappaji2019role,hamza2024mavlink}. To address these threats and ensure full protection against packet sniffing, encryption is essential.

Although several studies have focused on improving the efficiency of UAV-to-UAV communications using different communication technologies \cite{fakhreddine2022experiments,li2018uav}, the confidentiality of such communications remains relatively underexplored, particularly in real-world operational environments \cite{rodday2016exploring,yaacoub2020cyber,wang2024survey,lopez2021towards,olsson2023using}. Moreover, despite the widespread adoption of MAVLink in UAV systems, limited research has investigated the implementation and experimental evaluation of encryption mechanisms on hardware testbeds. While \cite{dong2021securing} proposed a secure communication framework for UAV swarms, it incurs significant computational complexity. 
Similarly, \cite{thompson2016confidential} demonstrated confidentiality in large UAV swarms, but reported that authenticated encryption techniques such as synthetic initialization vector (SIV) and EAX saturate the processor entirely at $50$ UAVs. 
These limitations highlight the need for lightweight and experimentally validated secure communication frameworks that ensure data confidentiality in real UAV systems without significantly impacting system performance.

In our previous work \cite{Dxtbhavya}, we proposed the \emph{MAVShield} cipher, which is  a lightweight encryption framework for MAVLink communications that was shown to achieve lower computational and energy overhead than other encryption algorithms, including  Advanced Encryption Standard in counter mode (AES-CTR) \cite{yustiarini2022comparative}, ChaCha20 \cite{bernstein2008chacha}, Speck in counter mode (Speck-CTR) \cite{beaulieu2017notes}, and Rabbit \cite{boesgaard2008rabbit}, for bidirectional RF communications between a UAV and a GCS. However, its application to secure multi-UAV swarm operations incorporating formation control and collision avoidance was not investigated. To address this limitation, the present work integrates the MAVShield algorithm within a real UAV swarm testbed operating over both RF and WiFi communication modes.

Also, we employ a leader-follower formation control strategy, which follows a widely adopted multi-agent coordination paradigm \cite{oh2015survey,bu2024advancement,lin2026leader}, where follower UAVs compute target positions directly from the leader's GPS telemetry in geodetic coordinates without intermediate transformations; safe separation during such flight is crucial. Since conventional artificial potential field (APF) methods \cite{zhang2022improved,ma2021artificial,Budiyanto,Jiayi} operate in Cartesian coordinates, GPS measurements must be converted to a Local Tangent Plane and the resulting commands transformed back for GPS-based flight controllers, introducing additional overhead. To address this, we propose a modified APF-based collision avoidance scheme that computes attractive and repulsive forces directly in geodetic coordinates, extending conventional Cartesian APF formulations to a multi-UAV geodetic framework \cite{9538804}. This eliminates coordinate transformations, while a gradually activated distance-threshold repulsive force reduces trajectory jitter caused by abrupt force changes \cite{Jiayi}, enabling smooth collision avoidance without local minimum trapping in the evaluated swarm scenarios.

In this study, we implement and evaluate the performance of five encryption algorithms-- \emph{MAVShield} \cite{Dxtbhavya}, AES-CTR \cite{yustiarini2022comparative}, Speck-CTR \cite{beaulieu2017notes}, ChaCha20 \cite{bernstein2008chacha}, and Rabbit \cite{boesgaard2008rabbit}-- within the communication framework of a custom-built UAV testbed. The encryption algorithms are individually deployed within both RF and WiFi communication architectures. The workflow of our communication
framework  is as follows:
\begin{itemize}
    \item In a UAV-to-UAV communication scenario, telemetry data is encrypted using each of the five encryption schemes prior to transmission and subsequently decrypted by the receiving UAVs upon packet arrival.
    
    \item For both communication modes, RF and WiFi, flight tests are conducted by integrating formation control and collision avoidance mechanisms. The formation control strategy supports the execution of three different aerial formations during flight, whereas collision avoidance is achieved through a modified APF-based algorithm, enabling safe UAV operations. This collision avoidance scheme operates directly in geodetic coordinates, eliminating the need for Cartesian transformations. A gradually activated, distance threshold based repulsive force enables smooth, proactive collision avoidance, with experimental validation on a real UAV swarm testbed.
    
    \item Key performance metrics such as CPU utilization, memory consumption, and packet delivery ratio (PDR) are measured for each encryption scheme. 
    
    \item A comparative analysis of different security algorithms is conducted to assess the performance of each encryption algorithm across different communication modes. 
    Our results show that despite incorporating additional security mechanisms, MAVShield maintains performance levels comparable to unencrypted communication, while outperforming the other four encryption schemes in terms of overall efficiency.
    
    \item To assess its cryptanalytic strength, MAVShield is subjected to algebraic attacks employing algebraic normal form (ANF) representations \cite{carlet2021boolean,kutsenko2021algebraic}, Linearization \cite{courtois2003algebraic}, Elimination \cite{courtois2012elimlin, gharib2015system}, and Boolean Satisfiability (SAT)-based solving \cite{audemard2018glucose,tseitin1983complexity}. Our results indicate that the cipher withstands these attacks effectively, preventing full secret key recovery across all tested rounds despite the availability of four known plaintext-ciphertext pairs to the adversary. 

    \item To verify the confidentiality of telemetry data exchanges, the Wireshark packet analyzer \cite{sanders2017practical} is employed to monitor network traffic and examine transmitted packets for potential leakage of sensitive information. The analysis confirms that MAVLink payloads remain encrypted during transmission among swarm members, thereby preserving the confidentiality of inter-UAV communications.

\end{itemize}

The rest of the paper is organized as follows. Section \ref{sec:Related_work} provides a review of related work. Section \ref{sec:system_enc_dec_model} outlines the system model, problem formulation,
and encryption-decryption model. Section \ref{sec:UAV2UAVCOMM} presents the UAV-to-UAV communication framework, which includes autonomous air formation strategies, the collision avoidance mechanism, and the design and integration of an encryption-decryption model compatible with RF and WiFi communication links. Section \ref{sec:security_algorithm} provides detailed descriptions of the security algorithms employed in this study-- MAVShield, AES-CTR, Speck-CTR, ChaCha20, and Rabbit. Section \ref{sec:algebraic_crypt} provides an algebraic cryptanalysis of MAVShield to show its resistance to key recovery attacks. Section \ref{sec:experimental_setup} describes the experimental hardware and software setup as well as implementation of encryption-decryption processes for both communication modes. Next, Section \ref{sec:implementation} details the integration of the security algorithms and the execution of mission scenarios for both RF and WiFi based UAV-to-UAV communications. Section \ref{sec:wireshark} presents a security evaluation using the packet analysis tool Wireshark to validate the confidentiality of transmitted data. A detailed comparative analysis of the various encryption schemes in UAV-to-UAV communication is provided in Section \ref{sec:performance_eval}, which highlights their performance differences and trade-offs under RF and WiFi communication. Lastly, Section \ref{sec:conclusions} concludes the paper and outlines some directions for future research.

\section{Related Work} \label{sec:Related_work}
UAV-enabled mesh networks operate in highly dynamic and infrastructure-less environments, often lacking centralized coordination, which poses significant challenges to both connectivity and security. Within the context of UAV swarm deployments, mesh networking is considered a viable approach to facilitate decentralized communication. However, these networks are susceptible to security vulnerabilities across multiple layers of the communication stack. To mitigate cyber threats at each of these layers, key defense mechanisms such as encryption, intrusion detection systems (IDS), and firewalls are employed, among which encryption plays a pivotal role \cite{lopez2021towards,wang2023survey}. 

The surveys in \cite{wang2024survey,javed2024state} systematically categorize major security threats in UAV swarms, arising primarily from the open and broadcast nature of wireless communication channels. This inherent vulnerability exposes UAV swarms to passive attacks such as eavesdropping, as well as active attacks including jamming, which can compromise both data confidentiality and control integrity. To mitigate these threats, the surveys outline a range of countermeasures, including cryptographic techniques, physical-layer security approaches, machine learning-based defenses, blockchain frameworks, and IDSs. Existing studies have further identified protocol-specific weaknesses in MAVLink based systems: \cite{rodday2016exploring} exposed the absence of payload encryption as a critical gap enabling real-time eavesdropping, while \cite{yaacoub2020cyber} cataloged jamming, spoofing, and data interception as the most prevalent threats across UAV cyber-physical system vulnerabilities, which directly establish a strong empirical basis for the encryption-centric approach adopted in this study.

In \cite{fakhreddine2022experiments}, the authors conducted real-world UAV-to-UAV experiments comparing WiFi (802.11ac), long term evolution-advanced (LTE-A), and 5G for throughput and latency. Their findings show that WiFi achieves significantly lower latency (approximately $7$ ms on average) and higher throughput when drones are in close proximity, while LTE-A and 5G provide more stable throughput of around $50$ Mbit/s over wider coverage areas. Although their work characterizes the communication medium comprehensively, it does not address the confidentiality of transmitted data. In contrast, our work focuses on securing MAVLink telemetry over the WiFi and RF channels identified as suitable for short-range swarm coordination, by integrating lightweight encryption algorithms to protect against eavesdropping attacks that such open wireless channels are inherently susceptible to. Further, \cite{li2018uav} highlighted that high throughput UAV communication cellular links introduce correspondingly higher exposure to interception if left unencrypted, reinforcing the need for application-layer encryption beyond physical-layer protections alone.

Another study \cite{dong2021securing} addressed secure UAV swarm communication by combining beamforming, artificial noise injection, power allocation, and UAV trajectory optimization to improve secrecy performance in the presence of eavesdroppers. This approach enhances the legitimate communication link while degrading the eavesdropper’s channel, achieving improved secrecy rates. However, the dependence on centralized optimization and trajectory planning increases the computational complexity, making it less suitable for lightweight and dynamic UAV swarm environments. In contrast, our work focuses on a more efficient security mechanism tailored for practical swarm deployments.

The work \cite{olsson2023using} presents an autonomous approach using a swarm of drones to detect and neutralize unauthorized intruders. The framework integrates reinforcement learning for distributed decision-making with behavior trees for coordinated mission execution, achieving a mission success rate of $93\%$ in simulated environments. In contrast, our work focuses on securing the communication layer by providing a cryptographic framework that protects RF and WiFi links from eavesdropping, thereby enabling robust and secure real-world autonomous operations.

On the scalability front, \cite{thompson2016confidential} analyzed authenticated encryption for large swarms of $50$ UAVs, concluding that standard AES modes and ChaCha20-Poly1305 can consume significant processor margins on generic hardware as traffic grows. Specifically, they found that specialized authenticated encryption techniques such as SIV and EAX impose prohibitively high computational overhead in a $50$ UAV swarm, while standard AES modes such as  
Galois counter mode (GCM) and chaining 
message authentication code (CCM) consume significant processing margins, leaving little room for other onboard tasks. A primary limitation of this study is that it relies on software-in-the-loop (SITL) simulations and standalone processor benchmarks rather than active flight testing. Similar to their architectural approach, our work implements security at the application layer. By moving beyond their theoretical and SITL-based foundation, our work addresses the need for a hardware-validated security framework that maintains data confidentiality without degrading real-world flight performance.

The paper \cite{hassija2019survey} surveyed security architectures for Internet of Things (IoT) devices, demonstrating that application-layer encryption with lightweight symmetric ciphers achieves adequate confidentiality with manageable overhead on embedded processors comparable to those used in UAV companion computers. This finding directly supports the architectural decision in our work to implement encryption at the application layer within the MAVLink communication stack, enabling security without hardware modification.

In our prior work \cite{Dxtbhavya}, we proposed MAVShield, a lightweight block cipher designed to secure MAVLink communications over UAV-to-GCS RF links ($433$ MHz). Security analysis demonstrated that MAVShield is resistant to differential cryptanalysis under a chosen-plaintext attack model, with the evaluated differential characteristics indicating an extremely low differential trail probability of approximately $2^{-157}$. Furthermore, over-the-air man-in-the-middle (MITM) and replay attacks were unsuccessful due to the integrated encryption and authentication mechanisms provided by MAVShield. Experimental evaluations showed that MAVShield achieves lower CPU utilization ($1.038\%$), memory consumption ($0.031\%$), and battery power overhead ($12.921\%$) than several widely used cryptographic algorithms while maintaining communication confidentiality. However, the study focused primarily on secure MAVLink communications between a GCS and a UAV and did not investigate the integration of secure communications with formation control and collision avoidance mechanisms in real UAV swarm operations across RF ($915$ MHz) or WiFi communication modes.

Collision avoidance has been extensively studied in UAV networks using APF based approaches. In \cite{Budiyanto}, a conventional APF framework was employed to avoid static and dynamic obstacles, including neighboring UAVs, and the approach was validated through a robot operating system (ROS)-based simulation environment. Similarly, \cite{ma2021artificial} applied APF based obstacle avoidance for quad rotors operating in dynamic environments. To address the limitations of conventional APF methods, \cite{zhang2022improved} proposed an APSO-VNP-APF-based path planning algorithm, while \cite{Jiayi} proposed an optimized APF algorithm to mitigate local minima, target unreachability, and trajectory oscillation issues in cooperative multi-UAV operations.
A comprehensive survey \cite{khargharia2025collision} reviews UAV swarm collision avoidance methods, confirming that deterministic reactive methods such as APF remain the foundational safety layer in swarm operations, while real-time feasibility on resource constrained hardware remains an open challenge for learning-based alternatives. Conventional APF implementations typically require conversion of GPS measurements into a local Cartesian reference frame, but our approach performs force computations directly in geodetic coordinates. This eliminates coordinate transformation overhead and enables seamless integration with GPS-based navigation systems. Also, in prior work, collision avoidance approaches are primarily validated in simulation environments. In contrast, our work is validated on a real UAV swarm testbed and enables early collision avoidance activation through a predefined safety threshold. Additionally, since our work employs a distance threshold-based repulsive force that gradually modifies the trajectory before a close encounter occurs, the UAVs do not experience local minimum trapping in the evaluated swarm scenarios.

\section{System Model, Problem Formulation, and Encryption-Decryption Model} \label{sec:system_enc_dec_model}

\begin{figure}[hbt]
    \centering
    \includegraphics[width=0.95\linewidth]{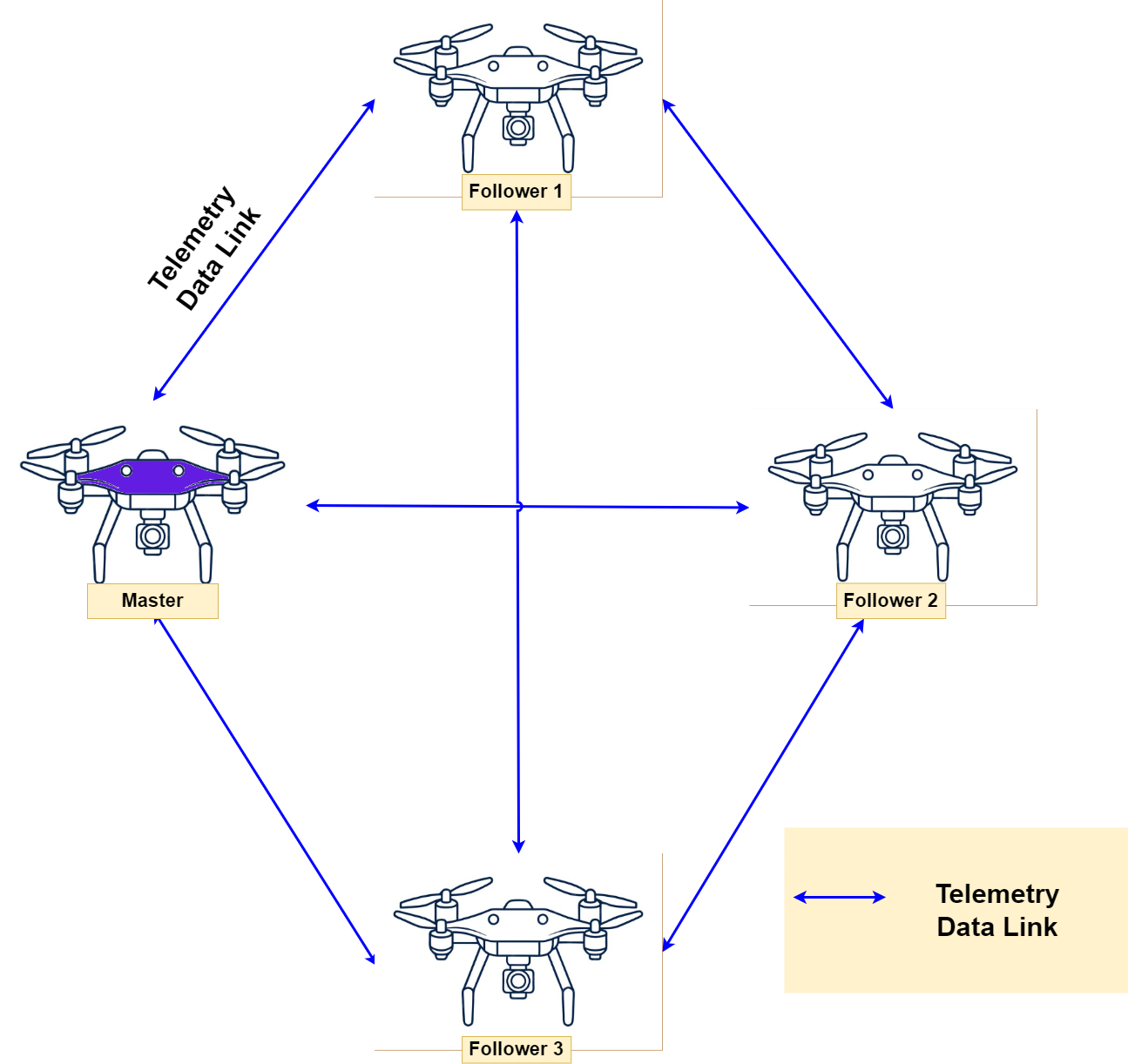}
    \caption{ The figure shows a swarm mesh network with one to all communication among UAVs. }
    \label{B1}
\end{figure}

\subsection{System Model }
As depicted in Figure \ref{B1}, communication within a UAV swarm is facilitated through air-to-air links. 
The periodic exchange of telemetry data between different UAVs of the swarm, including GPS coordinates, altitude, attitude (orientation), and proximity sensor readings, underpins the foundation for autonomous flight formation and collision avoidance. Formation control strategies leverage this telemetry to enable coordinated aerial configurations, such as linear, T-shaped, and Y-shaped patterns. Also, to maintain safe separation and coordinated motion, UAVs employ cooperative strategies based on APF, a path planning algorithm that allows UAVs to find a path to their destination without colliding with obstacles (other UAVs) \cite{messaoudi2023survey}. We consider a system in which UAV-to-UAV communication enables the dissemination of mission-critical information and telemetry messages over either RF or WiFi communication links. These multi-UAV mesh networks enable scalable, infrastructure-independent coordination for complex missions \cite{quaritsch2010networked}. The periodic telemetry exchange follows a broadcast-based paradigm, where each node disseminates its state information to all peers without requiring dedicated point-to-point connection management.

\subsection{Problem Formulation}
UAV-to-UAV wireless communications rely on the MAVLink protocol for data exchange due to its lightweight architecture and built-in mechanisms that help protect message integrity during transmission. However, as MAVLink communications are not encrypted by default, an adversary can passively intercept and access sensitive information without altering the transmitted data. Securing telemetry information, including UAV navigation data, across both communication modes (RF and WiFi) is essential, as any interception could compromise mission confidentiality, formation maintenance, and flight safety. Consequently, an important problem is to integrate a lightweight yet cryptographically secure protection mechanism to ensure the confidentiality of UAV communications against eavesdropping attacks. 

Each UAV continuously broadcasts telemetry data over both RF and WiFi communication channels, which follower UAVs utilize to maintain predefined geometric formations. To ensure safe flight during these maneuvers, we seek to design a modified APF-based collision avoidance framework operating directly in geodetic coordinates. We also require that the scheme should enable proactive collision avoidance through a predefined safety threshold and avoid local minimum trapping within the operating region of interest.

\subsection{Encryption-Decryption Model}
To safeguard against packet sniffing during data transmission, we implement a robust encryption-decryption framework for securing MAVLink communication over both RF and WiFi channels. For establishing reliable wireless communication between two aerial platforms, we utilize a Python-based communication module pre-deployed within the operating systems of companion computers onboard each UAV.

\begin{figure}[hbt]
    \centering
    \includegraphics[width=0.95\linewidth]{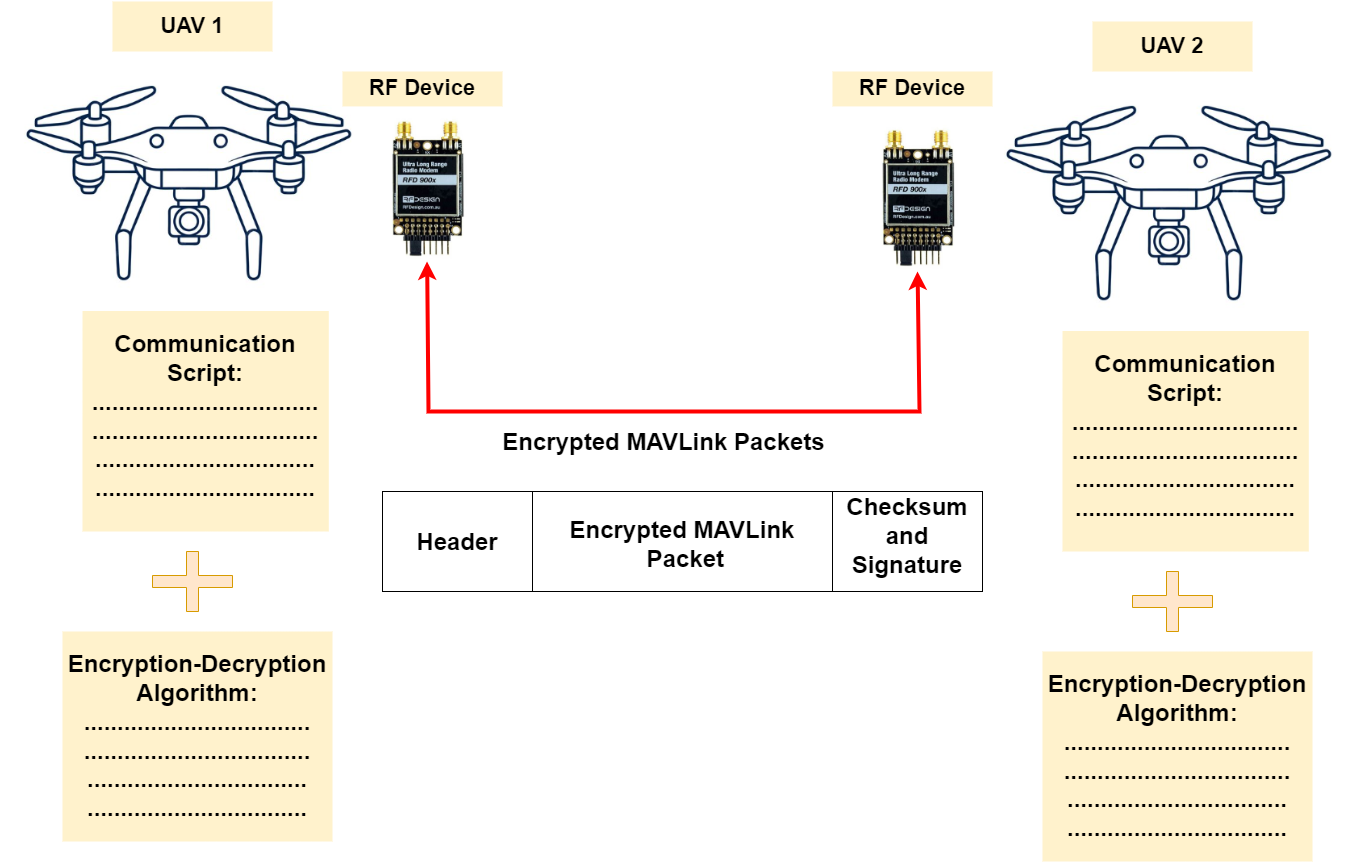}
    \caption{The figure shows integration of encryption-decryption algorithm in RF-based communication.}
    \label{B2}
\end{figure}

As depicted in Figure \ref{B2}, bidirectional communication is established via RF modules along with two RF antennas mounted on each UAV. To maintain the confidentiality of wireless communication, the encryption mechanism is seamlessly embedded within the Python-based communication module. In compliance with MAVLink protocol requirements, the packet header, comprising the system ID and message ID, remains unencrypted to ensure correct message routing and protocol compatibility. The payload of each MAVLink packet is encrypted, and a cryptographic checksum is computed over the encrypted content. Additionally, the complete packet is digitally signed using a SHA-256-based secure hash function in conjunction with a shared secret key to ensure data integrity and authenticity.

The encryption process is initiated dynamically at the point of message transmission, while decryption on the receiving UAV is executed immediately upon packet reception by the RF module, contingent upon successful verification of both the checksum and the digital signature. 

\begin{figure}[hbt]
    \centering
    \includegraphics[width=0.95\linewidth]{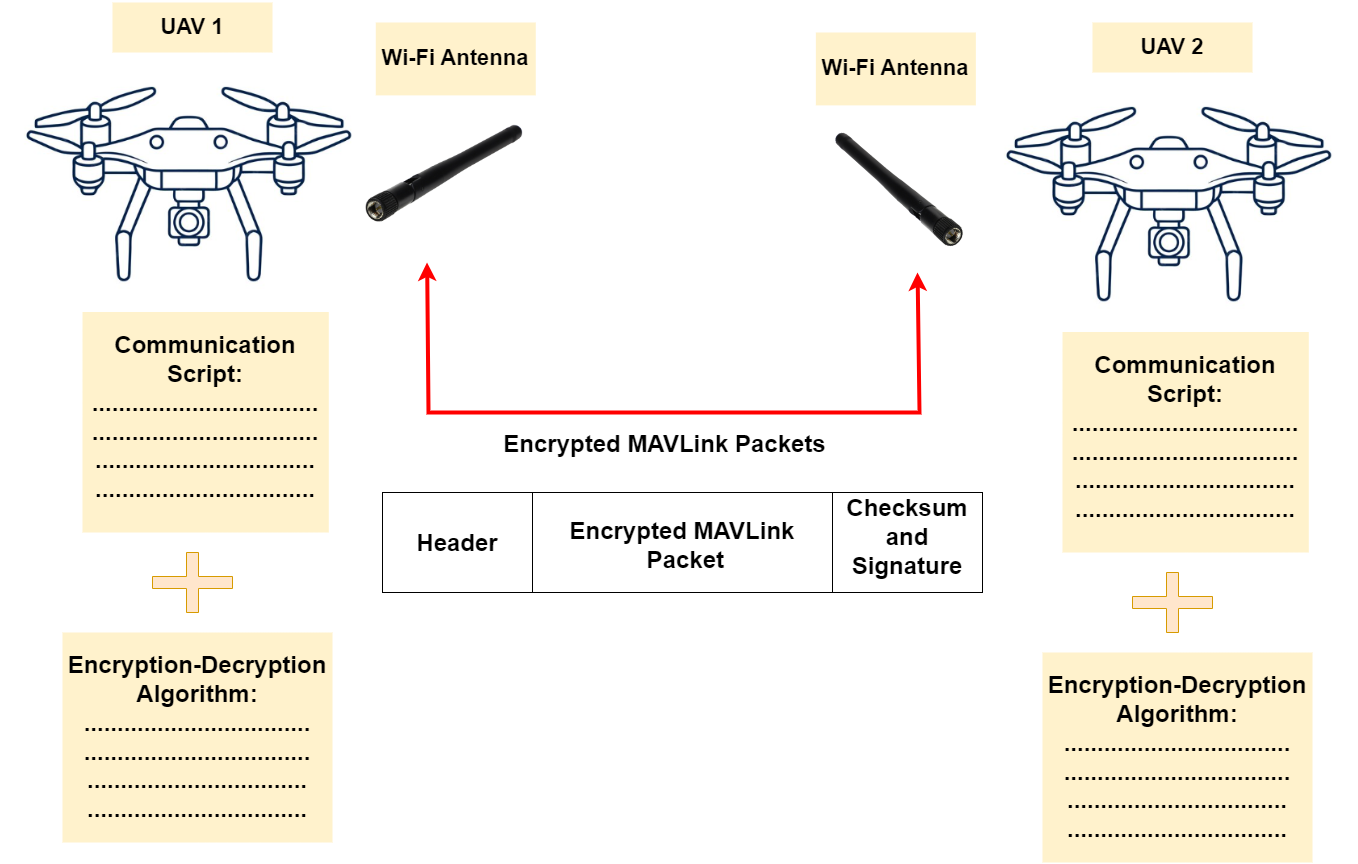}
    \caption{The figure shows integration of encryption-decryption algorithm in WiFi-based communication.}
    \label{B3}
\end{figure}

Analogously, as shown in Figure \ref{B3}, WiFi Modules along with two WiFi antennas are employed for inter-UAV communication, following the same encryption-decryption protocol to ensure secure wireless data exchange.

\section{UAV-to-UAV Communications} \label{sec:UAV2UAVCOMM}

In this section, we describe our UAV-to-UAV communication framework, including formation control strategies guided by telemetry data transmitted over RF and WiFi communication channels, a modified APF-based collision avoidance mechanism for safe inter-drone separation, and the RF and WiFi based communication architecture enabling secure and reliable swarm coordination.
\begin{figure}[hbt]
    \centering
    \includegraphics[width=0.95\linewidth]{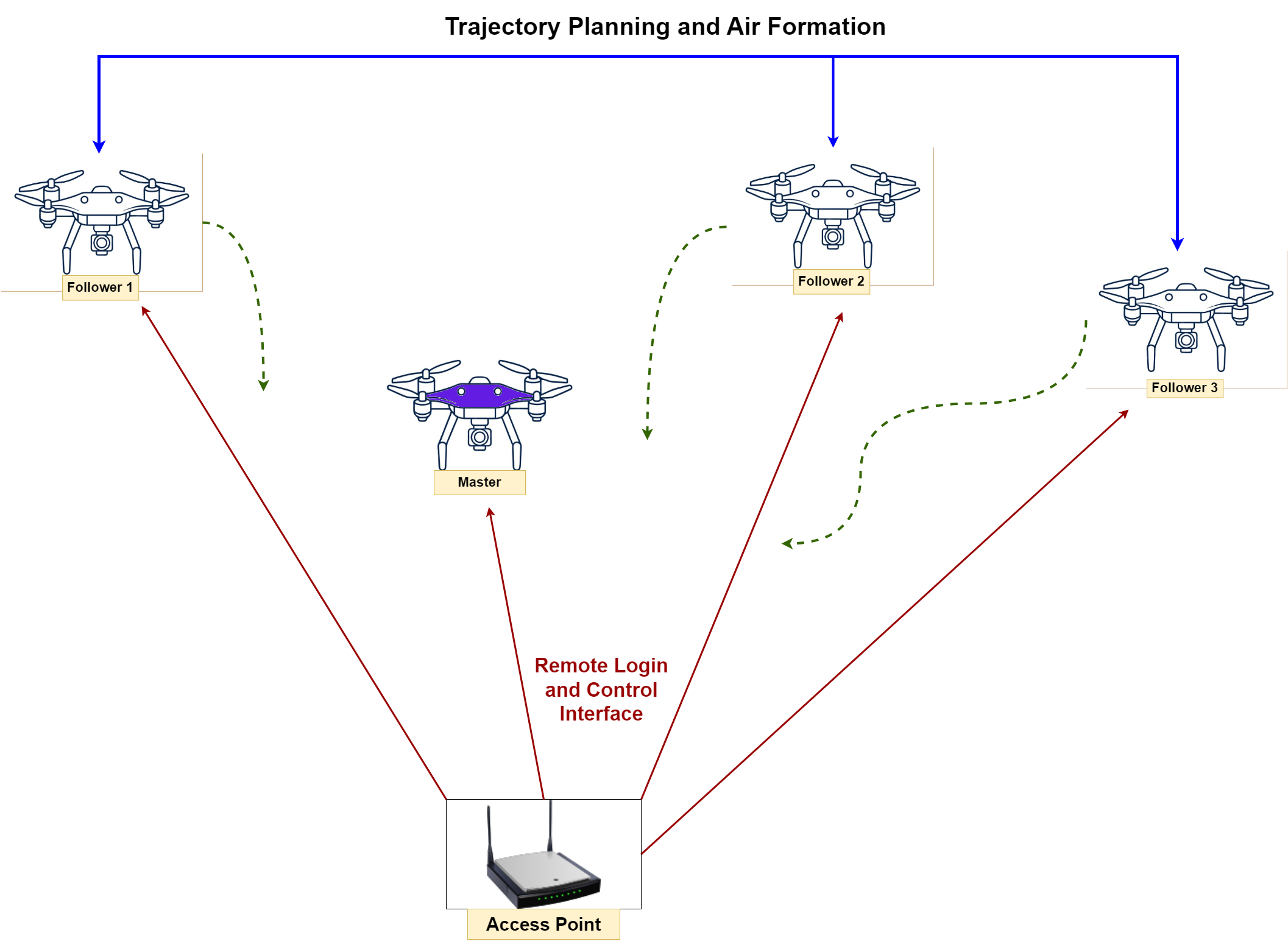}
    \caption{The figure shows our UAV-to-UAV communication framework.}
    \label{fig:UAV2UAV}
\end{figure}

\subsection{Formation}
In this study, we employ a swarm of four UAVs to execute coordinated flight formations \cite{oh2015survey,bu2024advancement,lin2026leader} as shown in Figure \ref{fig:UAV2UAV}, specifically in the shapes of a straight line, T-shape, and Y-shape (see Figures \ref{F1}, \ref{F2}, and \ref{F3}, respectively). Upon takeoff, each UAV assumes a precomputed position, equidistant from its neighbors, at a predetermined altitude. A formation control algorithm ensures that each UAV maintains its relative position and orientation within the formation, under the guidance of a designated leader. For each formation pattern, the leader communicates the telemetry data to the follower UAVs, which interpret this information to track their respective paths accordingly.

Algorithm \ref{alg1} outlines a pseudo-code for formation control in a drone swarm. The master drone is preloaded with the mission's planned waypoint, while the responsibility of creating the desired formation lies solely with the follower drones. For each of the three possible aerial formations, the follower drones are assigned predefined parameters such as formation angle and distance relative to the master drone, as well as altitude relative to the ground. The objective is to draw a specific pattern in the air using coordinated motion. 

At the outset of the process, a MAVLink connection is established between the flight control unit (FCU) and the companion computer (CC) on each follower drone (\emph{Line $1$}). The communication script provides the preassigned leader drone ID ($lead\_id$) and the follower drone ID ($flrs\_id$) as well as telemetry data broadcast by the master UAV, which is used to retrieve the leader's position information, including latitude, longitude, and heading (relative to true north) (\emph{Lines $4$-$5$}).

A position and distance dictionary is initialized for each follower drone (say $follower1$), using the drone ID (as key) and a counter starting at zero (\emph{Line $7$}). Once the leader’s position is available, the counter is incremented (\emph{Line $9$}) and the dictionary is updated accordingly (\emph{Line $10$}). Using the leader’s coordinates along with the follower’s predefined formation angle and distance, the target coordinates ($goal\_lat$, $goal\_lon$) for $follower1$ are calculated (\emph{Line $11$}).

To initiate the formation maneuver safely, when the counter reaches 1, $follower1$ is assigned a specific waypoint navigation speed (\emph{WPNAV$\_$SPEED}) and directed towards the target location (\emph{Lines $12$-$15$}). As it moves, its current position (\emph{Line $18$}) and relative distances to the goal (\emph{Line $20$}), leader (\emph{Line $21$}), and other followers are continuously monitored (\emph{Lines $23$-$27$}). If the distance between any two UAVs falls below a predefined safety threshold (\emph{Line $29$}), the target coordinates are dynamically recalculated  (\emph{Line $30$}). Also, its position and distance dictionary are updated (\emph{Line $31$}). The modified APF function (explained in Section \ref{ca}) then computes a corrective velocity vector ($v\_x$, $v\_y$) (\emph{Line $32$}) that ensures safe separation while guiding the follower towards its goal (\emph{Lines $33$-$36$}).

Finally, once the follower drone approaches the designated checkpoint, the algorithm introduces a brief delay (typically one second) to allow the UAV to stabilize at its position (\emph{Lines $39$-$42$}). This ensures that the swarm achieves the intended formation accurately and safely.

\begin{algorithm}
\small
\caption{Formation Control Algorithm (Follower UAV)}\label{alg1}
\textbf{Input:} Constants (angle, speed, distance, altitude) \\
\textbf{Output:} UAV control trajectory
\begin{algorithmic}[1]
\State \textbf{Init:} \texttt{connect\_FCU\_Jetson(MAVLink\_connection)}
\Function{Formation\_control}{\texttt{angle\_to\_lead, dist\_to\_lead}}
\While{ formation goal not reached}
     \State \text{Get \texttt{lead\_id,flrs\_id} from \texttt{RF\_comm}}
    \State \text{Get 
    \texttt{lead\_lat, lead\_lon, lead\_hdg}}
    \text{\hspace{9mm} from \texttt{RF\_comm}}
    \State \textbf{STEP I: Init position, distance}
    \State \texttt{pos = \{\}, dis = \{\},} \texttt{ counter = 0}
    \If{leader position available}
        \State \texttt{counter += 1}
        \State \texttt{pos[lead\_id]=lead\_lat, lead\_lon}
        \State \texttt{goal\_lat, goal\_lon = rel\_pos( lead\_lat, lead\_lon, dis\_to\_lead, lead\_hdg, angle\_to\_lead)}
        \If{\texttt{counter == 1}}
            \State Set \texttt{WPNAV\_SPEED} of \texttt{follower1}
            \State \texttt{goto\_waypoint(follower1, goal\_lat, goal\_lon, alt)}
        \EndIf
    \EndIf
    \State \textbf{STEP II: Get follower1 current position}
    \State \texttt{curr\_lat, curr\_lon, curr\_hdg = get\_global\_pos(follower1, home\_lat, home\_lon)}
    \State \textbf{STEP III: Calculate distances}
    \State \texttt{d\_goal = dis\_btw(curr\_lat,curr\_lon, goal\_lat, goal\_lon)}
    \State \texttt{dis[lead\_id] = dist\_btw(curr\_lat, curr\_lon, lead\_lat, lead\_lon)}
    \State \textbf{STEP IV: Other followers' position}
    \For{\texttt{id in flrs\_id}}
        \State Get \texttt{flr\_lat, flr\_lon} from \texttt{RF\_comm}
        \State \texttt{pos[id] = (flr\_lat, flr\_lon)}
        \State \texttt{dis[id] = dis\_btw(curr\_lat, curr\_lon, flr\_lat, flr\_lon)}
    \EndFor
    \State \textbf{STEP V: Collision avoidance}
    \While{\texttt{distance < threshold}}
        \State Recalculate \texttt{(goal\_lat, goal\_lon)}, \texttt{(curr\_lat, curr\_lon)}
        \State Update \texttt{pos, dis}
        \State \texttt{v\_x, v\_y = potential\_field( curr\_lat, curr\_lon, pos, goal\_lat, goal\_lon, d\_goal, dis)}
        \State Set velocity of \texttt{follower1}
        \If{All distances safe}
            \State \texttt{goto\_waypoint(follower1, goal\_lat, goal\_lon, alt)}
        \EndIf
    \EndWhile
    \State \textbf{STEP VI: Check formation}
    \If{\texttt{d\_goal} $<$ 1}
        \State \texttt{sleep(1)}
        \State Break loop (formation complete)
    \EndIf
\EndWhile
\EndFunction
\end{algorithmic}
\end{algorithm}

\subsection{Collision Avoidance}\label{ca}
The principle of the APF algorithm is based on a combination of the attractive potential field of the target point and the repulsive potential field of obstacles. The attraction generated by the target pulls the drone towards the target. The repulsive force generated by the obstacle point keeps the drone away from the obstacle. Finally, the UAV path is formed according to the combined force of the vectors \cite{zhang2022improved}.

We propose a modified APF-based framework for collision avoidance in a swarm, implemented using geodetic coordinates (latitude and longitude) rather than Cartesian coordinates. Unlike conventional APF \cite{ma2021artificial} approaches that operate in Cartesian space and are often limited to static or single-UAV scenarios, the proposed method addresses dynamic, real-time interactions among multiple UAVs in a two-dimensional operating space.
The algorithm leverages virtual attractive and repulsive forces calculated in geodetic space, enabling drones to dynamically adjust their paths while maintaining a safe separation from obstacles and other drones in the swarm. Our method incorporates real-time updates to the drones positions and velocities, ensuring collision-free navigation in dynamic environments.

\begin{figure}[hbt]
    \centering
    \includegraphics[width=0.95\linewidth]{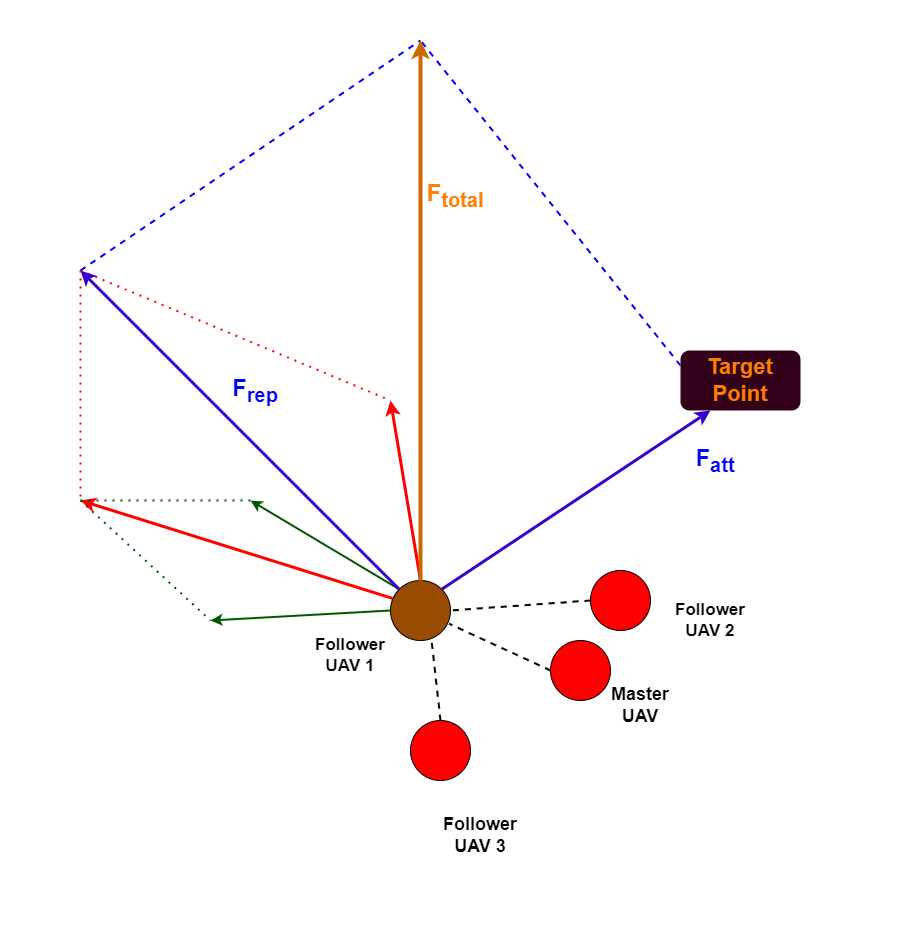}
    \caption{The figure shows a schematic of the modified APF-based method for collision avoidance.}
    \label{B8}
\end{figure}

To explain this mathematically, as illustrated in Figure \ref{B8},  we consider UAVs and target points as mass points and regard the working space of UAVs as a two-dimensional plane coordinate system. The core idea is to generate virtual forces that guide each drone safely while maintaining the desired swarm formation. Each UAV is influenced by the following force components:

\begin{itemize}
    \item \textbf{Attractive Force:} Guides the drone towards its target position based on the desired formation structure, denoted by $\mathbf{F}_{att}$.
    \item \textbf{Repulsive Force:} Pushes the drone away from neighboring drones to maintain safe separation, denoted by $\mathbf{F}_{rep}$.
\end{itemize}

The net force acting on each UAV is given by
\begin{equation}
    \mathbf{F}_{total} = \mathbf{F}_{att} + \mathbf{F}_{rep}.
\end{equation}

All force computations are carried out directly within the geodetic coordinate system using latitude and longitude values. In contrast to traditional APF techniques that rely on Cartesian coordinates, our approach eliminates the need for transformation into a Local Tangent Plane (LTP) or any intermediate reference frame. Instead, both attractive and repulsive forces are formulated explicitly in terms of latitude and longitude displacements.

\begin{enumerate}
    \item \textbf{Distance Calculation:} The inter-drone distance is calculated using the Haversine formula \cite{chan2020haversine}, as given in (\ref{haver}), which determines the great-circle distance between two points based on their latitude and longitude coordinates.
    
\begin{align}
a &= \sin^2\left(\frac{\Delta \varphi}{2}\right) + \cos(\varphi_1) \cos(\varphi_2) \sin^2\left(\frac{\Delta \lambda}{2}\right), \label{haver} \\
c &= 2 \tan^{-1} \left(\frac{\sqrt{a}}{\sqrt{1-a}} \right), \\
d &= R c,
\end{align}
where $a$ is an intermediate parameter representing the square of half the chord length between the two locations and $c$ denotes the angular distance (in radians) subtended by the two points at the center of the Earth. $\varphi_1$ and $\varphi_2$ denote the latitudes of the two points (in radians), $\lambda_1$ and $\lambda_2$ represent the corresponding longitudes (in radians), $\Delta \varphi = \varphi_2 - \varphi_1$ is the difference in latitudes, $\Delta \lambda = \lambda_2 - \lambda_1$ is the difference in longitudes, $R$ is the Earth's radius (typically $6371\,\text{km}$), and $d$ denotes the great-circle distance between the two points.

\item \textbf{Follower's Desired Position Calculation:} To determine the position of a follower UAV within the formation based on the leader's position and heading, we compute the heading offset (\ref{heading}), the positional offsets (\ref{position1})-(\ref{position2}), and the desired latitude (\ref{lat_leader}) and longitude (\ref{long_leader}) of the follower UAV.
\begin{align} 
    \theta' &= \theta + \alpha, \label{heading} \\
    (\Delta \varphi)_{FL} &= \frac{d_{FL} \cos(\theta')}{R}, \label{position1}\\
    (\Delta \lambda)_{FL} &= \frac{d_{FL} \sin(\theta')}{R \cos((\varphi_1)_L)}, \label{position2} \\
    (\varphi_2)_F &= (\varphi_1)_L + (\Delta \varphi)_{FL}, \label{lat_leader}\\
    (\lambda_2)_F &= (\lambda_1)_L + (\Delta \lambda)_{FL}. \label{long_leader}
\end{align}
Here, $d_{FL}$ denotes the distance between the follower and the master, $\alpha$ represents the angular offset of the follower relative to the master’s heading, $\theta$ is the master’s heading expressed in radians, and $(\varphi_1)_L$ and $(\lambda_1)_L$ correspond to the master’s latitude and longitude, respectively, in radians.

\item \textbf{Force Computation:} The attractive and repulsive forces are formulated as functions of the inter-drone distance and are inherently expressed in terms of latitude and longitude components. The computation of the attractive forces along the latitude and longitude directions are given in (\ref{att_lat}) and (\ref{att_long}), respectively:
\begin{align} 
F_{\text{att},\varphi} &= k_{\text{att}} . \Delta \varphi_{\text{goal}}.{d_{\text{goal}}^n}, \label{att_lat} \\
F_{\text{att},\lambda} &= k_{\text{att}}. \Delta \lambda_{\text{goal}}. \cos((\varphi_1)_{F_{cur}}).{d_{\text{goal}}^n}, \label{att_long}
\end{align}
where $k_{\text{att}}$ denotes the scaling factor for the attractive force, $\Delta \varphi_{\text{goal}}$ represents the latitude difference between the desired and current positions, $\Delta \lambda_{\text{goal}}$ is the corresponding longitude difference, $d_{\text{goal}}$ denotes the distance between the desired and current locations, $n$ is the exponent governing the attractive force, and $(\varphi_1)_{F_{\text{cur}}}$ represents the current latitude of the follower UAV.

\vspace{2mm}
Similarly, the repulsive force between the follower UAV and an obstacle and its components in the latitude and longitude directions are as in (\ref{follower_obst_force}), (\ref{rep_lat}), and  (\ref{rep_long}), respectively:
\begin{align} 
F_{\text{rep}1} &= -\left| \tan\left(\frac{d_{\text{obs}} - 8}{4} \right) \right|, \label{follower_obst_force} \\
F_{\text{rep},\varphi} &= k_{\text{rep}}.F_{\text{rep}1}. \frac{\Delta \varphi_{\text{obs}}}{d_{\text{obs}}}, \label{rep_lat} \\
F_{\text{rep},\lambda} &= k_{\text{rep}}.F_{\text{rep}} .\frac{\Delta \lambda_{\text{obs}} \cos((\varphi_1)_{F_{cur}})}{d_{\text{obs}}}, \label{rep_long}
\end{align}
where $k_{\text{rep}}$ denotes the scaling factor for the repulsive force, $\Delta \varphi_{\text{obs}}$ represents the difference in latitude between the obstacle and the follower UAV, $\Delta \lambda_{\text{obs}}$ is the corresponding longitude difference, and $d_{\text{obs}}$ denotes the distance between the obstacle and the follower’s current position. In (\ref{follower_obst_force}), the coefficient associated with $d_{\text{obs}}$, as well as the constants in the numerator and denominator, are determined by the threshold distance used to trigger collision avoidance.

To compute the net force components along the latitude and longitude of the follower UAV, we use (\ref{net_lat}) and (\ref{net_long}), respectively:

\begin{align} 
F_{\varphi} &= F_{\text{att},\varphi} + F_{\text{rep},\varphi}, \label{net_lat} \\
F_{\lambda} &= F_{\text{att},\lambda} + F_{\text{rep},\lambda}. \label{net_long}
\end{align}

The net force acting on the follower UAV is expressed as
\begin{equation}
F_{\text{net}} = \sqrt{F_{\varphi}^2 + F_{\lambda}^2}.
\end{equation}

\item \textbf{Velocity Decomposition:} The net force vector is decomposed into latitude and longitude components. Each component is then normalized and scaled by the desired velocity magnitude to generate velocity commands along the latitude and longitude directions. These commands are subsequently transmitted to the follower UAVs for execution. The resulting velocities along latitude and longitude are given in (\ref{lat_north}) and (\ref{long_east}), respectively:
\begin{align} 
v_{\text{lat}} &= \frac{F_{\varphi} \cdot v_{\text{max}}}{F_{\text{net}}}, \label{lat_north} \\
v_{\text{long}} &= \frac{F_{\lambda} \cdot v_{\text{max}}}{F_{\text{net}}}. \label{long_east}
\end{align}
Here, $v_{\text{max}}$ denotes the maximum allowable velocity for a follower UAV, which is set to $2\text{ m/s}$. The control parameters $k_{\text{att}}$, $n$, and $k_{\text{rep}}$ are set to $0.01$, $0.2$ and $1$, respectively.

\begin{figure}[hbt]
    \centering
    \includegraphics[width=0.97\linewidth]{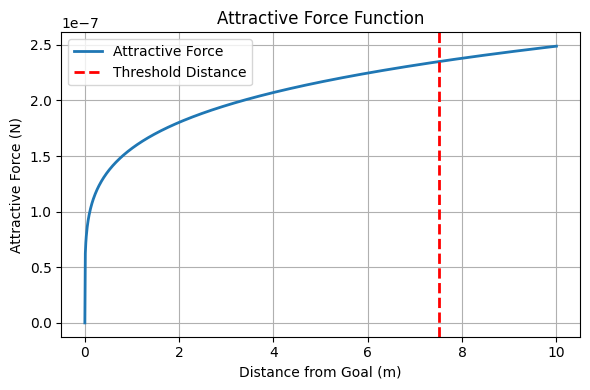}
    \caption{The figure shows the attractive force between the UAV and the goal, as given in (\ref{att_lat}).}
    \label{fig:attr_force}
\end{figure}

\begin{figure}[hbt]
    \centering
    \includegraphics[width=0.95\linewidth]{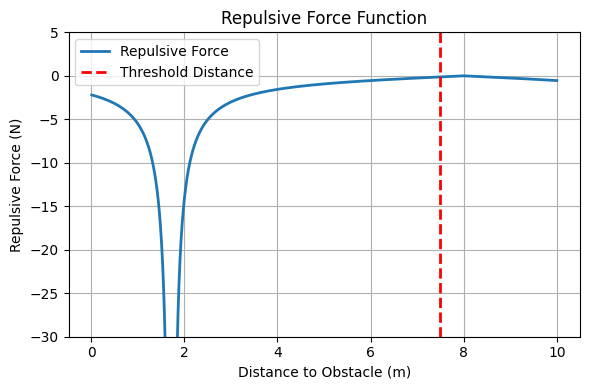}
    \caption{The figure shows the repulsive force between the UAV and the obstacle, as given in (\ref{follower_obst_force}).}
    \label{fig:rep_force}
\end{figure}

Figures \ref{fig:attr_force} and \ref{fig:rep_force} illustrate the attractive and repulsive force profiles employed by the proposed collision avoidance framework as a function of the distance between objects. Unlike conventional APF approaches \cite{Budiyanto,Jiayi}, where the repulsive force increases sharply when the UAV is in close proximity to an obstacle, the proposed repulsive force function (\ref{follower_obst_force}) initiates collision avoidance gradually at a predefined safety distance of $7.5$ m. Consequently, trajectory corrections begin at threshold distance, resulting in smoother avoidance maneuvers and reducing abrupt changes in velocity. Furthermore, we perform force computations directly in geodetic coordinates. Also, the proposed formulation produces attractive and repulsive force profiles that do not form an equilibrium point within the operating region of interest. As a result, no local minimum trapping was observed in the evaluated swarm scenarios, enabling UAVs to maintain safe inter-UAV separation while continuously progressing toward their formation objectives.
\end{enumerate}

\subsection{RF and WiFi-Based Communication}

The communication between the master and the follower UAVs is managed in two modes-- RF and WiFi, which we now describe.
\subsubsection{RF-Based Communication}
\begin{figure}[hbt]
    \centering
    \includegraphics[width=0.95\linewidth]{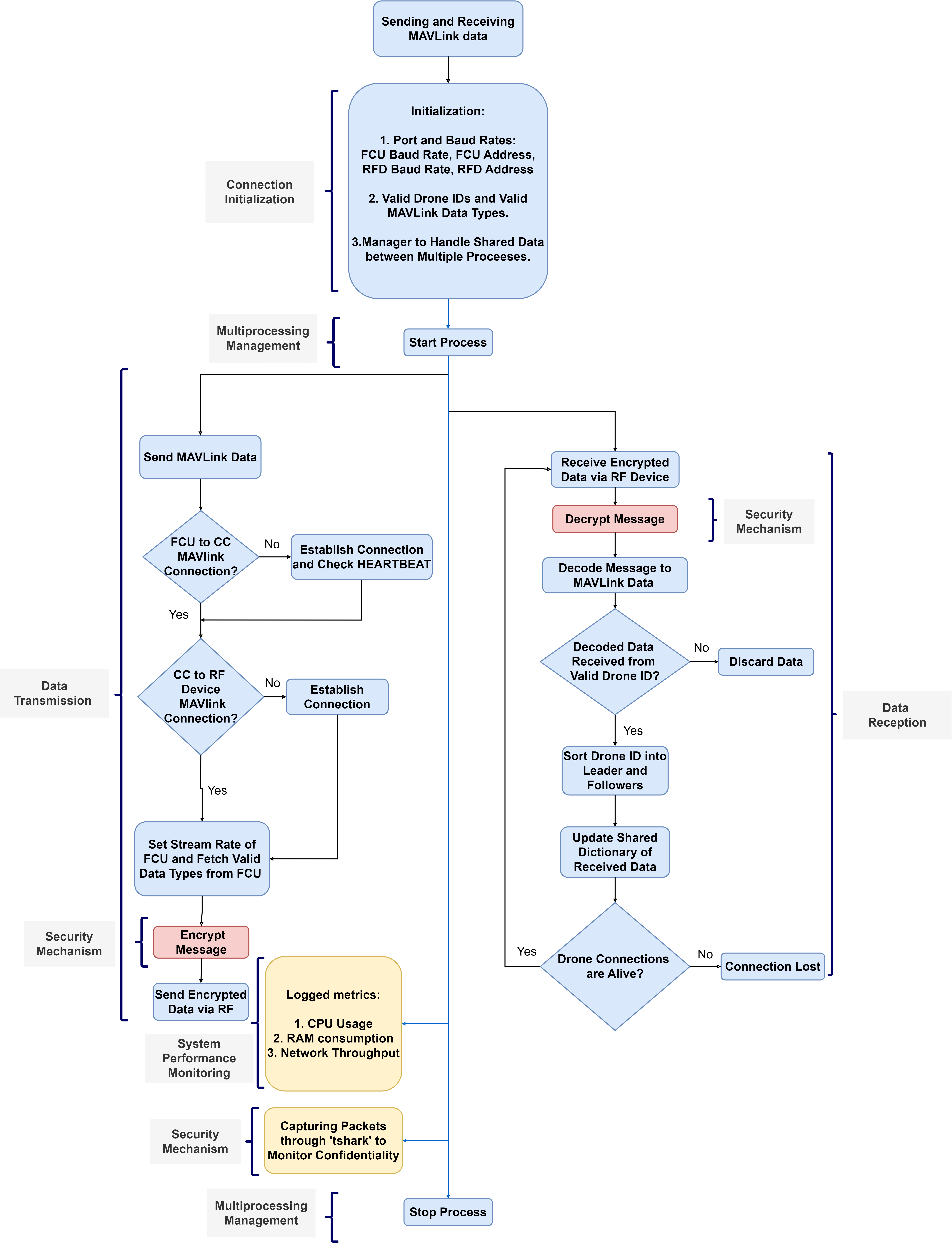}
    \caption{The figure shows a flowchart of RF-based UAV-to-UAV communication.}
    \label{B9}
\end{figure}

In this communication framework, each of the four UAVs is equipped with an RF module to facilitate bidirectional MAVLink based data exchange over the $915-928$ MHz band. As illustrated in Figure \ref{B9}, the RF-based MAVLink communication architecture is structured into five functional components: Connection Initialization, Multiprocessing Management, Data Transmission and Reception, Security Mechanisms, and System Performance Monitoring.

Each UAV broadcasts its telemetry data to the others, necessitating the management of parallel tasks across all drones. To enable concurrent execution of multiple processes on each UAV, the system employs Python's multiprocessing and concurrent modules. Internally, MAVLink communication is established via wired connections among the FCU, CC, and RF module within each UAV.

The software implementation begins by importing essential libraries: $mavutil$ from the pymavlink package for MAVLink protocol operations, $psutil$ for monitoring system resource usage (CPU and RAM), $tshark$ for packet capture and network traffic analysis, and $xcrypt$ for handling encryption and decryption tasks.

Subsequently, a dedicated class is defined to encapsulate all operations related to the transmission and reception of MAVLink messages. The FCU and the RF communication module are initialized with specified addresses and baud rates. A list of authorized drone IDs is defined to ensure that communication occurs only with trusted UAVs. To optimize communication efficiency and reduce bandwidth usage, only essential MAVLink message types are selectively transmitted from the FCU. Additionally, a process manager is leveraged to handle shared data between processes, ensuring robust parallel task execution.

Once the processes are initialized, the system begins continuous transmission and reception of MAVLink messages, while simultaneously monitoring key performance metrics such as CPU utilization, RAM consumption, and PDR.

Upon initiating MAVLink data transmission from a UAV, the system first verifies the connection between FCU and CC. If the connection is absent, it is established. A subsequent verification ensures that the MAVLink link between the CC and the RF module is active; if not, it is also initialized. After confirming connectivity, predefined MAVLink message types are retrieved from the FCU at a configured streaming rate. These messages are then encrypted using the selected security protocol; MAVShield, Speck-CTR, AES-CTR, ChaCha20, or Rabbit, before being broadcasted over the RF link to peer UAVs.

 The received data is decrypted using the same cryptographic scheme and subsequently parsed back into the standard MAVLink message format. A validation step follows, ensuring that the source UAV identifier matches a list of authorized drone IDs; packets from unknown sources are discarded. For validated UAVs, the system categorizes the UAVs into leader and followers and updates a shared dictionary that stores the received MAVLink telemetry. Concurrently, it monitors the liveness of UAV connections, if a UAV fails to respond within a $2$ second window, it is presumed to have either lost communication or exited the operational range and is temporarily marked as inactive. If no timeout is detected, the system continues to receive incoming MAVLink messages in a loop.

Finally, once the communication session concludes for a given UAV, all associated processes are gracefully terminated, and the established connections are properly closed.

\vspace{1mm}

\subsubsection{WiFi-Based Communication}

\begin{figure}[hbt]
    \centering
    \includegraphics[width=0.95\linewidth]{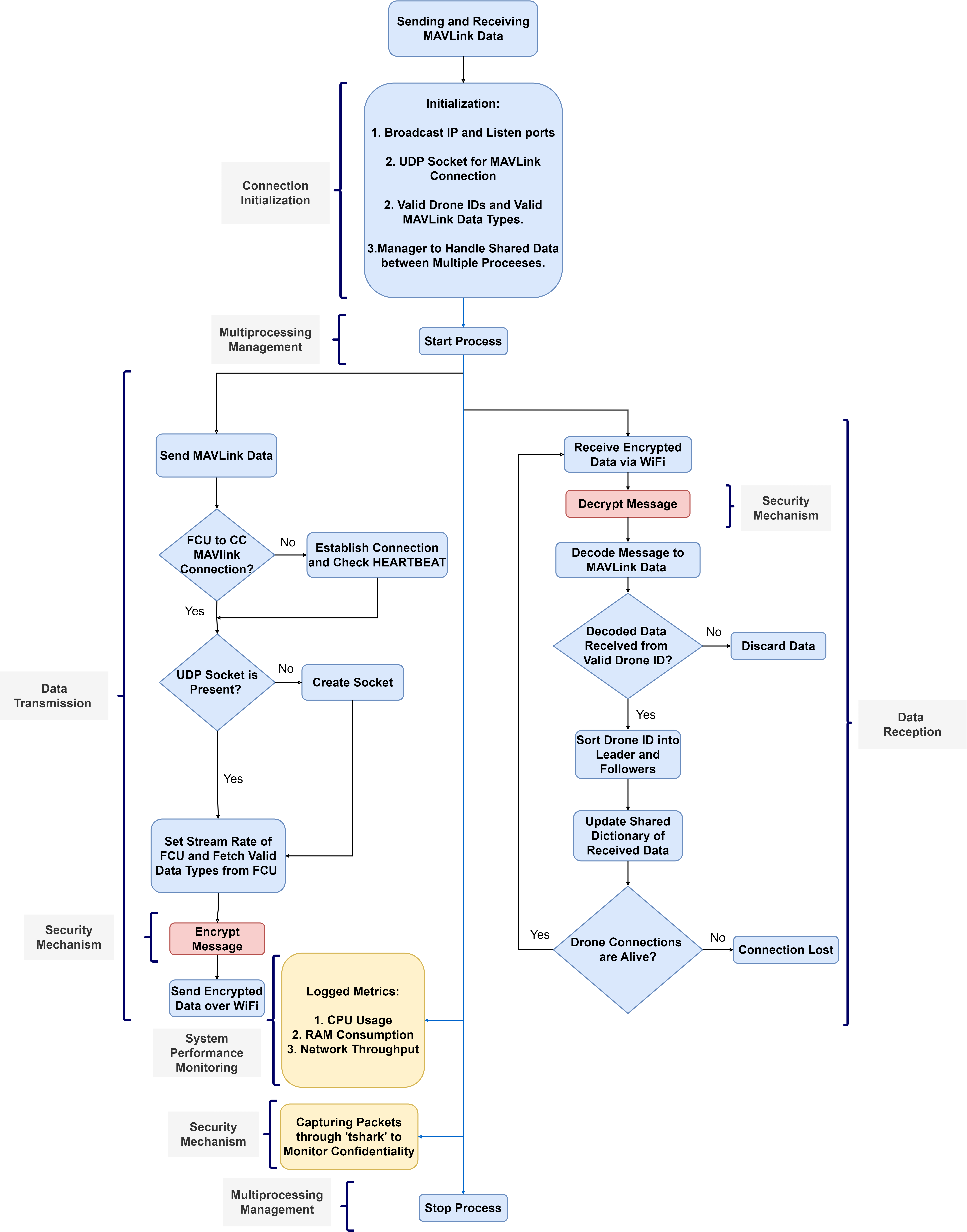}
    \caption{The figure shows a flowchart of WiFi-based UAV-to-UAV communication.}
    \label{B10}
\end{figure}

For WiFi-based communication among UAVs, each drone is equipped with a WiFi antenna and connected to a common dual-band ($2.4$ GHz / $5$ GHz) IEEE 802.11ac-based wireless network established through a centrally located access point. The network provides a coverage radius of approximately $100$ m, which is sufficient for the deployed testbed. This setup mirrors the architecture used in RF-based communication and comprises five primary components: Connection Initialization, Multiprocessing Management, Data Transmission and Reception, Security Mechanisms, and System Performance Monitoring.

Each UAV broadcasts its telemetry data over the WiFi network, allowing all other UAVs connected to the same access point to receive this data. This broadcast approach enables seamless integration of any UAV that joins the network, ensuring it can receive telemetry without requiring individual connections. 

As shown in Figure \ref{B10}, similar to the RF setup, we import essential libraries such as $mavutil$, $psutil$, and $xcrpt$ for communication, system resource monitoring, and cryptographic operations, respectively. Additionally, the $socket$ module is utilized to facilitate data transmission over Internet Protocol (IP) networks via either Transmission Control Protocol (TCP) or User Datagram Protocol (UDP). UDP is chosen for MAVLink telemetry due to its lower latency, no handshake, and reduced overhead, making it ideal for real-time communication. To handle concurrent tasks such as sending and receiving telemetry and logging system metrics such as CPU usage, RAM consumption, and PDR, we use Python's $multiprocessing$ module for parallel process execution.

Upon initializing the process, we define:
\begin{itemize}
    \item The broadcast IP ($255.255.255.255$) for telemetry distribution,
    \item A UDP socket for virtual communication,
    \item The FCU serial port address,
    \item A shared process manager for inter-process communication,
    \item Predefined filters to discard irrelevant FCU data.
\end{itemize}

Initially, a wired connection is established between the FCU and the CC onboard each UAV to share MAVLink data. The MAVLink packets are then wirelessly transmitted via a network interface card (NIC) with a WiFi antenna connected to CC. A UDP socket is then created for real-time, bidirectional communication, enabling shared data flows among all UAVs on the network. For ease of deployment and consistency, all UAVs listen on a common port ($14555$). Listening on a port implies that a UAV is passively waiting for incoming data at a specified network endpoint. Before transmission over the air, the message is encrypted using a predefined cryptographic scheme such as MAVShield, Speck-CTR, AES-CTR, ChaCha20, or Rabbit to ensure data confidentiality.

At the receiving end, a MAVLink connection is established at `$UDP$:$0.0.0.0$:$14555$' to bind to all available network interfaces and receive incoming messages on the given port. Upon receiving the encrypted packet, the other UAV:

\begin{itemize}
    \item Decrypts the message using the corresponding security algorithm,
    \item Parses the MAVLink message,
    \item Extracts the drone ID from the decoded data,
    \item Validates the ID, discarding messages from unknown sources,
    \item Classifies the drone as a leader or follower,
    \item Updates shared dictionaries used for coordination and decision-making.
\end{itemize}

Finally, once the communication task concludes, all processes are terminated, and associated virtual sockets and connections are closed cleanly.

\section{Security Algorithms} \label{sec:security_algorithm}

\subsection{Speck}
\begin{figure}[hbt]
    \centering
    \includegraphics[width=0.95\linewidth]{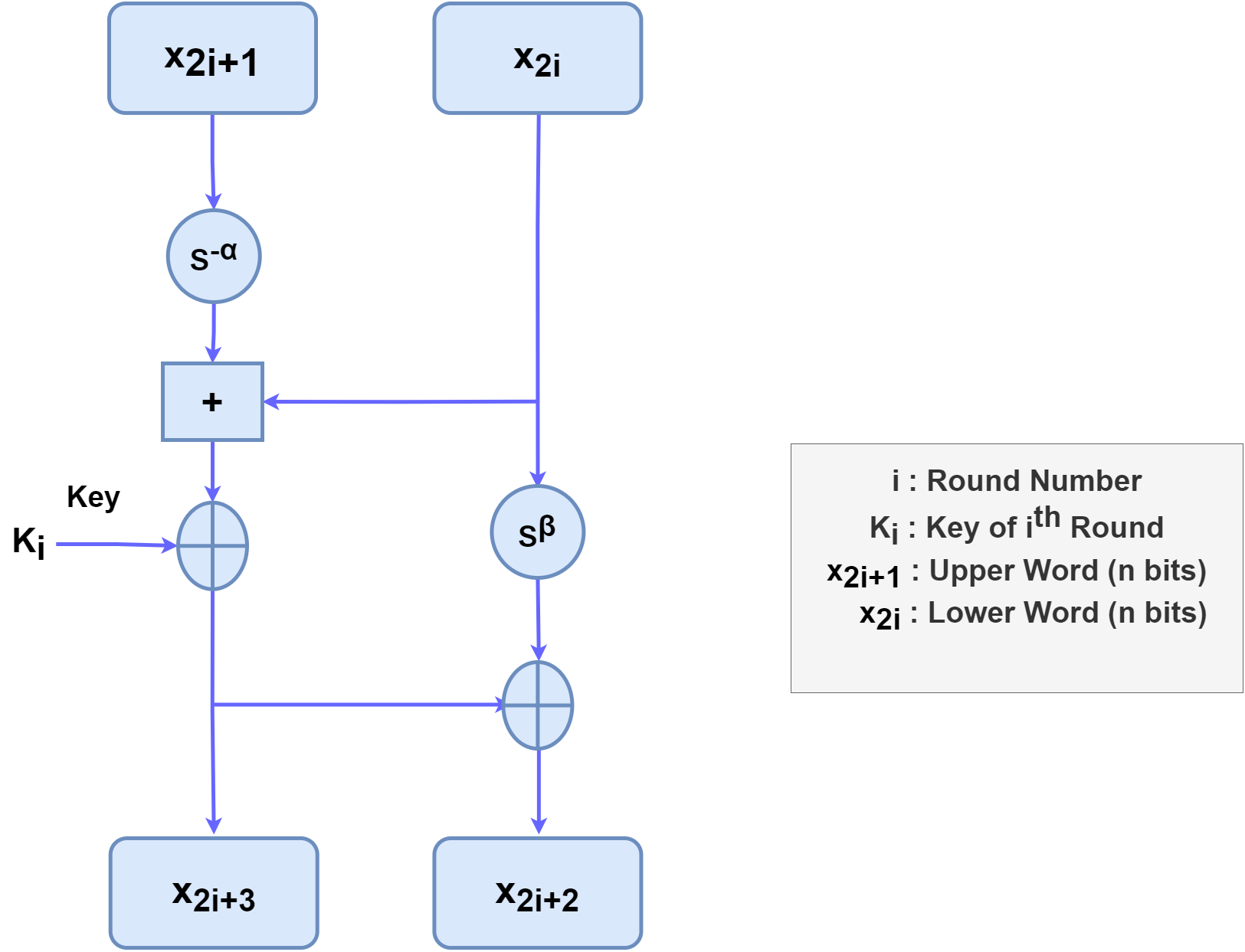}
    \caption{The figure shows the encryption process in the Speck cipher. }
    \label{B5}
\end{figure}

Speck is a lightweight block cipher based on the Add-Rotate-XOR (ARX) design paradigm and follows a Feistel-like structure \cite{beaulieu2015simon}. The algorithm operates on a $2n$-bit input, divided into two $n$-bit segments. As shown in Figure \ref{B5}, each iteration involves a right circular rotation of the first segment by $\alpha$ positions, followed by a modulo-$2$ addition (symbol $\boxplus$) with the second, and an XOR (symbol $\bigoplus$) with a round-specific key. Concurrently, the second part is rotated left by $\beta$ bits and XORed with the updated counterpart. The transformed pair proceeds to the next iteration. This process is repeated for a predefined number of rounds, determined by the selected parameter set. 

The Speck cipher incorporates a key scheduling process to derive a series of round keys from the original secret key. Notably, it reuses its core round function within the key expansion routine, which minimizes the implementation overhead and enhances computational efficiency in software-based systems that need round keys dynamically during encryption \cite{beaulieu2015simon}. 

In this paper, Speck is deployed in CTR mode, wherein a counter value initialized by an initialization vector (IV) is encrypted by the Speck block cipher to produce a keystream, which is then XORed with the plaintext to generate ciphertext. The IV is incremented by one for each successive plaintext block, ensuring that identical plaintexts produce distinct ciphertexts across different encryption instances.

\begin{table}[]
   \centering
   \caption{The table shows the Speck parameters.}
\begin{tabular}{|l|l|l|l|}
\hline
\rowcolor[HTML]{D9E1F2} 
\multicolumn{1}{|c|}{\cellcolor[HTML]{D9E1F2}{\color[HTML]{330001} \textbf{\begin{tabular}[c]{@{}c@{}}Block Size \\ (bits)\end{tabular}}}} & \multicolumn{1}{c|}{\cellcolor[HTML]{D9E1F2}\textbf{\begin{tabular}[c]{@{}c@{}}Key-Size\\ (bits)\end{tabular}}} & \multicolumn{1}{c|}{\cellcolor[HTML]{D9E1F2}\textbf{No. of Rounds}} & \multicolumn{1}{c|}{\cellcolor[HTML]{D9E1F2}\textbf{$\alpha$ and $\beta$ values}} \\ \hline
32                                                                                                                                         & 64                                                                                                              & 22                                                                  & 7,2                                                                                                         \\ \hline
\cellcolor[HTML]{F9F9F9}{\color[HTML]{330001} 48}                                                                                          & 72, 96                                                                                                          & 22, 23                                                              & 8, 3                                                                                                        \\ \hline
64                                                                                                                                         & 96, 128                                                                                                         & 26, 27                                                              & 8, 3                                                                                                        \\ \hline
96                                                                                                                                         & 96, 144                                                                                                         & 28, 29                                                              & 8, 3                                                                                                        \\ \hline
128                                                                                                                                        & 128, 192, 256                                                                                                   & 32, 33, 34                                                          & 8, 3                                                                                                        \\ \hline
\end{tabular}
\label{T1}
\end{table}

\subsection{MAVShield} 
\begin{figure}[hbt]
    \centering
    \includegraphics[width=0.95\linewidth]{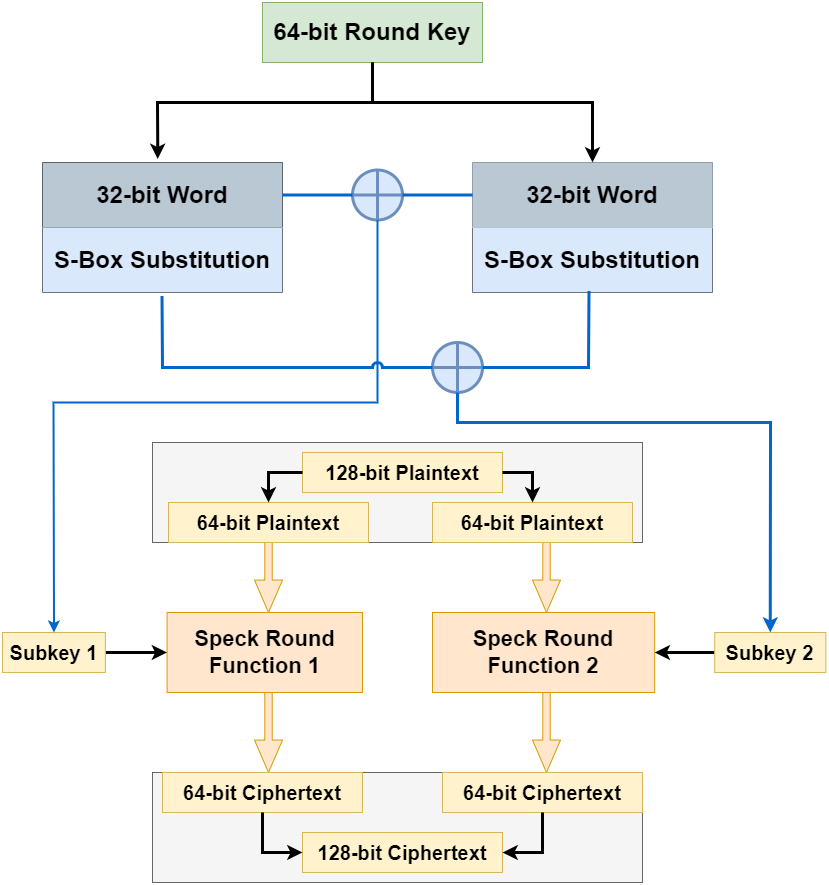}
    \caption{The figure shows the encryption process of the MAVShield algorithm.}
    \label{B6}
\end{figure}

The MAVShield encryption scheme proposed in our prior work \cite{Dxtbhavya} employs a lightweight symmetric block cipher with a $128$-bit block and key size. As illustrated in Figure \ref{B6}, it utilizes a dynamic key scheduling mechanism that iteratively updates each $64$-bit round key. The encryption process incorporates a series of structured operations, including word partitioning, S-box substitution, and bitwise XOR transformations. These operations ultimately yield two $32$-bit subkeys, each applied to a separate Speck round function.

The Speck cipher, operating on a $64$-bit block size with a $128$-bit key, transforms plaintext into ciphertext through a series of optimized rounds. A total of $10$ rounds are used for key generation and encryption, providing a favorable balance between computational efficiency, memory usage, and energy consumption, making the scheme well-suited for resource-constrained UAV platforms.
Similar to Speck, the MAVShield block cipher operates in CTR mode.

\subsection{AES}
AES is a symmetric-key encryption algorithm formally adopted as a global standard by the U.S. national institute of standards and technology (NIST) in 2001 \cite{abdullah2017advanced}. AES operates on a fixed plaintext block size of $128$ bits and supports key sizes of $128$, $192$, or $256$ bits. The secret key length determines the number of cryptographic transformation rounds ($N$): $10$ rounds for $128$-bit keys, $12$ rounds for $192$-bit keys, and $14$ rounds for $256$-bit keys.

\begin{figure}[hbt]
    \centering
    \includegraphics[width=0.95\linewidth]{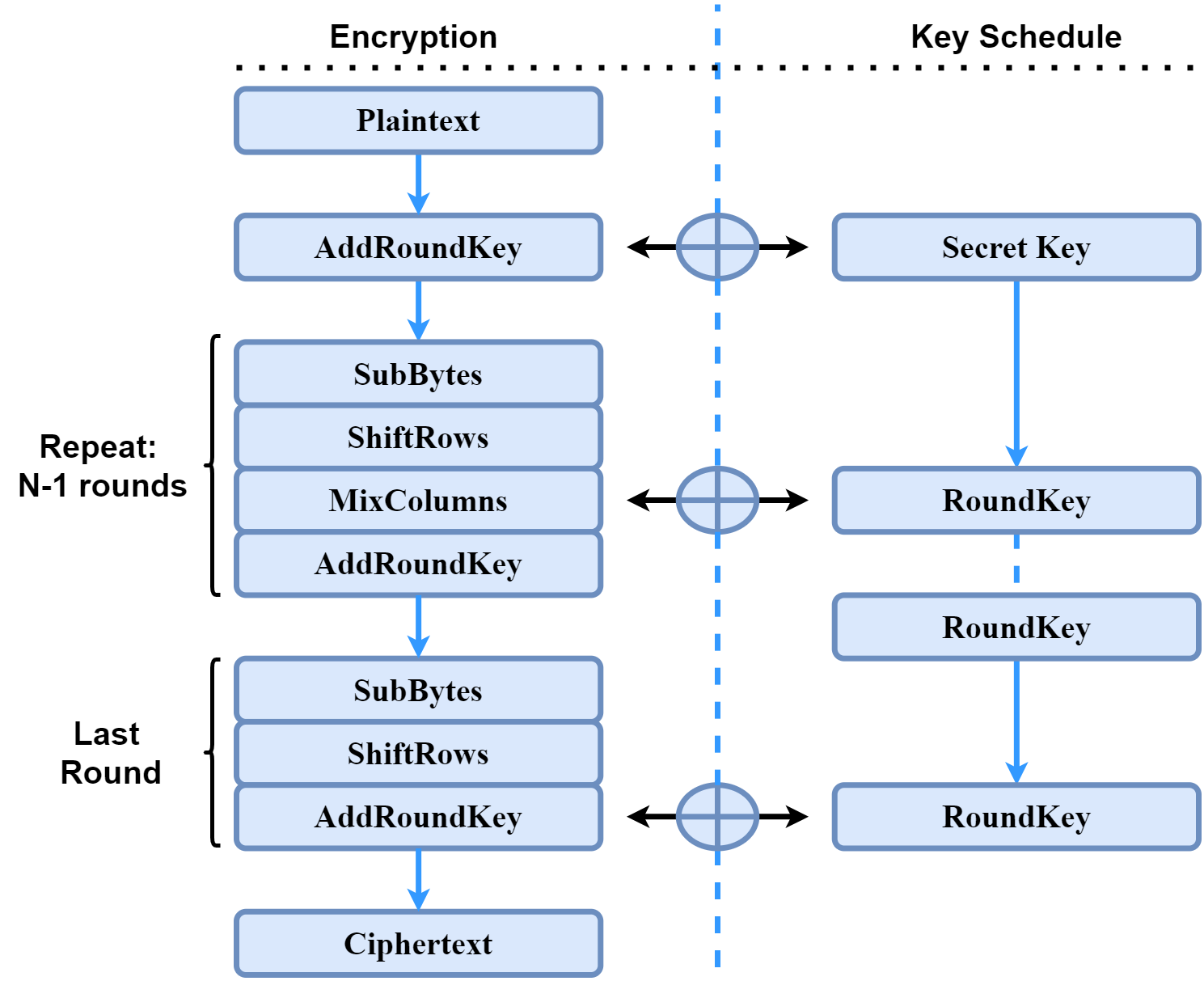}
    \caption{The figure shows the encryption process of the AES algorithm.}
    \label{B4}
\end{figure}

As shown in Figure \ref{B4}, each encryption round except for the final round consists of four core operations: SubBytes (nonlinear byte substitution using an S-box), ShiftRows (cyclic permutation of rows), MixColumns (column-wise mixing using linear transformation), and AddRoundKey (bitwise XOR with a round-specific subkey). The final round omits the MixColumns step \cite{abdullah2017advanced}. In this work, during AES-CTR encryption, the keystream derived from encrypted counter values is XORed with the plaintext to generate the ciphertext. Also, AES operates on data at the byte level, which enhances its suitability for both software and hardware implementations. It offers an effective balance between robust cryptographic security and computational efficiency, making it well-suited for a wide range of resource environments.

\subsection{ChaCha20}
ChaCha20 is a lightweight stream cipher, a variant of Salsa20 defined by Bernstein \cite{bernstein2008salsa20}. It utilizes CTR mode for faster symmetric encryption-decryption along with easy hardware and software implementations. 

\begin{figure}[hbt]
    \centering
    \includegraphics[width=0.95\linewidth]{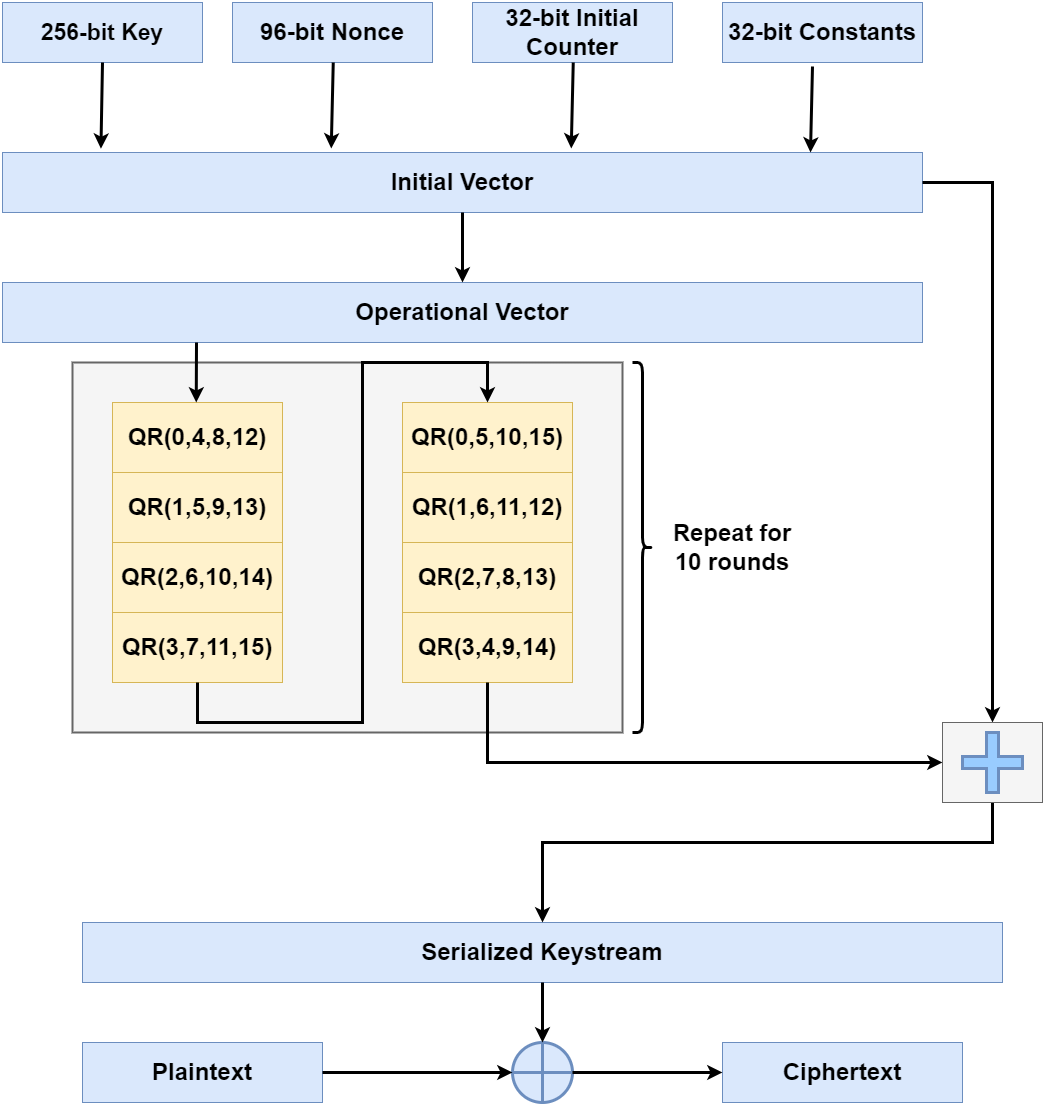}
    \caption{The figure shows the encryption process of the ChaCha20 algorithm.}
    \label{B7}
\end{figure}

As shown in Figure \ref{B7}, the algorithm utilizes a $256$-bit key, a $96$-bit nonce, a $32$-bit initial counter, and four predefined $32$-bit constants to construct a $512$-bit initial state vector $I$, composed of sixteen $32$-bit words. An operational vector $O$ is then initialized with the contents of $I$, arranged as a $4 \times 4$ grid as described below. 

\begin{equation*}
O =
    \begin{bmatrix}
        x_0 & x_1 & x_2 & x_3 \\
        x_4 & x_5 & x_6 & x_7 \\
        x_8 & x_9 & x_{10} & x_{11} \\
        x_{12} & x_{13} & x_{14} & x_{15}
    \end{bmatrix}
\end{equation*}

During encryption, the internal state of $O$ is iteratively transformed over $10$ double rounds. Each round consists of eight quarter-round (QR) functions that update four state variables by mixing the data through addition, bit-wise XOR, and rotation operations. Upon completion of all rounds, the final state is combined with the original vector $I$ via word-wise addition to produce the keystream. This keystream is then XORed with the plaintext to produce the ciphertext. Decryption follows an identical process in which the same keystream is regenerated and XORed with the ciphertext to recover the original plaintext \cite{kebande2023extended}.

\subsection{Rabbit}

Rabbit is a high-performance stream cipher inspired by the properties of real-valued chaotic maps, which exhibit extreme sensitivity to small input changes, resulting in an output that appears random and unpredictable over time \cite{boesgaard2008rabbit}. The cipher operates on a 128-bit secret key, producing 128-bit pseudo-random output blocks in each iteration. Its internal state comprises 513 bits, consisting of eight 32-bit state variables ($x_{j,i}$), eight 32-bit counters ($c_{j,i}$), and a single counter carry bit ($\phi_{j,i}$), where the subsystem index $j$ ranges from 0 to 7 and the iteration index $i$ ranges from 0 to 3. The state variables are updated through a system of coupled, nonlinear integer-valued functions.

\begin{figure}[hbt]
    \centering
    \includegraphics[width=0.95\linewidth]{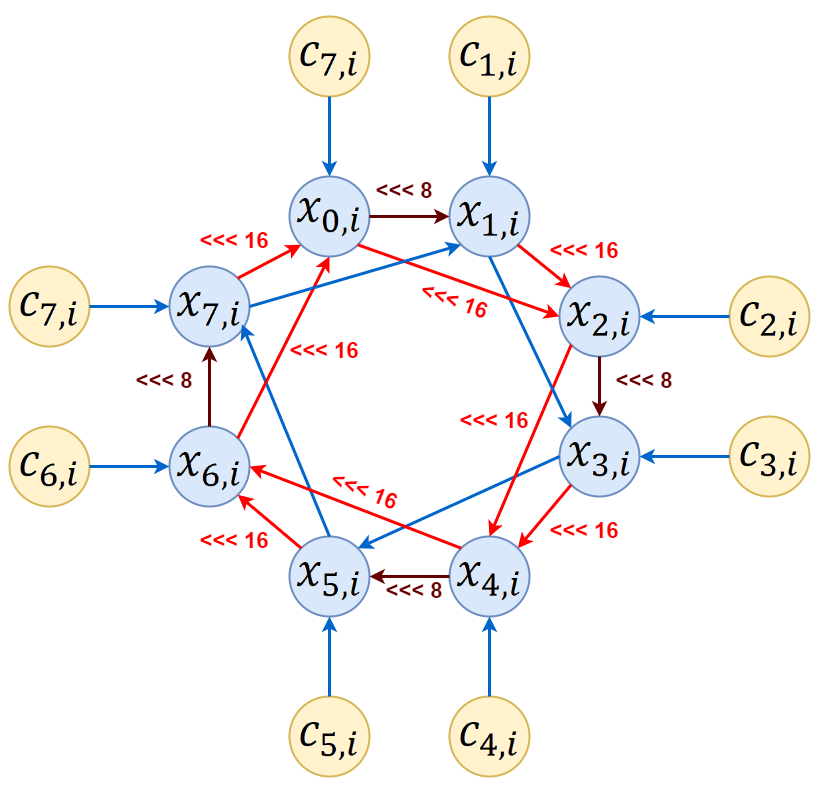}
    \caption{The figure shows the system architecture and data flow of Rabbit.}
    \label{B11}
\end{figure}

Figure \ref{B11} provides a visual representation of how the internal state is updated during encryption. Each state variable $x_{0,i}$ to $x_{7,i}$ is linked to a corresponding counter. In each round, state variables are updated using their previous values, associated counters, and neighboring state variables (typically two indices back), combined through nonlinear functions and bitwise rotations of $8$ or $16$ bits. This process introduces strong diffusion and ensures a complex and chaotic evolution of the internal state. Counters contribute to irregularity, preventing short cycles and enhancing security.

After four rounds, the generated pseudo-random data is XORed with the plaintext for encryption. The same data, when XORed with the ciphertext, recovers the original plaintext during decryption.

\section{Algebraic Cryptanalysis of MAVShield} \label{sec:algebraic_crypt}

In this section, we evaluate the resistance of MAVShield to algebraic cryptanalysis \cite{carlet2021boolean,kutsenko2021algebraic} by representing the cipher as a system of polynomial equations and analyzing its security against attacks based on linearization \cite{courtois2003algebraic}, Gaussian elimination \cite{courtois2012elimlin}, and SAT solving techniques \cite{audemard2018glucose,tseitin1983complexity}.

\subsection{MAVShield Algebraic Normal Form (ANF) over $\mathbb{F}_2$}


\begin{figure*}[htbp]
\centering

\begin{tikzpicture}[
    box/.style={
        draw=black, 
        fill=white, 
        rounded corners=4pt, 
        align=center, 
        font=\small,
        line width=0.7pt
    },
    inputNode/.style={
        box,
        minimum width=2.8cm, 
        minimum height=0.8cm, 
    },
    processNode/.style={
        box,
        minimum height=1.3cm, 
    },
    outputBox/.style={
        box,
        minimum width=6cm, 
        minimum height=0.8cm, 
    },
    arrow/.style={
        -{Stealth[scale=1.0]}, 
        thick,
        draw=black
    }
]

    \node (nu) at (0,4.5) [inputNode] {$\nu$ \textbf{(64 bits)}};
    \node (K)  at (6.0,4.5) [inputNode] {$K$ \textbf{(128 bits)}};
    \node (IV) at (10.5,4.5) [inputNode] {$IV^{(0)}$ \textbf{(128 bits)}};

    \node (layer1) at (0,2.5) [processNode, minimum width=3.4cm] {
        \textbf{Layer 1}\\ 
        \texttt{Round Value Generation}\\ 
    };

    \node (layer2a) at (6.0,2.5) [processNode, minimum width=4.0cm] {
        \textbf{Layer 2a}\\ 
        \texttt{Key Schedule Generation}\\ 
    };

    \node (layer3) at (10.5,2.5) [processNode, minimum width=4.2cm] {
        \textbf{Layer 3}\\ 
        CTR Counter Arithmetic\\ 
        $IV^{(j)} = IV^{(0)} + j$
    };

    \node (layer2b) at (6.0,0.3) [processNode, minimum width=3.8cm] {
        \textbf{Layer 2b}\\ 
        \texttt{Key Value Generation}\\ 
    };

    \node (layer4) at (7.6,-1.8) [processNode, minimum width=5.9cm] {
        \textbf{Layer 4}\\ 
        \texttt{Encryption}\\ 
    };

    \node (output) at (7.6,-3.5) [outputBox] {
        $\text{Keystream} \oplus \text{Plaintext} = \text{Ciphertext}$
    };

    
    \draw [arrow] (nu.south) -- (layer1.north);
    \draw [arrow] (K.south) -- (layer2a.north);
    \draw [arrow] (IV.south) -- (layer3.north);

    \draw [arrow] (layer1.east) -- (layer2a.west)
        node[midway, below, font=\tiny\bfseries, text=black, yshift=-1pt] {$\text{RV}[0\ldots19]$};

    \draw [arrow] (layer2a.south) -- (layer2b.north)
         node[midway, right, font=\tiny, text=black] {$\overline{KS}[0\ldots19]$};

    \draw [arrow] (layer3.south) -- (layer3.south |- layer4.north) 
        node[midway, right, font=\tiny, text=black] {$IV^{(j)}$};
        
   \draw [arrow] (layer2b.south) -- (layer2b.south |- layer4.north)
        node[pos=0.4, right, font=\tiny\bfseries, text=black] {$\text{KV}[0\ldots19]$};
    \draw [arrow] (layer4.south) -- (output.north);

\end{tikzpicture}
\caption{The figure shows a block schematic of MAVShield's data flow.}
\label{fig:mavshield_data_flow_bw}
\end{figure*}
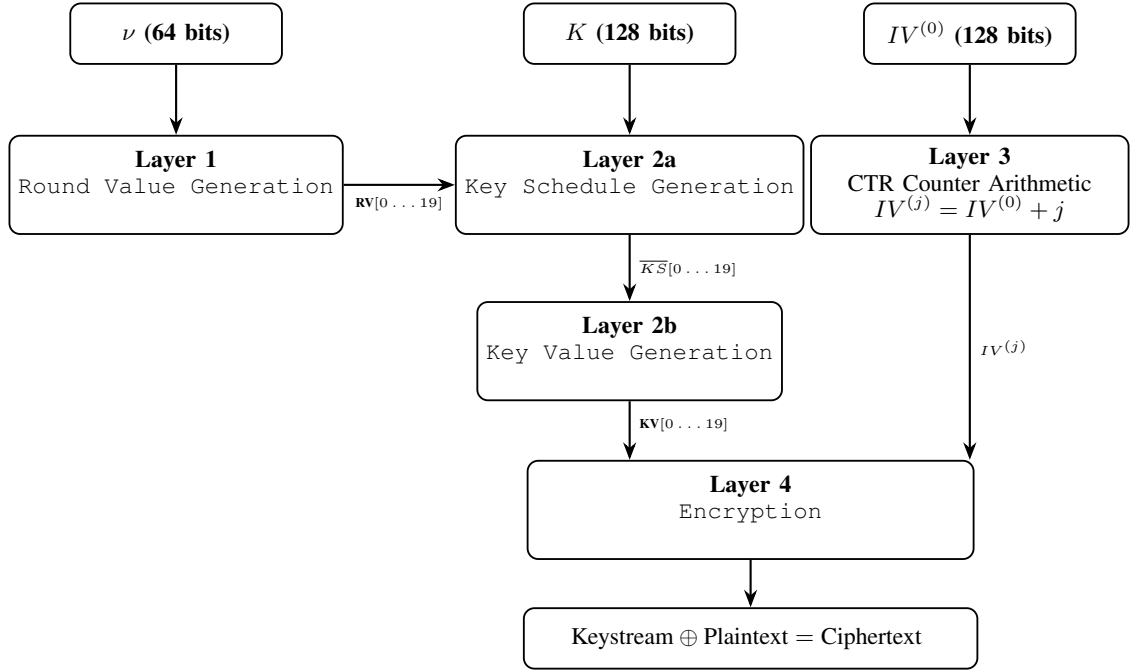

The data flow in the MAVShield block cipher is explained in Figure \ref{fig:mavshield_data_flow_bw}. The algebraic attack on MAVShield encodes the entire cipher as a system of polynomial equations over the two-element field $\mathbb{F}_2$ = \{$0, 1$\}. Every bit in the cipher-- nonce, key, IV, and all intermediate values-- becomes either a primary unknown or an auxiliary variable. The $320$ secret bits (64-bit nonce, 128-bit key, and 128-bit IV) are the primary unknowns as shown in Table \ref{tab:primary_unknowns}. All other bits are auxiliary variables introduced to keep every equation at degree at most 2.

\begin{table}[htbp]
\centering
\caption{The table shows the primary unknowns (total of 320 bits).}
\label{tab:primary_unknowns}
\begin{tabular}{lccl}
\toprule
\textbf{Variable group} & \textbf{Indices} & \textbf{Count} & \textbf{Used For} \\
\midrule
Nonce bits $\nu[0\ldots63]$  & $0 - 63$   & 64  & Layer 1 \\
Key bits $K[0\ldots127]$   & $64 - 191$  & 128 & Layers 2a and 2b \\
IV bits $IV[0\ldots127]$    & $192 - 319$ & 128 & Layers 3 and 4 \\
\bottomrule
\end{tabular}
\end{table}

Auxiliary variables start at index $320$ and are created on demand by each encoding function. Each Boolean variable or expression is stored as an ANF polynomial, which is a set of monomials, where each monomial is an immutable set of variable indices. Addition over $\mathbb{F}_2$ is symmetric difference (XOR). Multiplication is the union of variable sets (AND). The constant $1$ is represented as the empty monomial \{\}.

\subsubsection{Algebraic Encoding of Each Cipher Operation}

\paragraph{Rotation, NOT, XOR- Zero Cost Operations}
These operations introduce no new variables and add no equations to the system. They are implemented purely as index permutations or constant additions.

\paragraph{Modular Addition $x + y$ (mod $2^{32}$)}

Modular addition is the most expensive operation. It is encoded using a ripple-carry chain. For each bit position $b = 0, \cdots, 31$ with $c_{-1} = 0$ (no initial carry).

\begin{equation*}
\begin{aligned}
s_b &= x_b \oplus y_b \oplus c_{b-1} \\
c_b &= x_b \cdot y_b \oplus x_b \cdot c_{b-1}
      \oplus y_b \cdot c_{b-1}
\end{aligned}
\end{equation*}

Here, $s_b$ and $c_b$ are sum and carry variables responsible for linear and quadratic equations. Each $32$-bit addition introduces $63$ fresh variables ($32$ sum bits and $31$ carry bits; the final carry is discarded) and $63$ equations ($32$ linear and $31$ quadratic).

\paragraph{AES S-box Encoding}

The AES S-box has algebraic degree 7 over $\mathbb{F}_2$, which would normally produce high-degree equations. To maintain the global degree bound of at most $2$, each S-box output bit is first expressed in ANF using the Fast M\"obius Transform \cite{carlet2021boolean}.

Fresh output variables are then introduced for each output bit, and all higher-degree monomials are recursively decomposed into quadratic products using auxiliary or helper variables.
Each S-box at the byte-level introduces $8$ new output variables along with additional auxiliary variables depending on the ANF monomial structure.

\subsubsection{Layer-by-Layer Equation Construction}
\begin{figure*}
    \centering
    \includegraphics[width=0.95\linewidth]{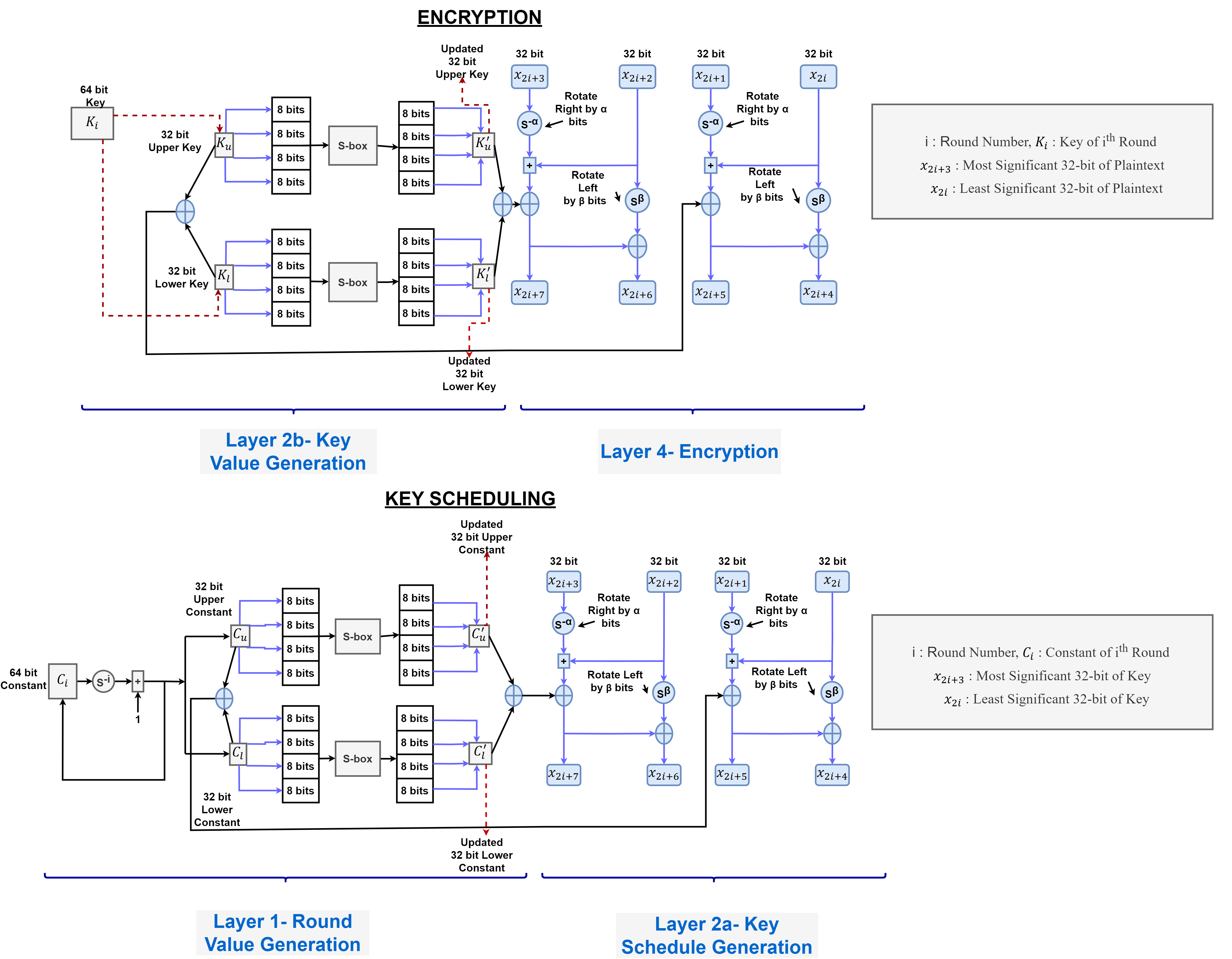}
    \caption{The figure shows the key scheduling and encryption schemes of the MAVShield cipher \cite{Dxtbhavya}.}
    \label{fig:MAVShield_full}
\end{figure*}

\begin{figure}[hbt]
    \centering
    \includegraphics[width=0.95\linewidth]{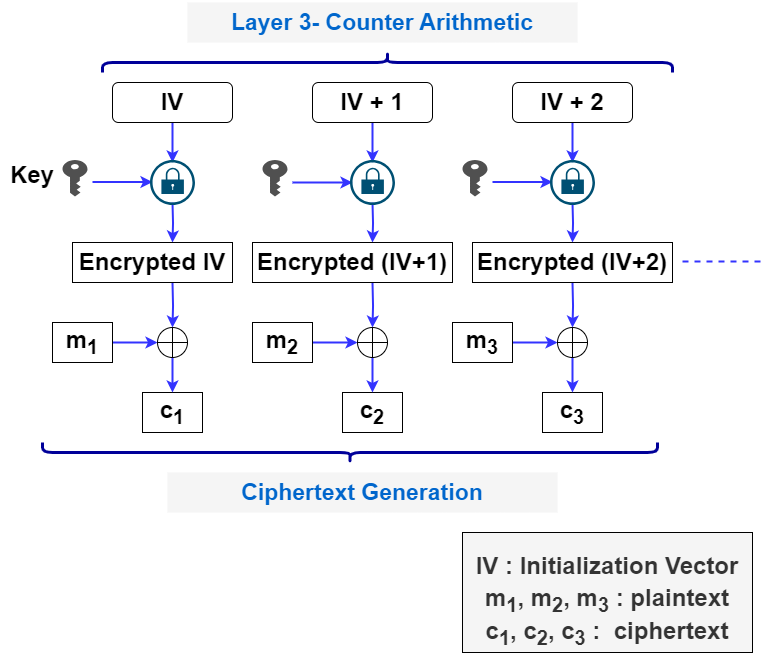}
    \caption{The figure shows a block schematic of the CTR mode used for the MAVShield block cipher \cite{Dxtbhavya}.}
    \label{fig:CTR_mode_ANF}
\end{figure}

Figure \ref{fig:MAVShield_full} shows the internal structure of the MAVShield cipher, which was detailed in our prior work \cite{Dxtbhavya} and consists of \emph{Layer $1$, Layer $2a$, Layer $2b$,} and \emph{Layer $4$}. \emph{Layer $3$} is explained in Figure \ref{fig:CTR_mode_ANF}.
The ANF builder assembles four layers sequentially. All intermediate symbolic words are passed between layers as lists of Poly objects.

\paragraph{Layer $1$- Round Value (RV) Generation}
\noindent


\begin{table}[t]
\centering
\caption{The Table shows Layer 1 counts (N rounds). The XOR, Rotation, NOT and split operations are inline and cost nothing. In each round of Layer 1, two $S_{32}$ operations are performed, one on H and one on L. Each $S_{32}$ processes $4$ bytes, giving a total of $8$ bytes  per round. The AES S-box linearization costs $254$ auxiliary variables and equations per byte, resulting in $2032$ variables and equations.}
\label{tab:layer1_counts}

\renewcommand{\arraystretch}{1.15}

\begin{tabular}{lccc}
\toprule
\textbf{Operation} &
\makecell{\textbf{Equations}\\\textbf{(Linear)}} &
\makecell{\textbf{Equations}\\ \textbf{(Quadratic)}} &
\textbf{New vars.} \\
\midrule

Rotation + NOT &
0 &
0 &
0 \\

Even RV XOR &
0 &
0 &
0 \\

\makecell{Odd RV S-box \\(2 words/round)} &
0 &
2032N &
2032N \\

Odd RV XOR &
0 &
0 &
0 \\

\midrule

\textbf{Total Layer 1} &
\textbf{0} &
\textbf{2032N} &
\textbf{2032N} \\

\bottomrule
\end{tabular}
\end{table}

Input: secret nonce $\nu\in\mathbb{F}_2^{64}$. Output: round-value array RV[0], $\cdots$, RV[19], each $\in$ $\mathbb{F}_2^{32}$.

For each $p = 0,1,\ldots,9$:

\begin{align}
\nu^{(p+1)}
&=
\sim\!\left(
\operatorname{ROR}_{64}\!\left(\nu^{(p)},\, p\right)
\right)\\
\left(H^{(p)},L^{(p)}\right)
&=
\operatorname{split}_{64}\!\left(\nu^{(p+1)}\right)
\\[6pt]
\mathbf{RV}[2p]
&=
H^{(p)} \oplus L^{(p)} \label{eq:RV_even}
\\[6pt]
\mathbf{RV}[2p+1]
&=
S_{32}\!\left(H^{(p)}\right)
\oplus
S_{32}\!\left(L^{(p)}\right) \label{eq:RV_odd}
\end{align}

$\operatorname{ROR}_{s}(q,r)$ denotes the right circular rotation of the $s$-bit variable $q$ by $r$ bit positions. In (\ref{eq:RV_even}), even round values $\mathbf{RV}[2p]$ depend on $\nu$ only through linear operations (rotation, NOT, XOR).  In  (\ref{eq:RV_odd}), odd round values $\mathbf{RV}[2p+1]$ are obtained through two $S_{32}$ applications (The 32-bit input word is partitioned into four 8-bit bytes, each byte is independently substituted using the S-box, and the resulting four substituted bytes are then recombined to form a 32-bit output word.), introducing algebraic degree $7$ per byte over $\mathbb{F}_2$. Table \ref{tab:layer1_counts} details the layer $1$ counts for $N$ rounds.

\paragraph{Layer $2a$ - ARX Key Schedule ($\overline{KS}$)}
\noindent

Input: master key $K = (k_0, k_1, k_2, k_3) \in (\mathbb{F}_2^{32})^4$,
round values ${RV}[0\ldots19]$. Output: pre-hardening subkeys $\overline{{KS}}[0], \ldots, \overline{{KS}}[19]$.

\medskip

Two independent chains run sequentially.


\begin{table}[t]
\centering
\caption{The table shows Layer 2a counts (2 chains $\times$ N rounds).}
\label{tab:layer2a_counts}

\renewcommand{\arraystretch}{1.15}

\begin{tabular}{lccc}
\toprule
\textbf{Operation} &
\makecell{\textbf{Equations}\\\textbf{(Linear)}} &
\makecell{\textbf{Equations}\\ \textbf{(Quadratic)}} &
\textbf{New vars.} \\
\midrule

Rotations \& XOR &
0 &
0 &
0 \\

Chain A: Modular additions &
32N &
31N &
63N \\

Chain B: Modular additions &
32N &
31N &
63N \\

\midrule

\textbf{Total Layer 2a} &
\textbf{64N} &
\textbf{62N} &
\textbf{126N} \\

\bottomrule
\end{tabular}

\end{table}

\noindent
\textbf{Chain A} \qquad
(uses $k_0, k_1$, and even round values):
For $i = 0, \ldots, 9$, let
$a = k_1^{(i)}$, $b = k_0^{(i)}$:
\begin{align}
a &\leftarrow \mathrm{ROR}_{32}(a,8) \label{eq:11} \\
a &\leftarrow a \boxplus b \label{eq:12} \\
a &\leftarrow a \oplus \mathbf{RV}[2i] \label{eq:13} \\
b &\leftarrow \mathrm{ROL}_{32}(b,3) \label{eq:14} \\
b &\leftarrow b \oplus a \label{eq:15}
\end{align}

\[
k_1^{(i+1)} = a,
\qquad
k_0^{(i+1)} = b
\]

\begin{equation}
\overline{\mathbf{KS}}[2i] = k_0^{(i+1)}
\label{eq:16}
\end{equation}

\vspace{1em}

\noindent
\textbf{Chain B} \qquad
(uses $k_2, k_3$, and odd round values):
For $i = 0, \ldots, 9$, let
$a = k_3^{(i)}$, $b = k_2^{(i)}$:
\begin{align}
a &\leftarrow \mathrm{ROR}_{32}(a,8) \label{eq:17} \\
a &\leftarrow a \boxplus b \label{eq:18} \\
a &\leftarrow a \oplus \mathbf{RV}[2i + 1] \label{eq:19} \\
b &\leftarrow \mathrm{ROL}_{32}(b,3) \label{eq:20} \\
b &\leftarrow b \oplus a \label{eq:21}
\end{align}

\[
k_3^{(i+1)} = a,
\qquad
k_2^{(i+1)} = b
\]

\begin{equation}
\overline{\mathbf{KS}}[2i + 1] = k_2^{(i+1)}
\label{eq:22}
\end{equation}

Chains A and B share no intermediate state and are fully independent. Table \ref{tab:layer2a_counts} details the layer $2a$ count for N rounds.
\\


\begin{table}[t]
\centering
\caption{The table shows the Layer 2b counts (N rounds).}
\label{tab:layer2b_counts}

\renewcommand{\arraystretch}{1.15}

\begin{tabular}{lccc}
\toprule
\textbf{Operation} &
\makecell{\textbf{Equations}\\\textbf{(Linear)}} &
\makecell{\textbf{Equations}\\ \textbf{(Quadratic)}} &
\textbf{New vars.} \\
\midrule

Even KV XOR &
0 &
0 &
0 \\

Odd KV S-box (2 words/round) &
0 &
2032N &
2032N \\

Odd KV XOR &
0 &
0 &
0 \\

\midrule

\textbf{Total Layer 2b} &
\textbf{0} &
\textbf{2032N} &
\textbf{2032N} \\

\bottomrule
\end{tabular}

\end{table}

\paragraph{Layer $2b$- Key Value (KV) Generation}
\noindent

For each $p = 0,1,\ldots,9$:
\[
u^{(p)} = \overline{\mathbf{KS}}[2p],
\qquad
v^{(p)} = \overline{\mathbf{KS}}[2p + 1]
\]

\begin{equation}
\mathbf{KV}[2p]
=
u^{(p)} \oplus v^{(p)}
\label{eq:23}
\end{equation}

\begin{equation}
\mathbf{KV}[2p + 1]
=
S_{32}\!\left(u^{(p)}\right)
\oplus
S_{32}\!\left(v^{(p)}\right)
\label{eq:24}
\end{equation}

The resulting array $\mathbf{KV}[0]$, $\ldots$, $\mathbf{KV}[19]$ is the final key schedule used in encryption. Table \ref{tab:layer2b_counts} details the layer $2b$ count for $N$ rounds.

\subsubsection{Layer $3$- CTR counter arithmetic}
\noindent

For block index $j = 0,1,2,\ldots$
(a public known integer), the counter passed to the block cipher is
\begin{equation}
IV^{(j)}
=
IV^{(0)} + j
\pmod{2^{128}}.
\label{eq:25}
\end{equation}

Table \ref{tab:layer3_counts} details the layer $3$ count for $N$ rounds and $L$ known plaintext-ciphertext pairs.

\begin{table}[t]
\centering
\caption{The table shows the Layer 3 counts (N rounds). Layer 3 equations are the sparsest in the system. It has at most
3 variables per carry equation (IV bit, carry, and their product).
For small $j<16$, the carry chain is non-trivial only in the
lowest 4 bits, reducing the actual equation count further.}
\label{tab:layer3_counts}

\renewcommand{\arraystretch}{1.15}

\begin{tabular}{lccc}
\toprule
\textbf{Operation} &
\makecell{\textbf{Equations}\\\textbf{(Linear)}} &
\makecell{\textbf{Equations}\\ \textbf{(Quadratic)}} &
\textbf{New vars.} \\
\midrule

j=0 &
0 &
0 &
0 \\

j = 1\ldots L each &
128 + 3 &
128 &
128 \\

\midrule

\makecell{\textbf{Total Layer 3:}\\ \textbf{L=4, (j=0,1,2,3)}} &
128(L-1) + 3 &
128(L-1) &
128(L-1) \\

\bottomrule
\end{tabular}
\end{table}


\begin{table}[H]
\centering
\caption{The table shows the Layer 4 counts (2 sub-blocks $\times$ N rounds).}
\label{tab:layer4_counts}

\renewcommand{\arraystretch}{1.15}

\begin{tabular}{lccc}
\toprule
\textbf{Operation} &
\makecell{\textbf{Equations}\\\textbf{(Linear)}} &
\makecell{\textbf{Equations}\\ \textbf{(Quadratic)}} &
\textbf{New Vars.} \\
\midrule

Rotations \& XOR &
0 &
0 &
0 \\

\makecell{Left sub block:\\ Modular additions} &
32N &
31N &
63N \\

\makecell{Right sub block:\\ Modular additions} &
32N &
31N &
63N \\

Output constraints&
128 &
0 &
0 \\

\midrule

\textbf{Total Layer 4: L=4} &
\textbf{(64N + 128)L } &
\textbf{(62N)L} &
\textbf{(126N)L} \\

\bottomrule
\end{tabular}

\end{table}

\subsubsection{Layer $4$- Encryption}

We perform $10$-round ($N=10$) double-block ARX encryption.

\vspace{0.8em}

Parse $IV^{(j)}=(v_0,v_1,v_2,v_3)$ as four 32-bit words.

Initialise state:
\[
(c_0,c_1,c_2,c_3)\leftarrow(v_0,v_1,v_2,v_3).
\]

For $i=0,1,\ldots,9$:

\vspace{0.4em}

\textit{Left sub-block:}

\begin{align}
c_1 &\leftarrow \mathrm{ROR}_{32}(c_1,8) \tag{26} \\
c_1 &\leftarrow c_1 \boxplus c_0 \tag{27} \\
c_1 &\leftarrow c_1 \oplus \mathbf{KV}[2i] \tag{28} \\
c_0 &\leftarrow \mathrm{ROL}_{32}(c_0,3) \tag{29} \\
c_0 &\leftarrow c_0 \oplus c_1 \tag{30}
\end{align}

\vspace{0.3em}

\textit{Right sub-block:}

\begin{align}
c_3 &\leftarrow \mathrm{ROR}_{32}(c_3,8) \tag{31} \\
c_3 &\leftarrow c_3 \boxplus c_2 \tag{32} \\
c_3 &\leftarrow c_3 \oplus \mathbf{KV}[2i+1] \tag{33} \\
c_2 &\leftarrow \mathrm{ROL}_{32}(c_2,3) \tag{34} \\
c_2 &\leftarrow c_2 \oplus c_3 \tag{35}
\end{align}

Table \ref{tab:layer4_counts} details the layer $4$ counts for $N$ rounds and $L$ known plaintext-ciphertext pairs.

\subsubsection{Ciphertext Generation}
The keystream block for index $j$ is
\[
Z^{(j)} = (c_0,c_1,c_2,c_3) \in \mathbb{F}_2^{128}.
\]

For Ciphertext byte $i$=$0,1, \ldots ,15$:
\begin{equation}
C[i] = M[i] \oplus Z^{(j)}[i].
\tag{36}
\end{equation}

Finally, Table \ref{tab:combined_equation_counts} summarizes the total counts of equations and variables corresponding to $1$, $2$, and $3$ rounds, respectively, under the assumption that the adversary possesses four known plaintext-ciphertext pairs.

\begin{table*}[t]
\centering
\caption{The table shows ANF combined equation counts produced for $L=4$ known plaintext-ciphertext blocks.
}
\label{tab:combined_equation_counts}

\renewcommand{\arraystretch}{1.15}

\begin{adjustbox}{width=\textwidth}

\begin{tabular}{|l|l|c|c|c|c|c|c|}
\hline

\textbf{Layer} &
\textbf{Function} &
\textbf{N=1 eqs.} &
\textbf{N=1 vars.} &
\textbf{N=2 eqs.} &
\textbf{N=2 vars.} &
\textbf{N=3 eqs.} &
\textbf{N=3 vars.} \\

\hline

Primary variables&
 $\nu, K, IV^{(0)}$ &
--- &
320 &
--- &
320 &
--- &
320 \\
\hline

Layer 1 &
Round val generation &
2,032 &
2,032 &
4,064 &
4,064 &
6,096 &
6,096 \\
\hline

Layer 2a &
ARX key schedule &
126 &
126 &
252 &
252 &
378 &
378 \\
\hline

Layer 2b &
Key value generation &
2,032 &
2,032 &
4,064 &
4,064 &
6,096 &
6,096 \\
\hline

Layers 3 &
CTR arithmetic &
765 &
765 &
765 &
765 &
765 &
765 \\
\hline

Layers 4 &
Encryption &
1,016 &
504 &
1,520 &
1,008 &
2,024 &
1,512 \\
\hline

TOTAL &
 &
5,971 &
5,779 &
10,665 &
10,473 &
15,359 &
15,167 \\
\hline

\end{tabular}

\end{adjustbox}

\end{table*}

\subsection{Strategy: Linearization, ElimLin, and SAT-Based Algebraic Cryptanalysis}
To evaluate the algebraic attack resilience of MAVShield, an attack strategy based on \emph{linearization} \cite{courtois2003algebraic, canteaut2016lecture}, \emph{ElimLin reduction} \cite{courtois2012elimlin,gharib2015system}, and \emph{SAT-assisted solving} \cite{audemard2018glucose} was implemented. This strategy targets the multivariate quadratic ANF system generated from the cipher and attempts to iteratively simplify the nonlinear equation structure before invoking a SAT solver.

The objective of this analysis was not only to investigate potential structural weaknesses, but also to assess whether the algebraic complexity introduced by MAVShield is sufficient to resist advanced equation solving techniques employed in modern algebraic cryptanalysis \cite{kutsenko2021algebraic}.

\subsubsection{Linearization of Degree-2 Monomials}

The ANF representation of MAVShield contains nonlinear quadratic monomials of the form:
\begin{equation}
x_i x_j,
\end{equation}
which make direct solving computationally difficult. Following the classical linearization approach introduced in algebraic cryptanalysis \cite{courtois2003algebraic}, each degree-2 monomial was substituted with a new auxiliary variable
\begin{equation}
y_{ij} = x_i x_j.
\end{equation}

This transforms the original nonlinear system into a significantly larger, but purely linear system over Galois Field $\mathrm{GF}(2)$. The resulting linearized system can then be processed using Gaussian elimination techniques.

Although linearization increases the total number of variables, it enables efficient extraction of linear dependencies and rank properties of the equation system \cite{canteaut2016lecture}.

\subsubsection{Gauss-Jordan Elimination}
After linearization, the generated Boolean system was represented as a dense binary matrix over $\mathrm{GF}(2)$. Full Gauss-Jordan elimination \cite{gharib2015system} was then applied to identify:

\begin{itemize}
    \item linearly dependent equations,
    \item singleton variable assignments,
    \item rank deficiencies, and
    \item directly solvable variables.
\end{itemize}

The elimination process attempts to reduce the system into row-echelon form while preserving equivalence under Boolean arithmetic. Any row containing a single variable directly reveals the value of that variable.

The rank obtained after elimination serves as an important indicator of algebraic hardness. A low-rank system generally indicates excessive linear dependency and potential structural weakness, whereas a high rank system suggests strong diffusion and resistance against algebraic simplification.

\subsubsection{ElimLin Iterative Reduction}

To further simplify the equation system, the ElimLin algorithm proposed in \cite{courtois2012elimlin} was employed.

ElimLin operates iteratively as follows:

\begin{enumerate}[label=(\roman*)]
    \item Linearize the nonlinear ANF equations,
    \item Apply Gauss-Jordan elimination,
    \item Extract newly determined variables,
    \item Substitute these variables back into the original nonlinear system,
    \item Repeat until no additional variables can be recovered.
\end{enumerate}

The primary advantage of ElimLin is that every newly recovered variable may simplify multiple quadratic monomials, thereby progressively reducing the algebraic complexity of the system.

For weakly nonlinear ciphers, ElimLin may rapidly collapse the equation system and expose large portions of the secret state. However, for cryptographically strong constructions, the iterative reductions typically stagnate after recovering only a small subset of variables.

\subsubsection{SAT-Based Residual Solving}
After completion of ElimLin iterations, the residual ANF system was translated into conjunctive normal form (CNF) using Tseitin encoding \cite{tseitin1983complexity}. The resulting CNF instance was then processed using the \emph{Glucose4} SAT solver \cite{audemard2018glucose}.

The SAT solver attempts to determine a satisfying assignment for the remaining unknown variables while respecting all Boolean constraints generated by the cipher equations.

This final phase evaluates whether the remaining nonlinear structure of MAVShield can still be efficiently resolved after extensive algebraic preprocessing.

\subsection{Security Interpretation}

Experimental evaluation demonstrated that although linearization and ElimLin were capable of recovering a limited subset of variables, the majority of the secret-dependent nonlinear structure remained unresolved. Even after iterative elimination and SAT-assisted reduction, complete recovery of the nonce, key, and IV was not achieved. 

Here, we use $L=4$ blocks, which means the adversary has $4$ known plaintext-ciphertext pairs, all encrypted under the same master key, IV, and nonce, just with the counter (IV) incrementing for each block.

These observations indicate that MAVShield preserves substantial algebraic complexity even under aggressive preprocessing attacks. 

Table \ref{tab:elimlin_strategy_b_comparison} shows the execution traces of Strategy (Linearization, ElimLin, and SAT Solver) against MAVShield for different round configurations ($N=1,2,3$) using $L=4$ known plaintext-ciphertext blocks. Increasing the number of rounds significantly enlarges the algebraic system, resulting in higher monomial growth, increased CNF complexity, longer elimination times, and substantially reduced variable recovery efficiency. In particular, for $N=3$, ElimLin converges after recovering only a single variable, indicating strong resistance of MAVShield against iterative algebraic simplification and SAT-assisted cryptanalysis.

The inability of the solver pipeline to converge towards full key reconstruction suggests that MAVShield maintains strong resistance against advanced algebraic cryptanalysis under the evaluated attack configuration.

\begin{table*}[t]

\centering
\caption{The table shows the  results of the algebraic cryptanalysis of MAVShield using Strategy (Linearization, ElimLin, and SAT Solver) for $L=4$ known plaintext-ciphertext pairs under different round (N) configurations.}
\label{tab:elimlin_strategy_b_comparison}
\renewcommand{\arraystretch}{1.2}

\tiny
\begin{adjustbox}{width=\textwidth}
\begin{tabular}{|l|c|c|c|}
\hline
\textbf{Parameter} & \textbf{$N=1$, $L=4$} & \textbf{$N=2$, $L=4$} & \textbf{$N=3$, $L=4$} \\
\hline

Initial ANF Equations & 5,971 & 10,665 & 15,359 \\
Initial ANF Variables & 5,779 & 10,473 & 15,167 \\
Degree-2 Monomials & 7,688 & 21,710 & 40,512 \\
Linearized Variables & 13,467 & 32,183 & 55,679 \\
Linear System Size & $5,971 \times 13,467$ & $10,665 \times 32,183$ & $15,359 \times 55,679$ \\
\hline

Initial GF(2) Rank & 5,965 & 10,665 & 15,359 \\
Initial Singleton Solutions & 643 & 13 & 1 \\
Initial Linearization Time & 0.02 s & 0.06 s & 0.09 s \\
Initial Gauss--Jordan Time & 7.95 s & 46.96 s & 147.48 s \\
\hline

ElimLin Iterations & 25 & 25 & 2 \\
Variables Determined by ElimLin & 697 & 154 & 1 \\
Percentage Variables Recovered & 12.1\% & 1.5\% & 0.0\% \\
Residual Equations & 5,342 & 10,526 & 15,358 \\
Total ElimLin Time & 163.98 s & 1194.33 s & 300.16 s \\
\hline

Residual CNF Variables & 36,447 & 92,294 & 163,764 \\
Residual CNF Clauses & 124,330 & 324,899 & 581,941 \\
Unit Clauses Added & 68 & 0 & 0 \\
SAT Solver & Glucose4 & Glucose4 & Glucose4 \\
SAT Solving Time & 0.044 s & 23.297 s & Not Recovered \\
SAT Result & SAT & SAT & UNSAT \\
\hline

Recovered Nonce Correctly & $\times$ & $\times$ & $\times$ \\
Recovered Key Correctly & $\times$ & $\times$ & $\times$ \\
Recovered IV Correctly & $\times$ & $\times$ & $\times$ \\
Overall Recovery Status & Partial Recovery & Partial Recovery & No Recovery \\
\hline

\end{tabular}
\end{adjustbox}
\end{table*}

\begin{table*}[t]
    \centering
    \caption{The table shows a comparison of attacks on Speck vs. MAVShield.}
    \label{tab:speck_mavshield_comp}
    \renewcommand{\arraystretch}{1.2}

    \begin{adjustbox}{width=\textwidth}
        \begin{tabular}{|l|c|c|}
        \hline
        Phase & Speck (R=3) & MAVShield (R=3) \\
        \hline
        Known plaintext-ciphertext pair & 1 & 4 \\
        \hline
        Secret space & 64-bit & 320-bit \\
        \hline
        Linearization & 1,236 linearized vars.& 55,679 linearized vars.\\
        \hline   
        Residual equations after ElimLin & 307 eqs. &  15,358 eqs.\\
        \hline
        SAT solver input size & 11,020 clauses & 5,81,941 clauses \\
        \hline
        SAT solving time & 0.2-1.0 s (all solvers) & Not recovered \\
        \hline
         Secret recovery & Full key recovered for 3 rounds & No round configuration yielded complete key recovery \\
         \hline
        
        \end{tabular}
    \end{adjustbox}
\end{table*}

\subsection{Comparative Algebraic Analysis: MAVShield vs. Speck}

As discussed in \cite{kutsenko2021algebraic}, Speck with $3-4$ rounds is broken by the proposed attack pipeline because its only nonlinear operation is modular addition, which generates very sparse equations (at most $6$ variables each). ElimLin can propagate substitutions across the sparse structure, and the resulting SAT instance is small enough for a SAT solver to resolve within seconds to minutes. Full key recovery is confirmed in \cite{kutsenko2021algebraic} for $R \leq 4$. Notably, the Speck attack uses a single known plaintext-ciphertext pair ($L=1$) throughout. one pair already provides enough constraints to break low-round variants because the $64$-bit key space is small and the equation system is sparse enough for the solver to navigate.

As shown in Tables \ref{tab:elimlin_strategy_b_comparison} and \ref{tab:speck_mavshield_comp}, MAVShield at even one round resists the same pipeline, because the AES S-box encoding introduces $254$ auxiliary equations per byte, creating a dense interlocked nonlinear system where ElimLin stagnates after recovering only $12\%$ of variables ($N=1$) or effectively nothing ($N \geq 2$). The SAT instance it leaves behind is far larger than anything Speck produces, and the solver either returns a partial result or reports UNSAT for the recovery problem. Critically, $L=4$ known plaintext-ciphertext pairs are required even to mount a meaningful attack on MAVShield, because the $320$ secret bits ($64$-bit nonce, $128$-bit key, and $128$-bit IV) are shared across all CTR blocks, each incremented IV block presents a new encryption instance over the same unknowns, piling on constraints without introducing new secrets. Even with this four-fold data advantage ($L=4)$ over Speck's single-pair setup ($L=1)$, full secret recovery fails at every tested round count.

\section{Experimental Setup} \label{sec:experimental_setup}
In this section, we describe the software and hardware components used and the experimental setup for our drone testbed.

\subsection{Hardware Setup}

\begin{figure*}[hbt]
    \centering
    \includegraphics[width=0.95\linewidth]{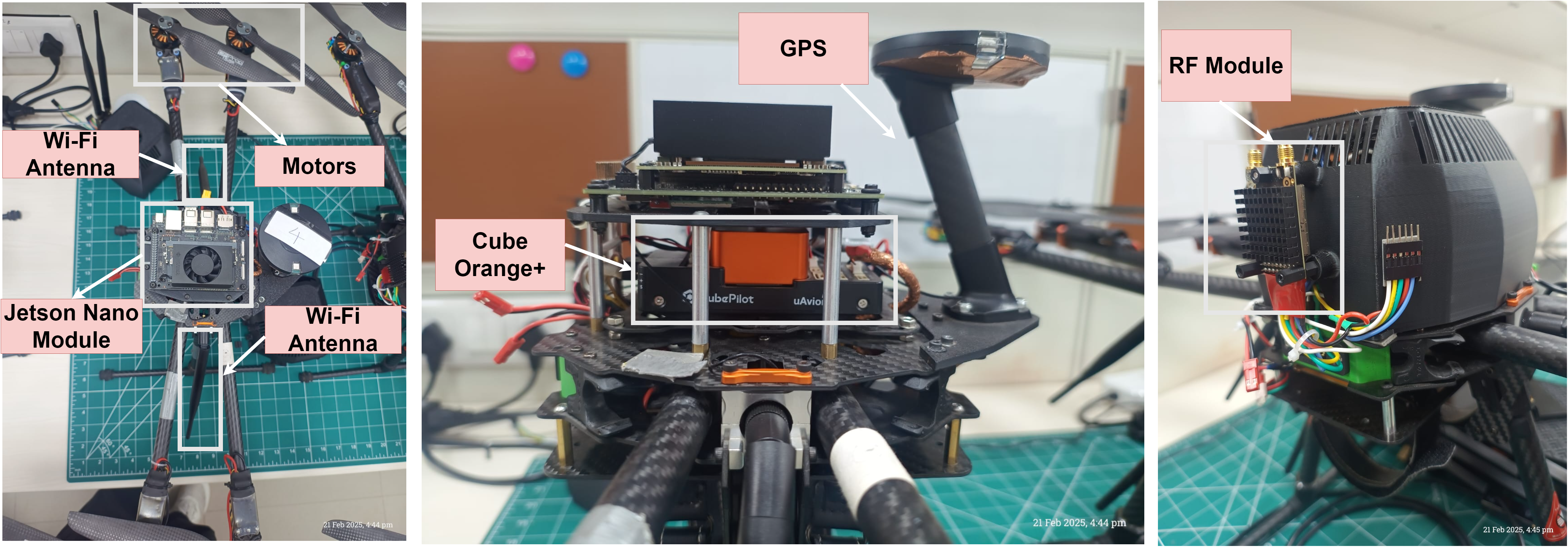}
    \caption{The figure shows the components integrated on each UAV platform.}
    \label{fig:swarm_drone_setup}
\end{figure*}

Each UAV in the swarm is based on a custom-built quadcopter platform equipped with a comprehensive suite of hardware components to support autonomous flight and inter-node coordination as shown in Figure \ref{fig:swarm_drone_setup}. The core flight control is managed by a Pixhawk Cube Orange$^+$ flight controller \cite{cube}, integrated with a Here4 dual-band high-precision real-time kinematic (RTK) GNSS module \cite{GPS} for centimeter-level positioning accuracy. An NVIDIA Jetson Orin Nano embedded computing module \cite{orin_nano} serves as the CC, enabling onboard processing for tasks such as real-time perception, decision-making, and communication handling. For long-range telemetry, an \emph{RFD$900$x} radio modem \cite{RFD} is employed, while a dual-band ($2.4$ GHz / 5 GHz) IEEE 802.11ac-based Intel BE$200$ wireless NIC with WiFi antennas \cite{NIC} facilitates peer-to-peer data exchange. The system is powered by a $6$-cell (6S) $10,000$ mAh Lithium Polymer (LiPo) battery \cite{LiPo}, providing sufficient energy for extended flight durations. Propulsion is achieved using four brushless DC motors, each controlled via dedicated electronic speed controllers (ESCs) \cite{motor}. A ground-based WiFi access point- TP Link Archer C60, which supports dual-band ($2.4$ GHz / $5$ GHz) operation and complies with IEEE 802.11ac, provides additional communication support for controlled mission execution and monitoring.

The Jetson Orin Nano module used for testing has the following characteristics:
\begin{itemize}
    \item CPU: 6-core, 1.7 GHz, Arm Cortex-A78AE 64-bit central processor, 
    \item GPU: 512-core, 1020 MHz, NVIDIA Ampere architecture graphical processor,
    \item RAM: 4 GB 64-bit LPDDR5 51 GB/s,
    \item Storage: 500 GB, Samsung SN770 SSD.
\end{itemize}

\subsection{Software Setup}
For the software setup, we perform an upgrade of the factory version of the Jetson Orin Nano by installing the latest firmware (QSPI image) and simultaneously flashing the JetPack $6.2$ image. This process is conducted using a host x$86$ personal computer (PC) with Ubuntu 22.04.5 and the necessary software components loaded onto a solid state device (SSD) \cite{jetson_setup}. Upon completion of the setup, we proceed to install the requisite Python libraries essential for executing the RF or WiFi-based communication script. This script encompasses a formation control algorithm, a collision avoidance mechanism utilizing the APF algorithm, and the $xcrypt$ files necessary for all encryption and decryption schemes, which are stored on the CC of each UAV. 

A WiFi access point is utilized to establish a WiFi network, enabling remote connectivity to the Jetson Orin Nano module, which has the Ubuntu 22.04.5 operating system, via the secure shell (SSH) protocol. This configuration allows for remote access and management of the system, thereby negating the need for direct physical interaction with the device. Through the SSH connection, users can efficiently execute commands, monitor system performance, and carry out administrative tasks from a distance.

\section{Implementation} \label{sec:implementation}

\subsection{Software Implementation}

Initially, each UAV is preloaded with Python scripts for RF and WiFi-based communication, integrated with selected encryption algorithms-- MAVShield, Speck-CTR, AES-CTR, ChaCha20, or Rabbit, and equipped with formation control and collision avoidance functionalities.

A ground-based WiFi router establishes a secured network to which all participating UAVs are preassigned to connect. The network is password-protected to prevent unauthorized UAVs from joining. The SSH protocol is utilized to remotely access the Ubuntu 22.04.5 operating system running on the Jetson Orin Nano module onboard each UAV. Access to each UAV requires a distinct set of SSH credentials, which helps safeguard the system from external access and prevents potential tampering with the communication algorithms during flight.

\subsection{Mission Execution}

At the onset of the mission, all UAVs are powered via batteries and positioned on the ground in a straight line with an inter-vehicle spacing of approximately $10$ meters. Each UAV is assigned a unique drone ID from the set $\{1,2,3,4\}$, with UAV $2$ designated as the master drone.

A ground-based access point connects all UAVs to a user-operated PC, which is used to initiate Secure Shell sessions into the onboard Ubuntu operating systems of each UAV via the command-line interface. Once connectivity is established and the relevant control scripts are launched on each UAV, the master drone initiates takeoff from its home position at a speed of $2.5$ meters per second. The subsequent process is detailed below:
\begin{itemize}
\item Telemetry data is continuously broadcast by each UAV to the rest of the swarm; however, only the follower UAVs utilize the received information to achieve coordinated formation flight. Once the master UAV exceeds an altitude of $2$ meters ascertained via its onboard ($DISTANCE\_SENSOR$) telemetry, the follower UAVs (drone ID: $1, 3,$ and $4$) initiate their takeoff. All UAVs ascend to a designated formation altitude of $20$ meters, maintaining a lateral separation of $10$ meters, and followers align themselves at a fixed formation angle of $90^\degree$ relative to the leader UAV, forming a predefined geometric configuration of \emph{straight line} (Formation $1$) as illustrated in Figure \ref{F1}. Next, the UAVs maintain their Formation $1$ positions in a hover state for $10$ seconds. The UAVs transition to different altitudes to mitigate collision risks during the subsequent change of formation.

\begin{figure}[hbt]
    \centering
    \includegraphics[width=0.95\linewidth]{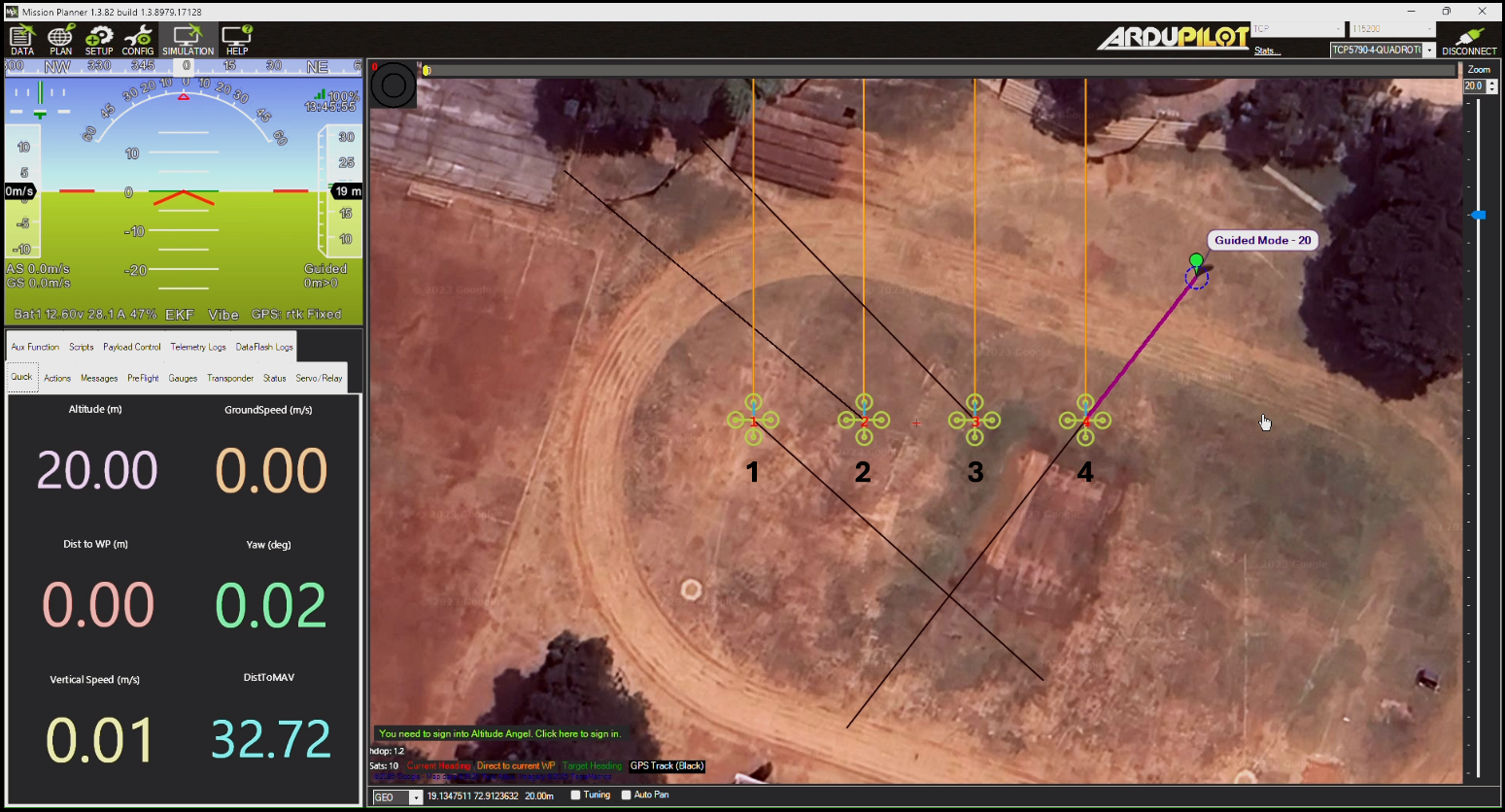}
    \caption{The figure shows Formation 1, in which there is a straight line geometric pattern of the UAV swarm.}
    \label{F1}
\end{figure}

\item As illustrated in Figure \ref{F2}, during Formation $2$ (\emph{T-shape}), all follower UAVs (drone ID: $3$, $4$, and $1$) move to the designated formation altitude of $20$ meters and assume positions $10$ meters away from the master UAV. Each follower aligns at angular offsets of $90\degree$, $180\degree$, and $270\degree$ relative to the leader's true north orientation. After a $10$ second hover in Formation $2$, the UAVs depart to different altitudes to provide the necessary vertical clearance for the upcoming formation transition.

\begin{figure}[hbt]
    \centering
    \includegraphics[width=0.95\linewidth]{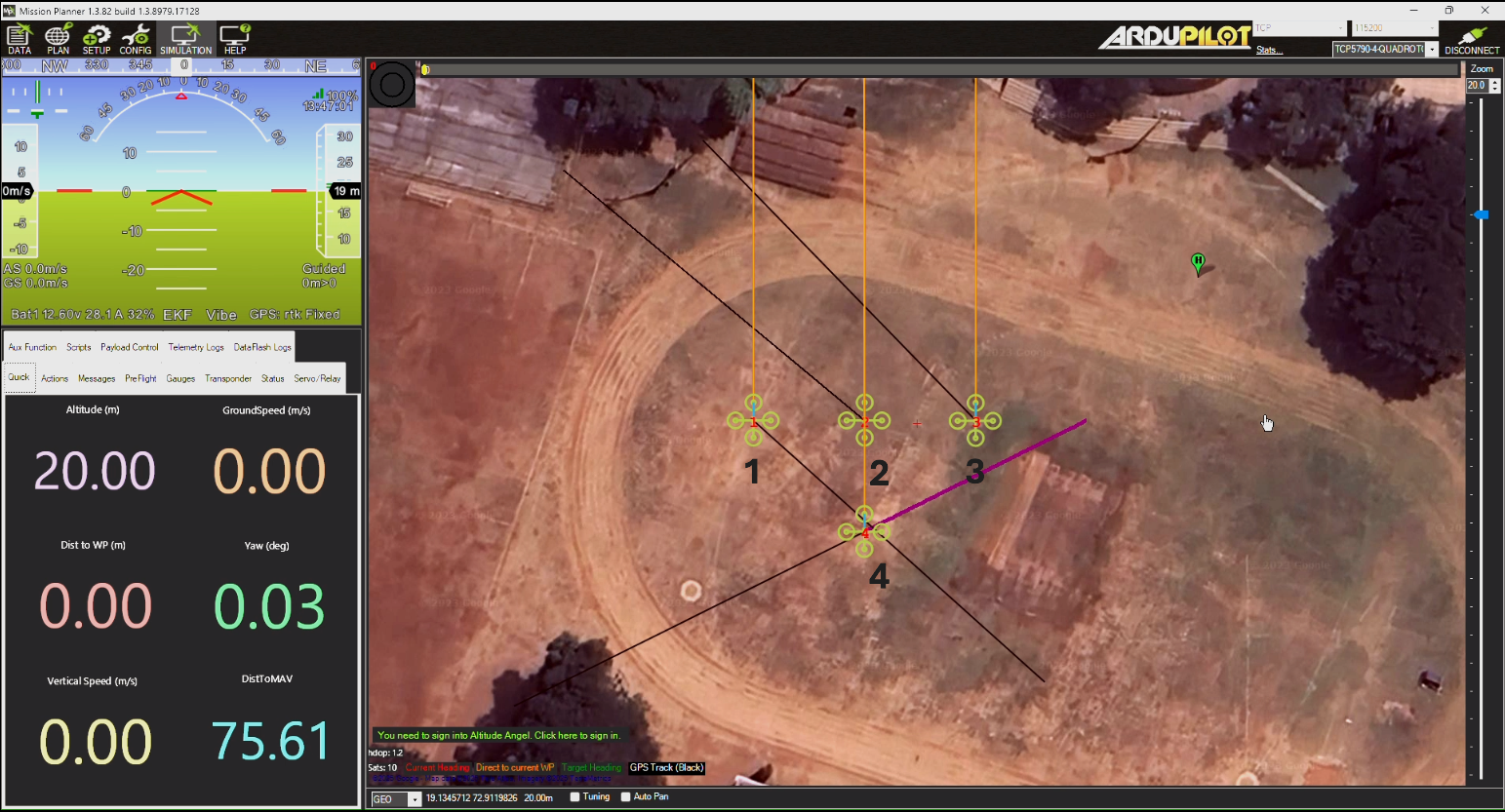}
    \caption{The figure shows Formation  2, in which there is a T-shaped geometric pattern of the UAV swarm.}
    \label{F2}
\end{figure}

\item In Formation $3$ (\emph{Y-shape}), as depicted in Figure \ref{F3}, all UAVs reach the formation altitude of $20$ meters. The follower drones (ID: $3$, $4$, and $1$) position themselves at a radial distance of $15$ meters from the leader UAV, forming angles of $45\degree$, $180\degree$, and $315\degree$, respectively, with respect to the leader’s true north reference and remain in their goal positions for $10$ seconds.

\begin{figure}[hbt]
    \centering
    \includegraphics[width=0.95\linewidth]{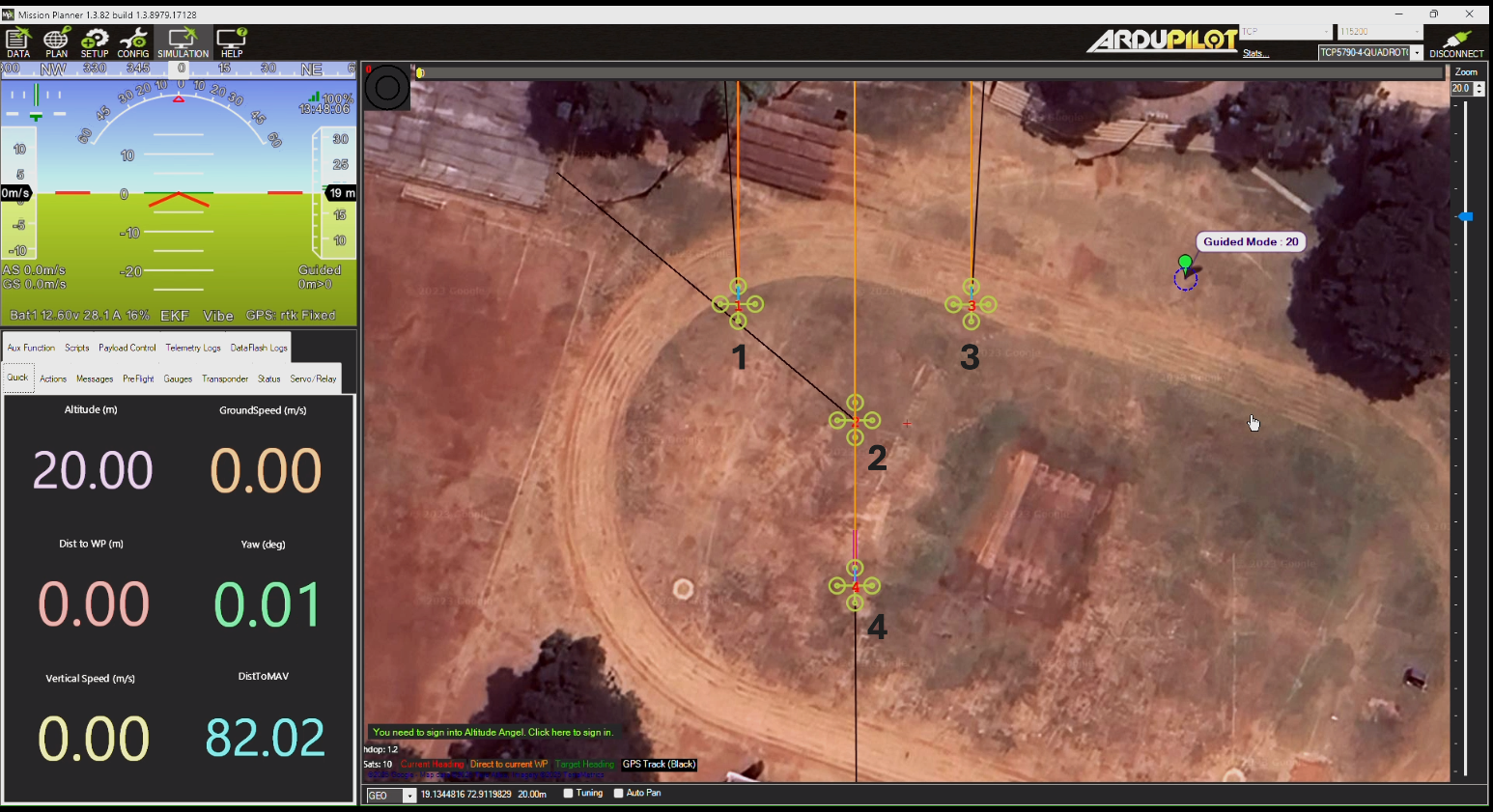}
    \caption{The figure shows Formation 3, in which there is a Y-shaped geometric pattern of the UAV swarm.}
    \label{F3}
\end{figure}

\item During each formation process, if the inter-UAV distance falls below a threshold of $7.5$ meters, the collision avoidance mechanism is triggered. This algorithm calculates the current positions of the leader and follower UAVs using MAVLink telemetry data ($GLOBAL\_POSITION\_INT$), ensuring a safe separation among them. The follower UAVs then proceed towards their respective goal positions with a dynamically adjusted horizontal velocity-- initially set to $1$ meter per second, guided by the modified APF algorithm to avoid collisions.

\item Upon completion of the formation maneuver, all UAVs initiate a vertical descent at a speed of $2.5$ meters per second and successfully return to their respective home positions.
\end{itemize}

\section{Security Analysis using Wireshark Packet Analyzer} \label{sec:wireshark}

Wireshark \cite{sanders2017practical} is a network packet analyzer that operates on the same network as the communicating swarm UAVs. It captures packets in real time, allowing detailed inspection of all data exchanged over the network. It supports a wide range of protocols, facilitating in-depth analysis of network communications. Captured packets are presented in a structured format, including information such as timestamps, source and destination IP addresses, protocol types, and payload contents. To enable MAVLink protocol analysis within the Wireshark interface, the $mavlink\_2\_common.lua$ script is used as a plugin, which allows proper parsing and interpretation of MAVLink messages.

MAVLink packets broadcast by each swarm UAV can be intercepted by any node connected to the same WiFi network. In this setup, the UAVs are assigned IP addresses $192.168.0.110$, $192.168.0.111$, $192.168.0.112$, and $192.168.0.113$, where $192.168.0.110$ corresponds to the master UAV and the remaining addresses belong to follower UAVs.  

\begin{figure}[hbt] 
    \centering
    \subfloat[]{
        \includegraphics[width=0.98\linewidth]{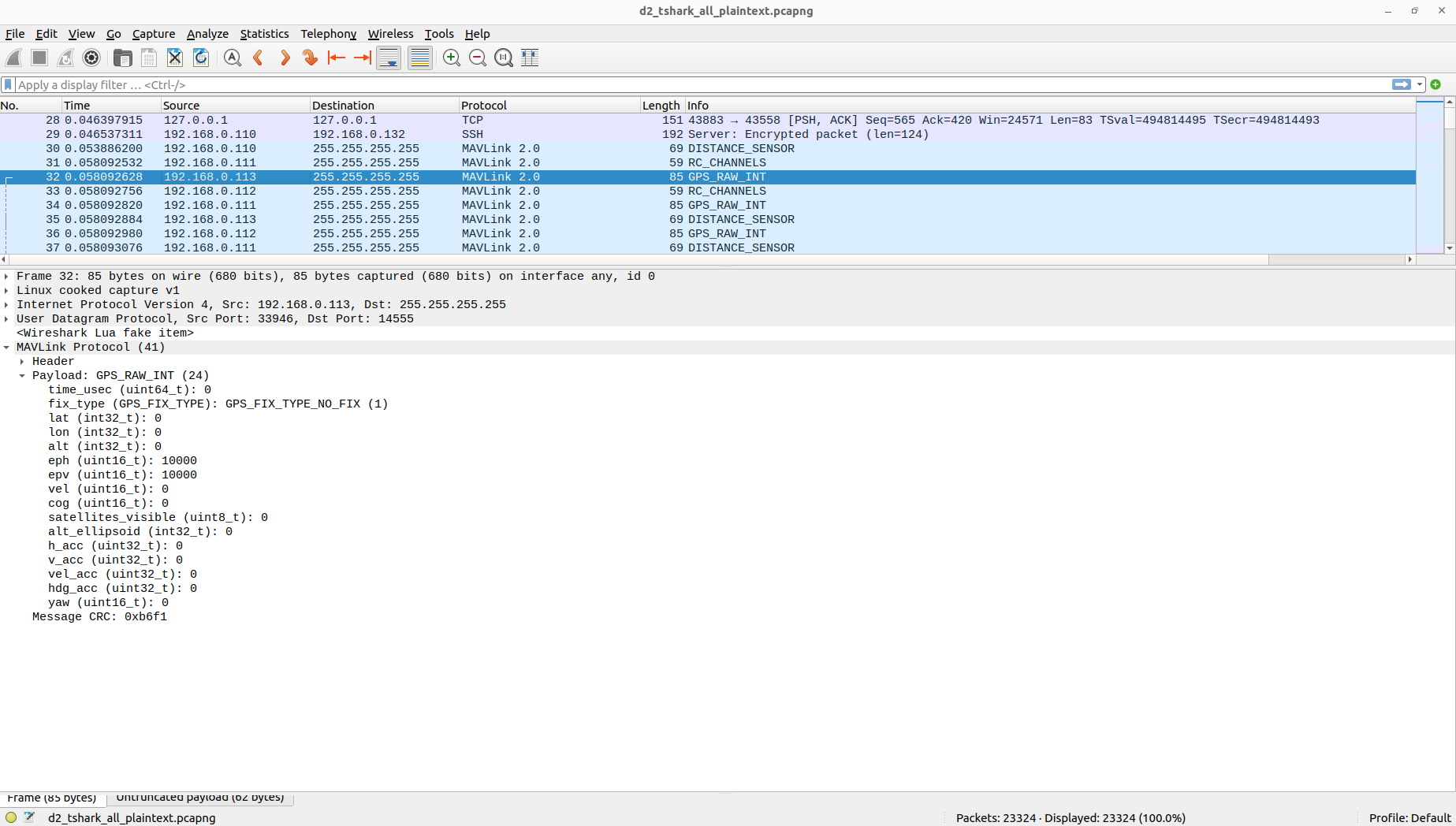}
        \label{fig:tshark_pt1}
    }
    \hfill
    \subfloat[]{
        \includegraphics[width=0.98\linewidth]{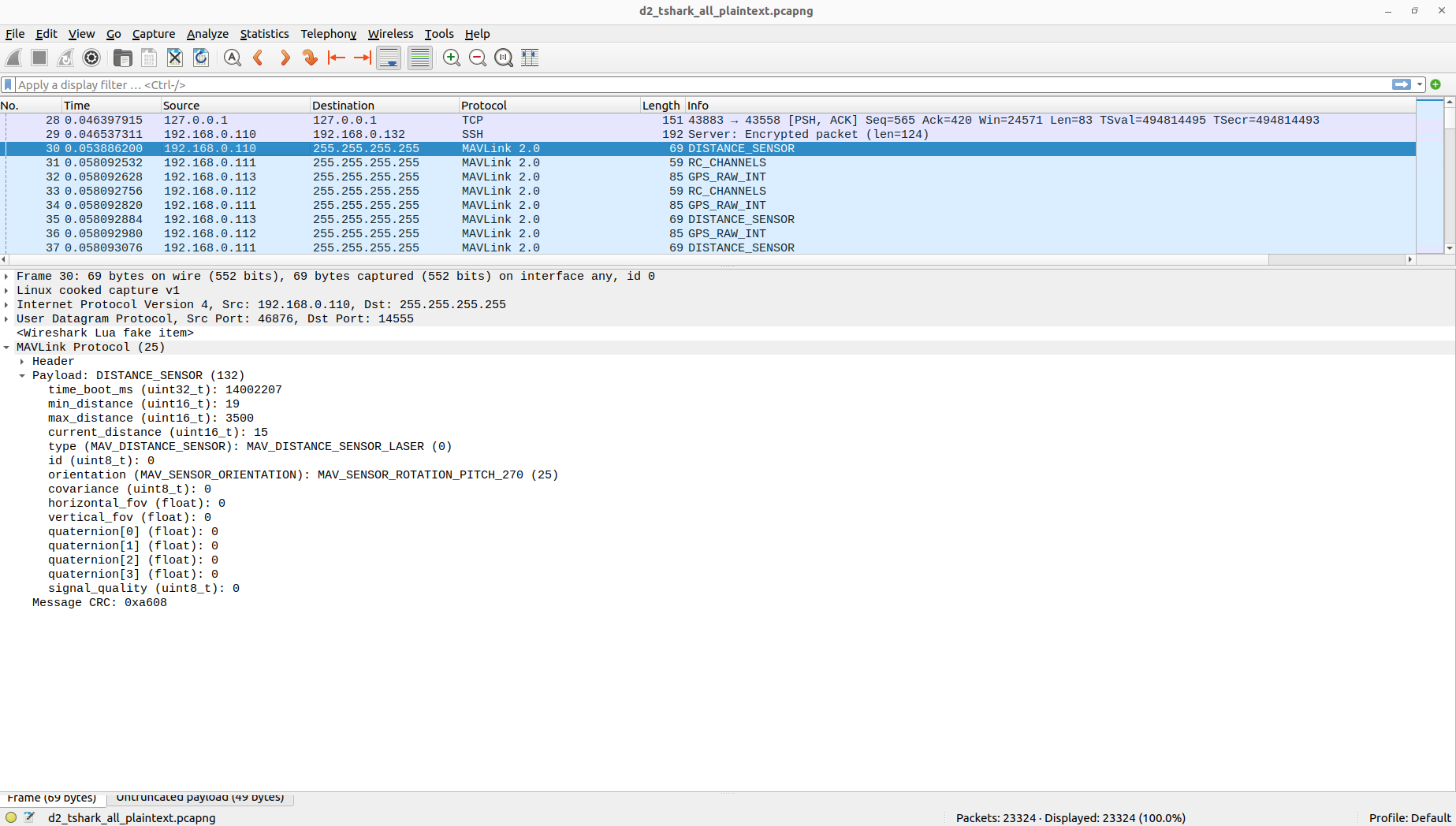}
        \label{fig:tshark_pt2}
    }
    \caption{The figures show real-time Wireshark captures of unencrypted MAVLink broadcast traffic originating from swarm UAVs 1 and 2. Figure (a) (respectively, (b)) shows drone 1 (respectively, 2) broadcasting unencrypted telemetry data to the other UAVs in the swarm.}
\end{figure}

\begin{figure}[hbt] 
    \centering
    \subfloat[]{
        \includegraphics[width=0.98\linewidth]{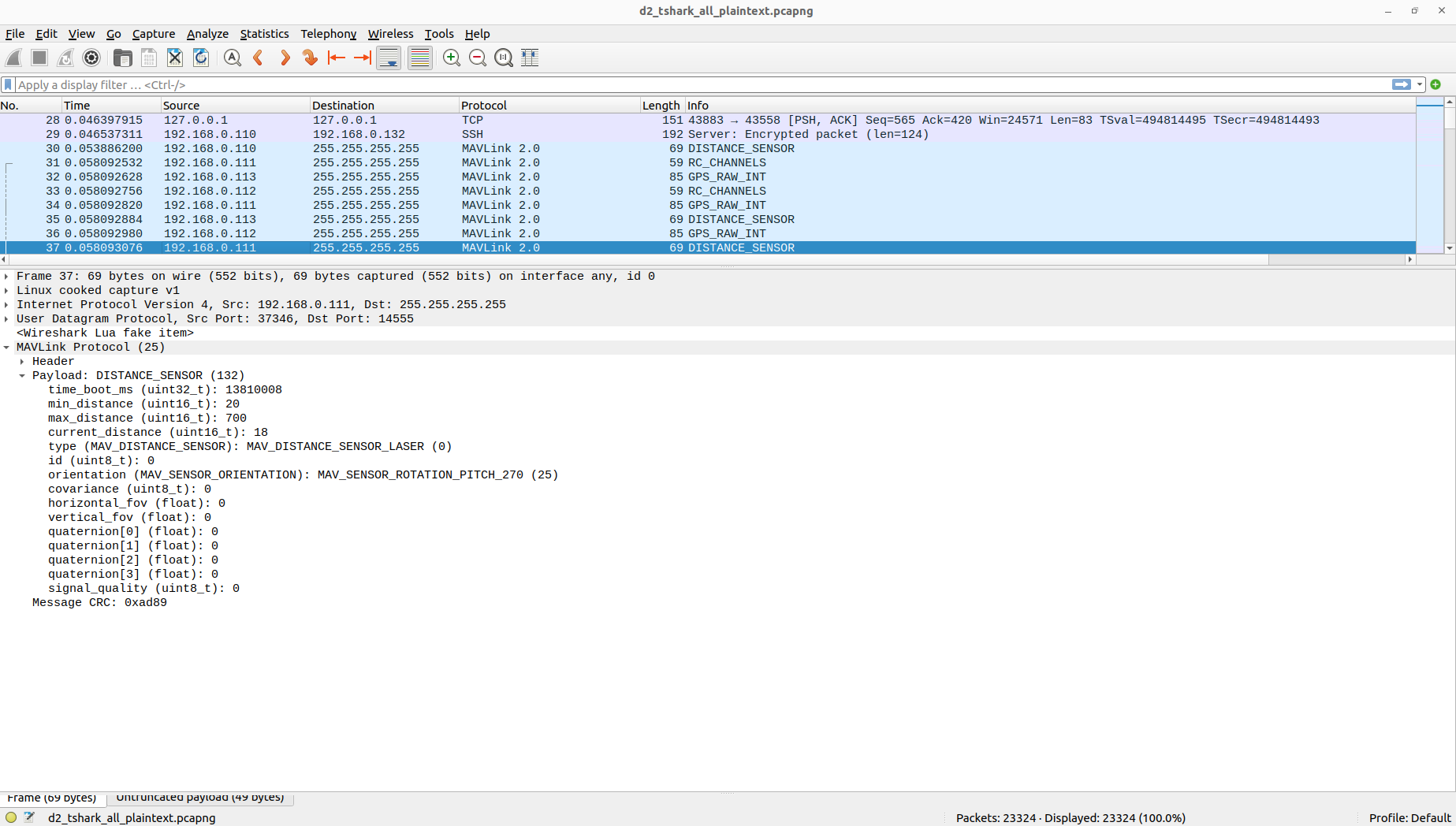}
        \label{fig:tshark_pt3}
    }
    \hfill
    \subfloat[]{
        \includegraphics[width=0.98\linewidth]{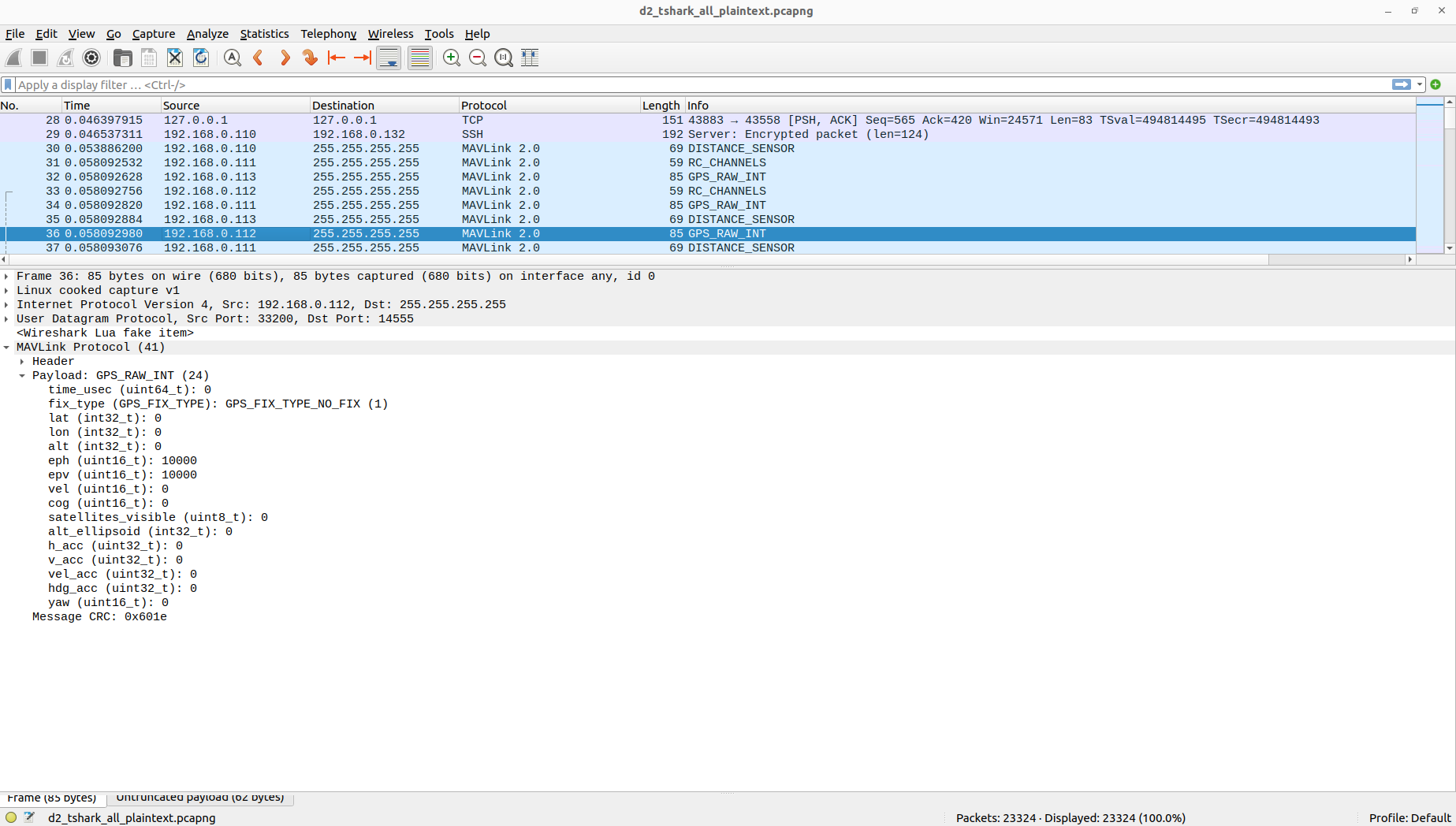}
        \label{fig:tshark_pt4}
    }
    \caption{The figures show real-time Wireshark captures of unencrypted MAVLink broadcast traffic originating from swarm UAVs 3 and 4. Figure (a) (respectively, (b)) shows drone 3 (respectively, 4) broadcasting unencrypted telemetry data to the other UAVs in the swarm.}
\end{figure}

When communication is unencrypted, real-time packet interception is demonstrated in Figures \ref{fig:tshark_pt1}, \ref{fig:tshark_pt2}, \ref{fig:tshark_pt3}, and \ref{fig:tshark_pt4}. As shown, the payload is transmitted in plaintext and can be readily decoded using the MAVLink library, thereby revealing sensitive information exchanged among the participating drones.

\begin{figure}[hbt] 
    \centering
    \subfloat[]{
        \includegraphics[width=0.98\linewidth]{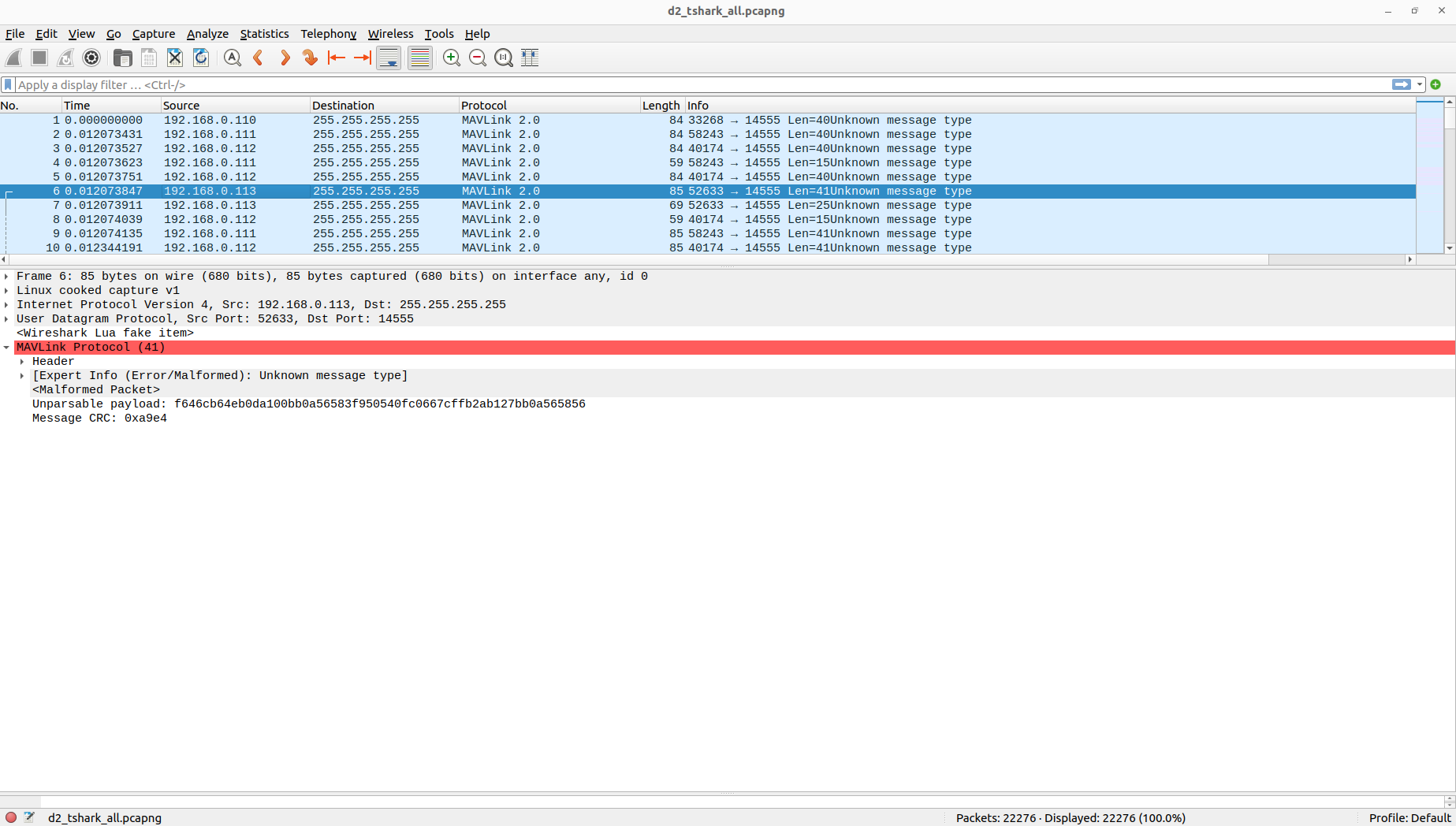}
        \label{fig:tshark_mavshield1}
    }
    \hfill
    \subfloat[]{
        \includegraphics[width=0.98\linewidth]{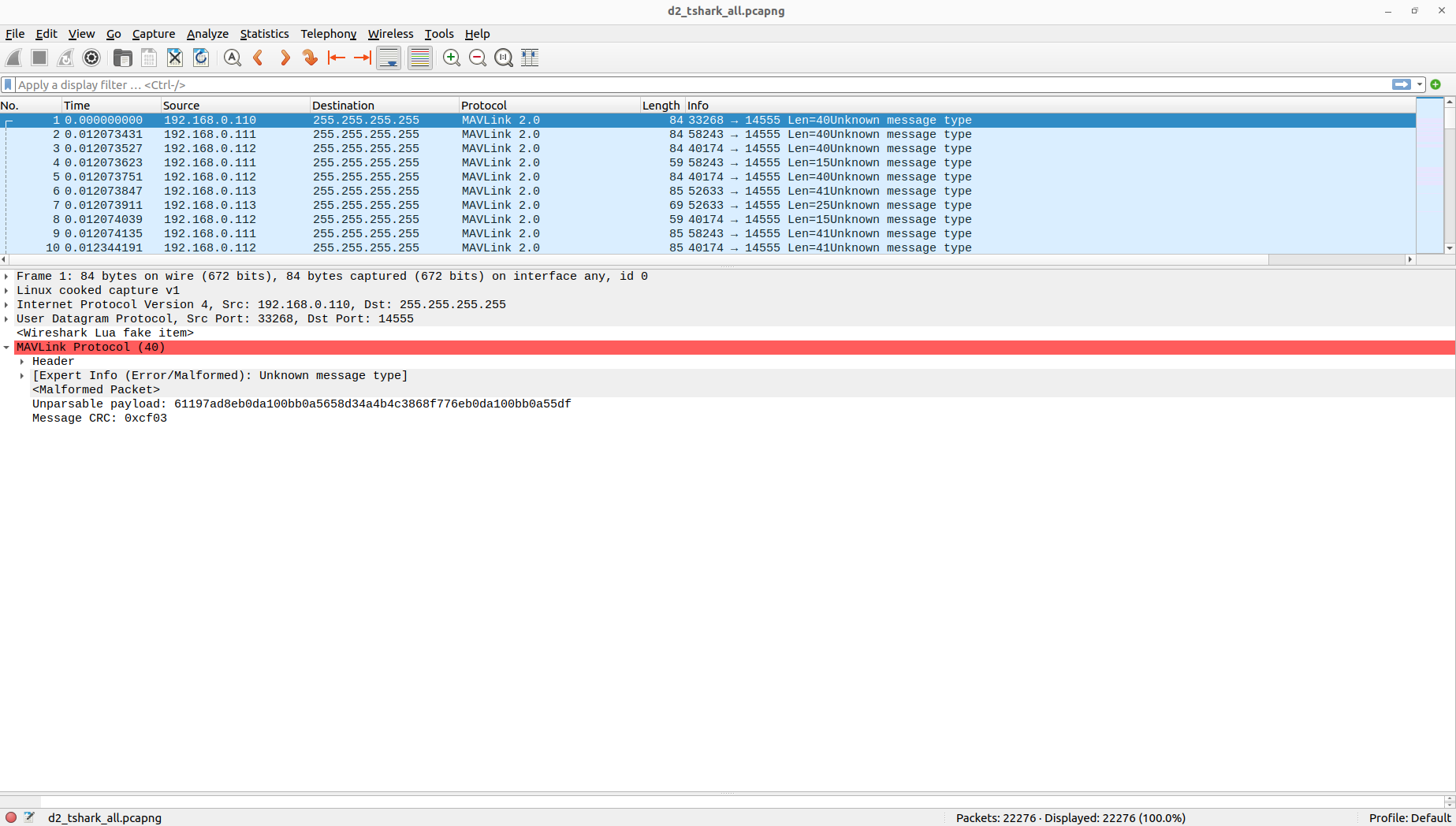}
        \label{fig:tshark_mavshield2}   
    }
    \caption{The figures show real-time Wireshark captures of encrypted MAVLink broadcast traffic originating from swarm UAVs 1 and 2. Figure (a) (respectively, (b)) shows drone 1 (respectively, 2) broadcasting encrypted telemetry data to the other UAVs in the swarm.}
\end{figure}

\begin{figure}[hbt] 
    \centering
    \subfloat[]{
        \includegraphics[width=0.98\linewidth]{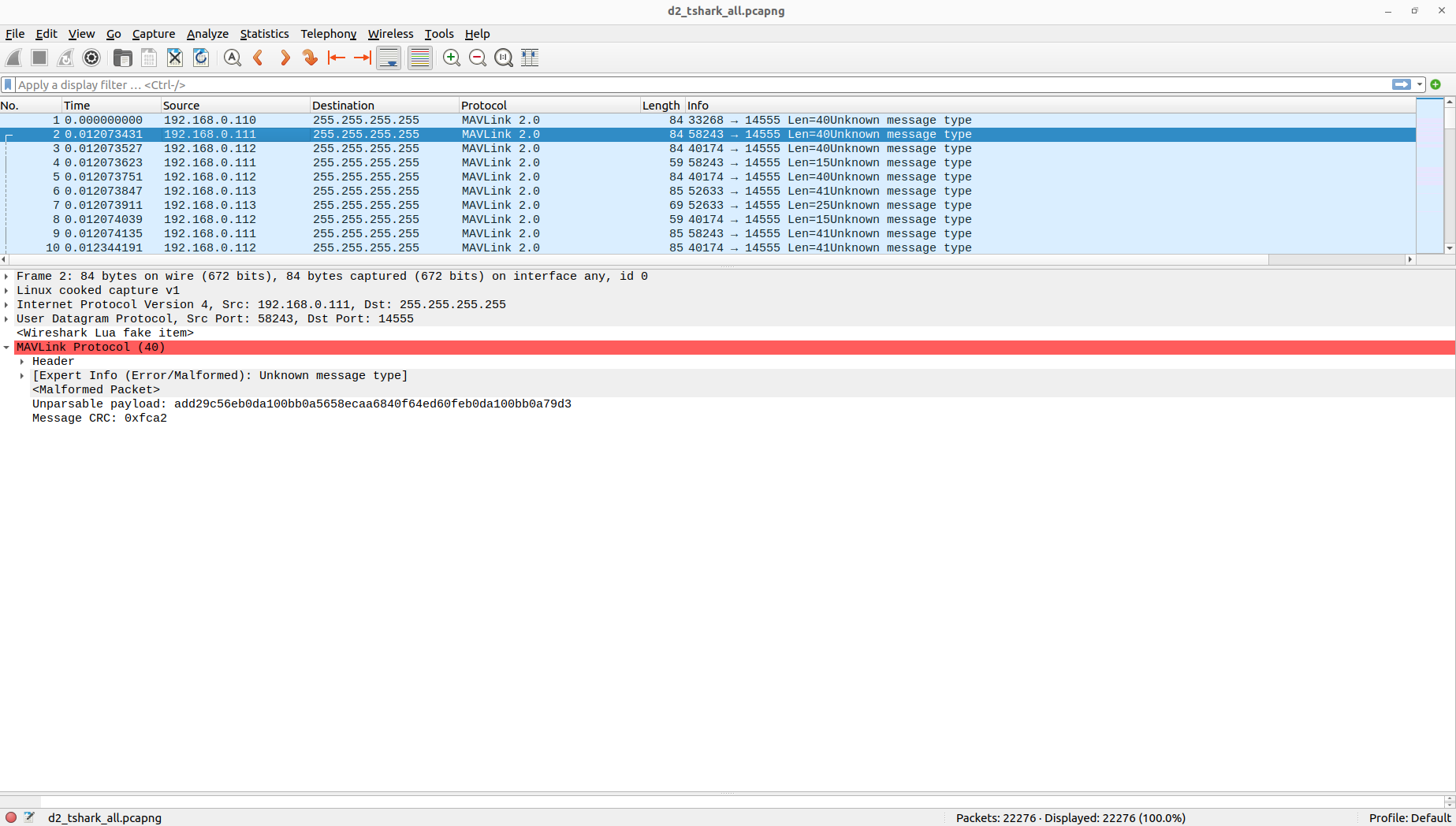}
        \label{fig:tshark_mavshield3}   
    }
    \hfill
    \subfloat[]{
        \includegraphics[width=0.98\linewidth]{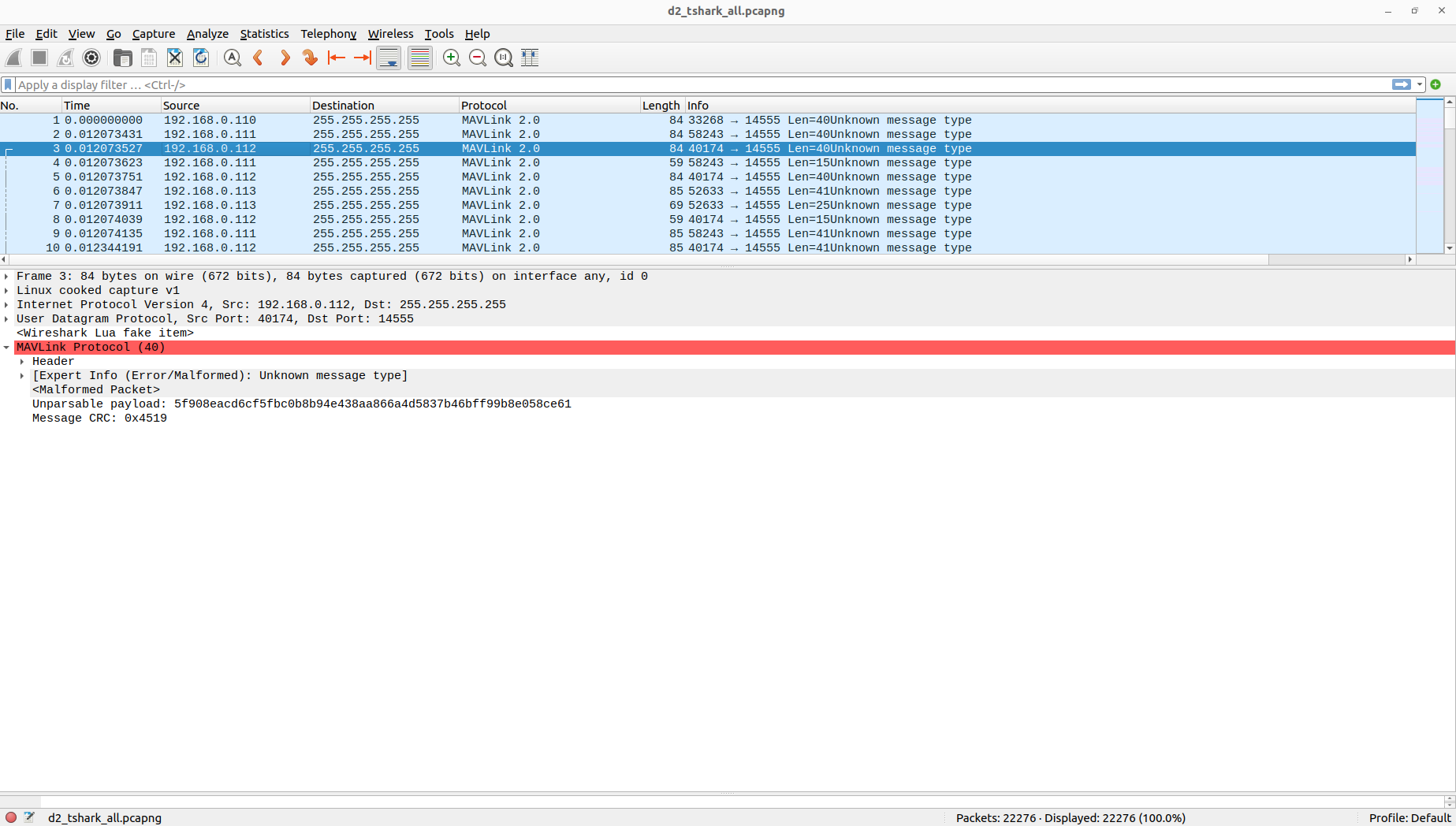}
        \label{fig:tshark_mavshield4}   
    }
    \caption{The figures show real-time Wireshark captures of encrypted MAVLink broadcast traffic originating from swarm UAVs 3 and 4. Figure (a) (respectively, (b)) shows drone 3 (respectively, 4) broadcasting encrypted telemetry data to the other UAVs in the swarm.}
\end{figure}

In contrast, when communication is encrypted, real-time packet interception is illustrated in Figures \ref{fig:tshark_mavshield1}, \ref{fig:tshark_mavshield2}, \ref{fig:tshark_mavshield3}, and \ref{fig:tshark_mavshield4}. In this case, encryption is applied at the application layer through a Python script running on the CC of each swarm UAV. It can be seen that the applied cipher effectively conceals the payload data across all broadcast sources, preventing MAVLink from parsing the encrypted content. These results demonstrate that no meaningful information can be extracted from the exchanged packets, and hence the confidentiality of mission-critical information among the drones is preserved. 

\section{Performance Evaluation} \label{sec:performance_eval}

To assess the efficiency and reliability of the proposed communication frameworks, a comparative performance analysis was conducted among various encryption algorithms against the unencrypted communication baseline across both WiFi and RF-based channels. The evaluation is characterized by three primary metrics: CPU load ($\%$), RAM usage ($MB$), and PDR ($\%$). These metrics provide insights into the computational overhead, memory footprint, and communication reliability associated with each scheme.

\subsection{Experiment Methodology}
The evaluation utilizes a swarm of four UAV nodes configured in a decentralized wireless mesh network. In this many-to-many architecture, MAVLink telemetry packets are broadcasted from each node to the remaining three peers simultaneously, simulating high-density traffic typical of coordinated swarm operations. Data collection was aggregated over four independent flight rounds to ensure statistical significance. For each UAV and each round, system-level metrics were captured over a $180$ second stabilized window, resulting in $181$ data points per flight round per node to isolate the algorithmic performance from network induced jitter.

\subsection{Comparative Analysis among Security Algorithms} 
\begin{figure}[hbt]
    \centering
    \includegraphics[width=0.95\linewidth]{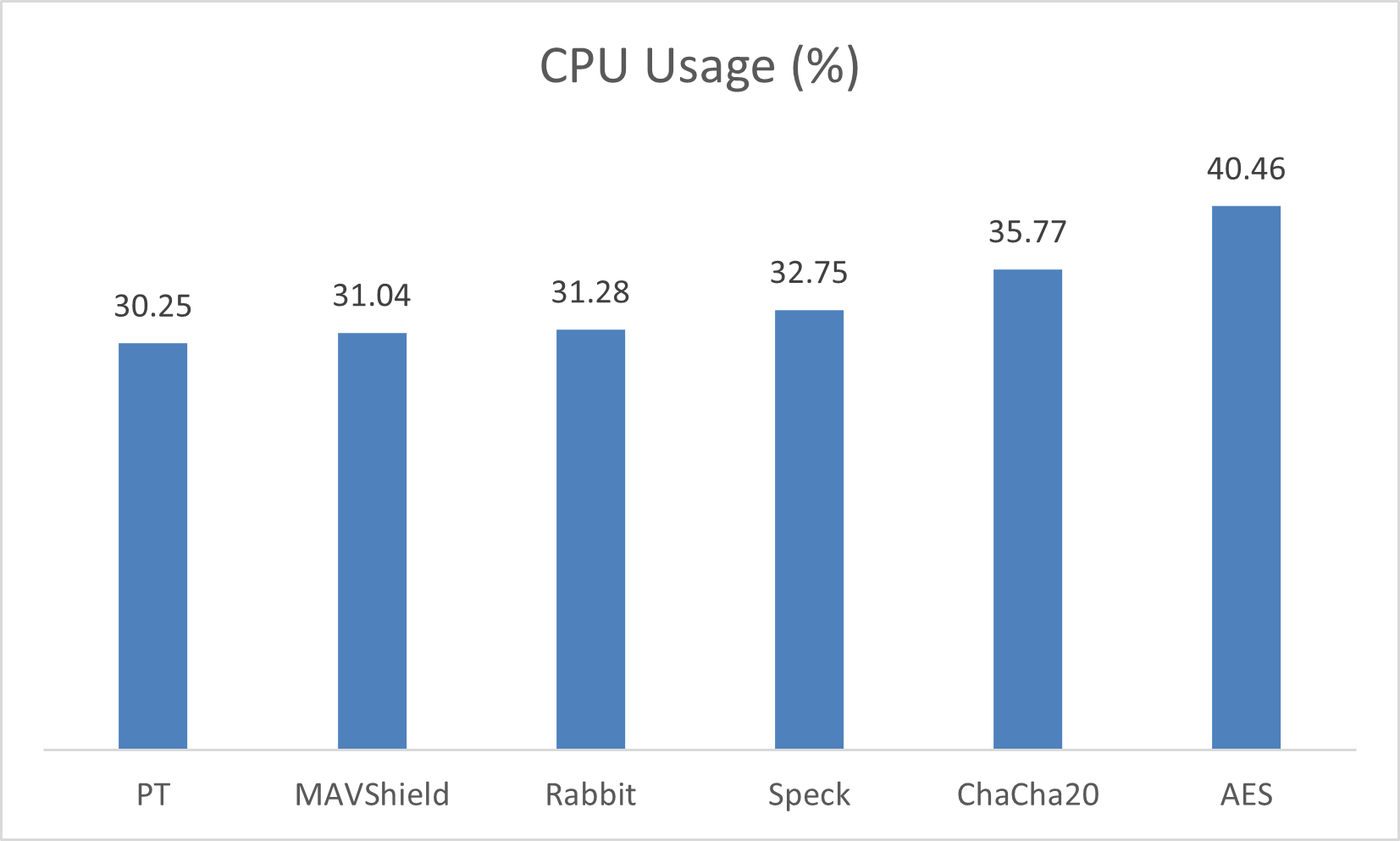}
    \caption{The figure shows the CPU load (\%) distribution across the WiFi-based mesh network for follower UAV node 3.}
    \label{fig:WiFi_CPU_orin2}
\end{figure}

\begin{figure}[hbt]
    \centering
    \includegraphics[width=0.95\linewidth]{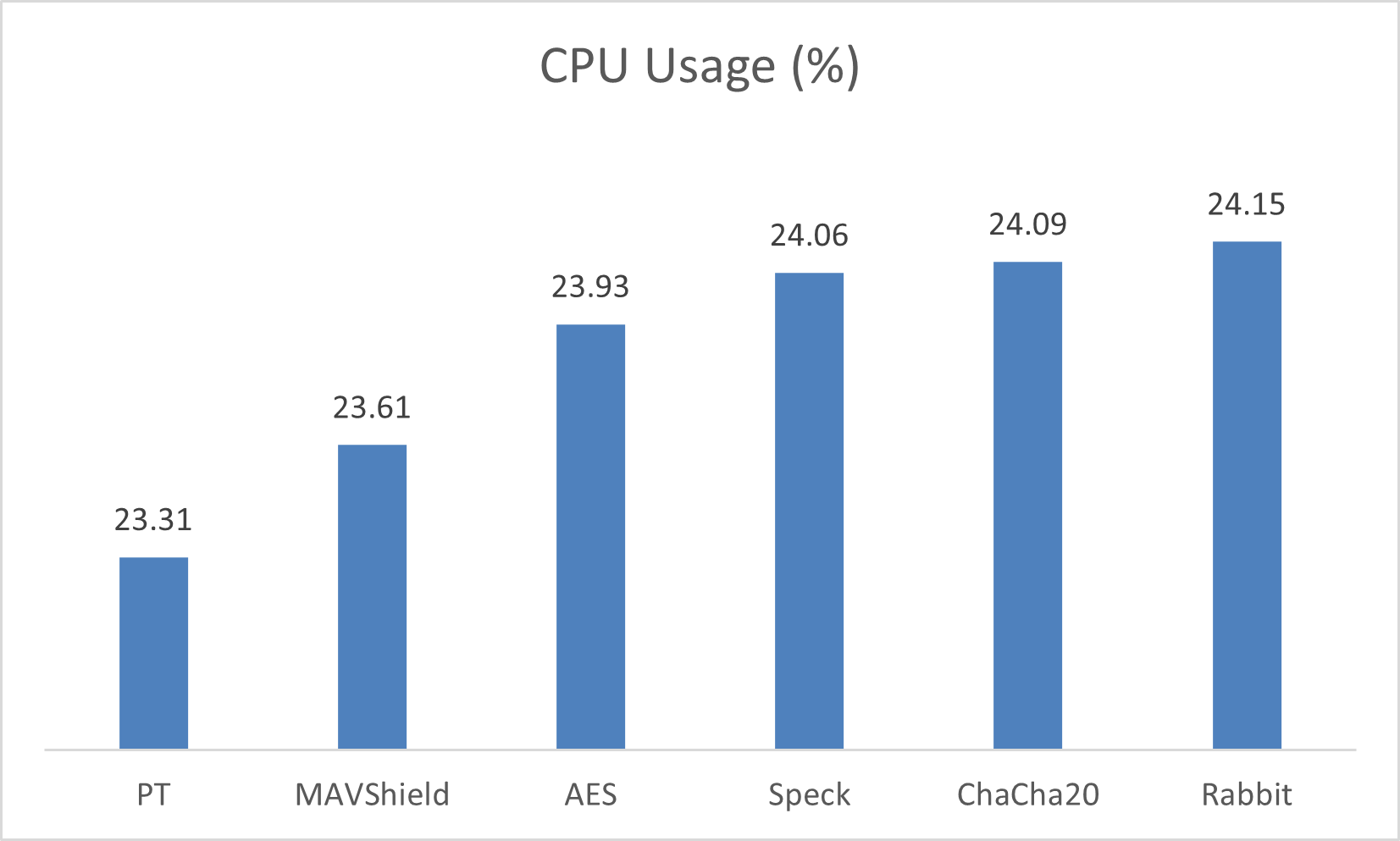}
    \caption{The figure shows the CPU load (\%) distribution across the RF-based mesh network for follower UAV node $3$.}
    \label{fig:RF_CPU_orin2}
\end{figure}

\paragraph{CPU Load ($\%$)} 
It quantifies the computational load on each UAV for concurrently handling data multi-processing, encryption, and mesh routing across three simultaneous data streams. Figures \ref{fig:WiFi_CPU_orin2} and \ref{fig:RF_CPU_orin2} present the temporal CPU utilization profiles for follower UAV node $3$ under WiFi and RF communication, respectively.  These results provide a comparative benchmark of system load across the five security schemes relative to an unencrypted communication baseline, highlighting the processing overhead introduced by securing the decentralized mesh network.

In the WiFi-based swarm communication environment, the CPU utilization is significantly higher across all schemes due to high data rates, frequent packet generation, and concurrent multi-node communication. As shown in Figure \ref{fig:WiFi_CPU_orin2}, unencrypted communication exhibits the lowest CPU usage due to the absence of cryptographic overhead, while MAVShield proposed in our prior work \cite{Dxtbhavya}, achieves a comparable CPU utilization of $31.04\%$, highlighting its lightweight design and efficient integration. In contrast, other cryptographic algorithms incur higher computational costs due to more complex operations and increased per-packet processing.

Under RF communication, the CPU utilization is considerably lower across all schemes as compared to WiFi due to reduced data rates and packet frequency. As shown in Figure \ref{fig:RF_CPU_orin2}, MAVShield achieves a CPU utilization of $23.61\%$, which is close to that of unencrypted communication, and outperforms other algorithms that introduce relatively higher overhead even in constrained conditions.
\\

\begin{figure}[hbt]
    \centering
    \includegraphics[width=0.95\linewidth]{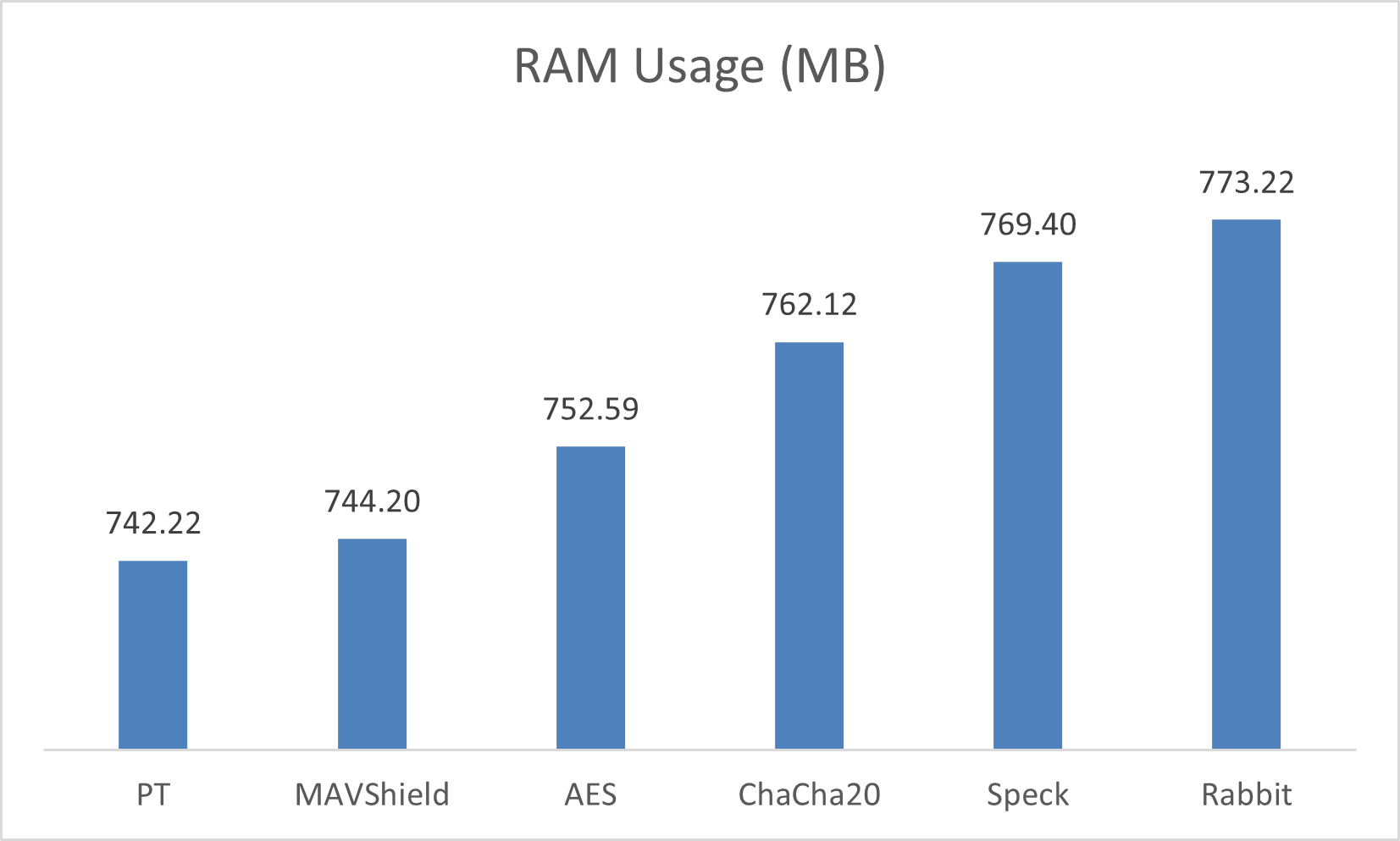}
    \caption{The figure shows the RAM usage (MB) distribution across the WiFi-based mesh network for follower UAV node $3$.}
    \label{fig:WiFi_RAM_orin2}
\end{figure}

\begin{figure}[hbt]
    \centering
    \includegraphics[width=0.95\linewidth]{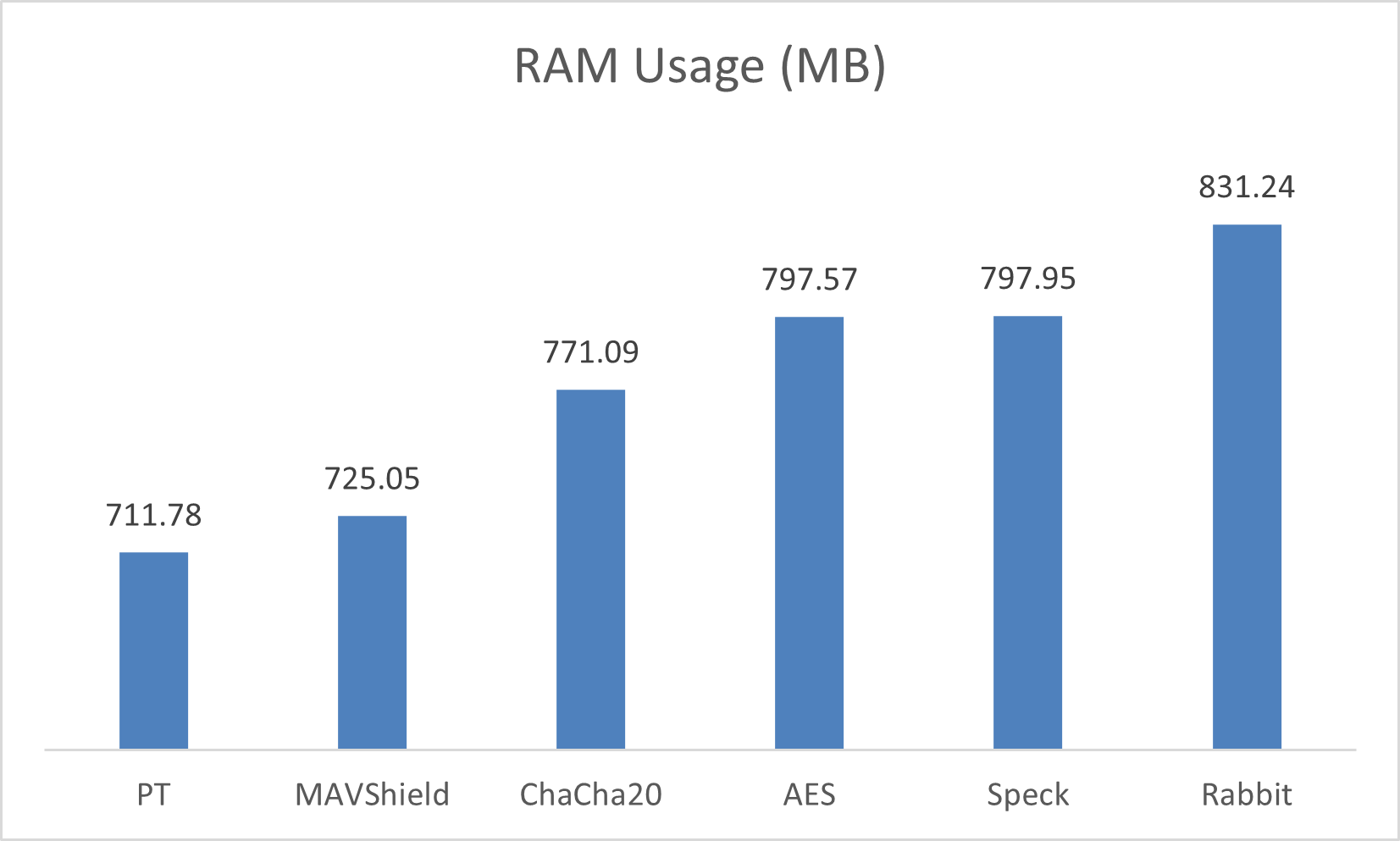}
    \caption{The figure shows the RAM usage (MB) distribution across the RF-based mesh network for follower UAV node $3$.}
    \label{fig:RF_RAM_orin2}
\end{figure}

\paragraph{RAM Usage ($MB$)}
It evaluates the memory footprint of the communication stack and buffering requirements for managing multiple incoming and outgoing telemetry streams. Figures \ref{fig:WiFi_RAM_orin2} and \ref{fig:RF_RAM_orin2} illustrate the RAM utilization for follower UAV node $3$ under WiFi and RF communication, respectively. As illustrated in Figure \ref{fig:WiFi_RAM_orin2}, in the case of WiFi, MAVShield maintains a low memory footprint of $744.20$~MB, which is close to that of unencrypted communication, whereas the other schemes require additional memory for intermediate states and processing overhead. Similarly, in the case of RF communication, MAVShield maintains an efficient memory footprint of $725.05$~MB as shown in Figure \ref{fig:RF_RAM_orin2}, which is again close to that of unencrypted communication, and outperforms all the other cryptographic approaches.

\begin{figure}[hbt]
    \centering
    \includegraphics[width=0.95\linewidth]{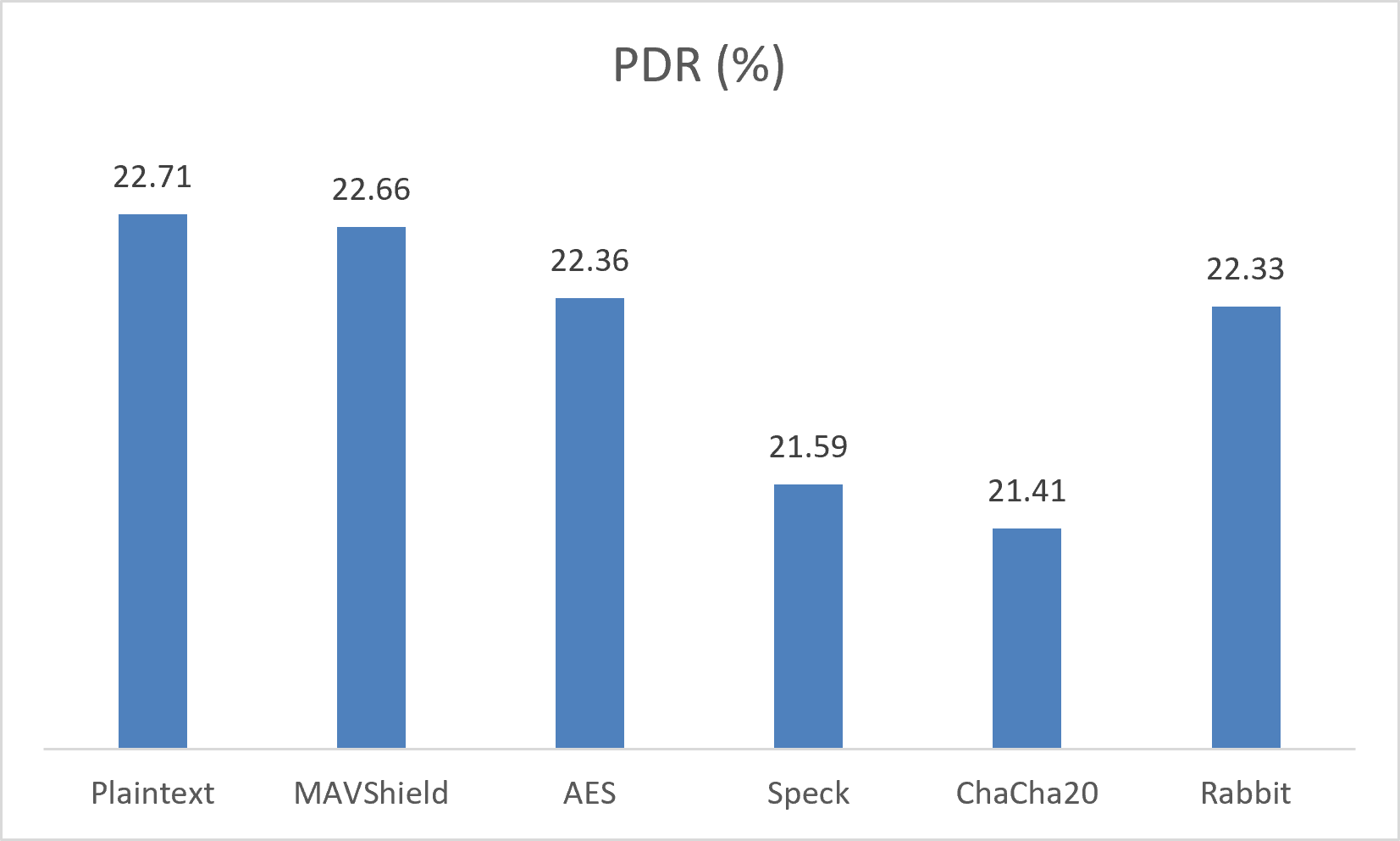}
    \caption{The figure shows the PDR ($\%$) across the WiFi-based mesh network among all UAVs.}
    \label{fig:WiFi_meanPDR_all}
\end{figure}

\begin{figure}[hbt]
    \centering
    \includegraphics[width=0.95\linewidth]{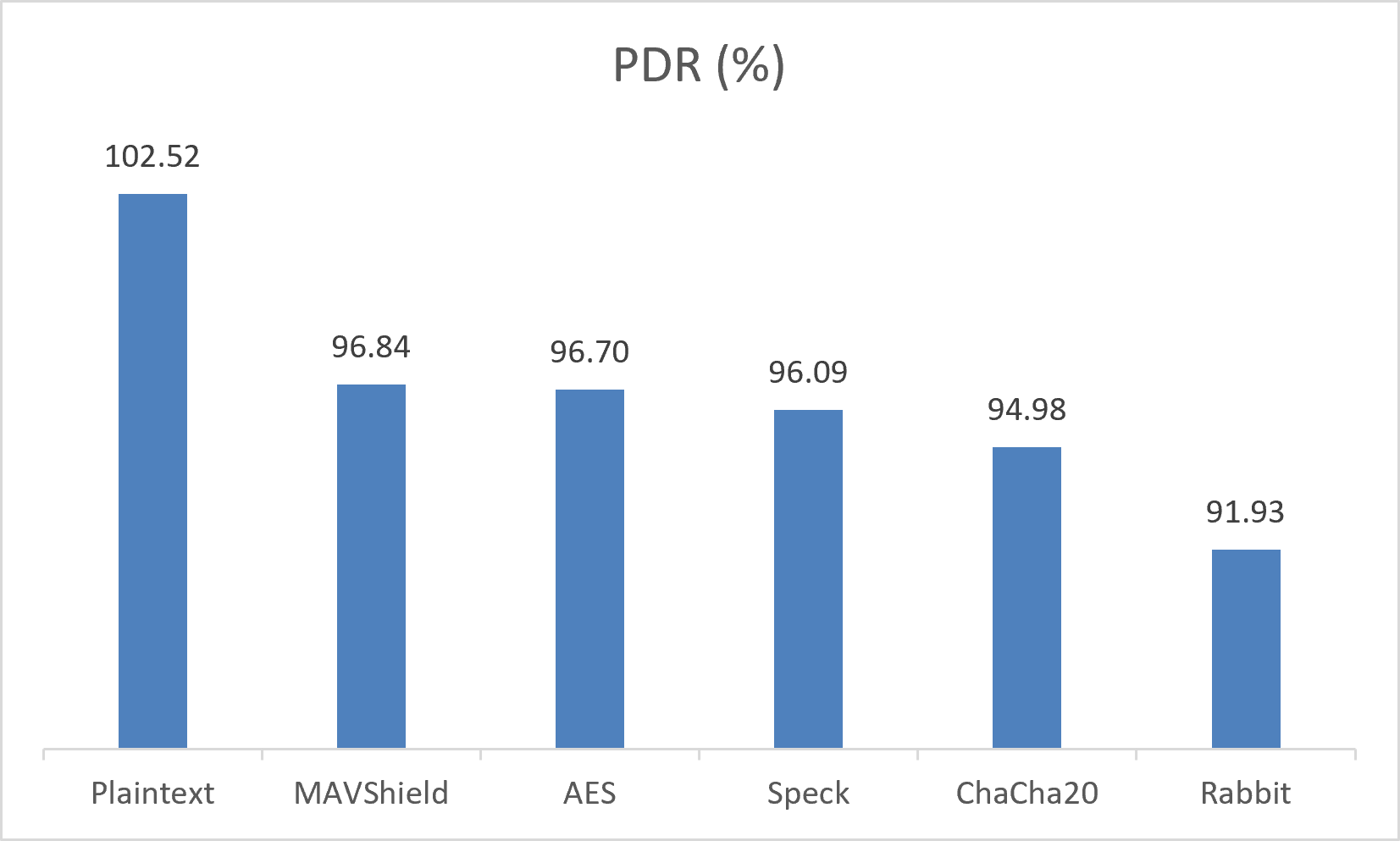}
    \caption{The figure shows the PDR ($\%$) across the RF-based mesh network for all UAVs.}
    \label{fig:RF_maxPDR_all}
\end{figure}

\paragraph{PDR}
It is used to evaluate the effectiveness and reliability of data dissemination within the UAV swarm network. The PDR for a given drone is calculated as the ratio of its actual received data rate to the total transmission rate of all other drones in the network. The transmit rate is the aggregate rate at which a node broadcasts encrypted telemetry to the rest of the swarm, whereas the receive rate represents the effective throughput at which a UAV ingests and validates incoming telemetry packets from its three peer nodes within the mesh. Consequently, PDR reflects the combined effects of transmission efficiency, packet losses, network congestion, and cryptographic overhead on successful packet delivery. Figures \ref{fig:WiFi_meanPDR_all} and \ref{fig:RF_maxPDR_all} illustrate the PDR ($\%$) for all UAVs under WiFi and RF communication, respectively.

As illustrated in Figure \ref{fig:WiFi_meanPDR_all}, unencrypted communication achieves the highest PDR ($22.71\%$), since it involves no cryptographic processing. Among the secure approaches, MAVShield attains the highest PDR ($22.66\%$), indicating that it preserves network performance close to the baseline while still providing security. Note that the transmit rate under WiFi consistently exceeds the receive rate across all schemes, primarily due to network contention, packet collisions, and buffering limitations.

In terms of throughput, RF communication exhibits a higher receive rate than transmit rate, as each UAV aggregates data from multiple nodes while transmitting only its own stream, under strict bandwidth constraints. Figure \ref{fig:RF_maxPDR_all} presents the PDR across all UAVs, where unencrypted communication achieves the highest PDR ($102.52\%$) due to the absence of cryptographic overhead. Among the secure schemes, MAVShield attains a PDR of $96.84\%$, slightly outperforming the other algorithms and having minimal packet loss despite added security. This analysis confirms MAVShield's ability to maintain reliable communication performance.

\begin{table*}[h]
\centering
\caption{The table shows a comparative analysis of WiFi and RF Communication.}
\begin{tabular}{|l|p{7cm}|p{7cm}|}
\hline
\textbf{Metric} & \textbf{WiFi} & \textbf{RF} \\
\hline

\textbf{CPU Usage} & High (due to high rate) & Low (due to limited data rate) \\
\hline

\textbf{RAM Usage} & High (buffering) & Low (minimal buffering)\\
\hline

\textbf{PDR} & Low (due to collisions, router packet loss) & High (multi-node aggregation) \\
\hline

\textbf{Overall Behavior} & Resource-intensive, high throughput & Resource-efficient, bandwidth-constrained \\
\hline

\end{tabular}
\label{tab:wifi_rf_comparison}
\end{table*}

\paragraph{Comparison between WiFi and RF Communication}
WiFi imposes significantly higher computational demands due to increased throughput and concurrency, whereas RF operates under more constrained, yet stable conditions with lower resource utilization. Across both environments, MAVShield consistently achieves efficiency comparable to the unencrypted communication baseline in terms of CPU and RAM usage, while ensuring secure communication and maintaining the highest PDR. This establishes MAVShield as the best-performing algorithm among the evaluated schemes. Table \ref{tab:wifi_rf_comparison} presents a comparative analysis of WiFi and RF-based swarm communication.

\section{Conclusions and Future Work} \label{sec:conclusions}
In this work, five encryption algorithms-- MAVShield, AES-CTR, Speck-CTR, ChaCha20, and Rabbit-- were integrated into the communication architecture of four custom built UAVs to establish robust and secure communication links for UAV-to-UAV interactions over both RF and WiFi channels. The performance of the encryption algorithms was evaluated by conducting extensive flight tests using a UAV swarm hardware testbed. The encrypted telemetry data was utilized to achieve autonomous formation control and collision avoidance during flight operations conducted using the testbed. Our proposed geodetic APF-based collision avoidance framework eliminates the coordinate transformations required by conventional APF methods and enables smooth, proactive collision avoidance in real UAV swarm operations without local minimum trapping in the evaluated scenarios. Key performance metrics such as CPU utilization, memory consumption, and PDR were measured for each encryption scheme.  Our results show that despite incorporating additional security mechanisms, MAVShield maintains performance levels comparable to unencrypted communication, while outperforming the other four encryption schemes in terms of overall efficiency. Furthermore, algebraic cryptanalysis and Wireshark-based traffic analysis confirmed its strong resistance to key recovery attacks and its ability to preserve telemetry confidentiality. Overall, the results indicate that MAVShield is an efficient and secure solution for UAV swarm communication. Some directions for future work include scaling the UAV swarm to larger sizes and integrating an IDS to enhance resilience against active cyber threats.

\bibliographystyle{IEEEtran} \bibliography{ref}
\end{document}